\documentclass[11pt,twoside,english,brazil, timesnewroman]{report}
\usepackage[T1]{fontenc}
\usepackage[latin9]{inputenc}
\usepackage{etex}
\usepackage{varioref}
\usepackage{geometry}
\usepackage{caption}
\usepackage{fancyhdr}
\usepackage{babel}
\usepackage{calc}
\usepackage{setspace}
\usepackage{lipsum}
\usepackage[unicode=true,pdfusetitle,
 bookmarks=true,bookmarksnumbered=false,bookmarksopen=false, breaklinks=false,pdfborder={0 0 1},backref=false,colorlinks=false]
 {hyperref}
\makeatletter
\usepackage{nextpage}
\newcommand\myclearpage{\cleartooddpage[\thispagestyle{empty}]}
\usepackage[fit]{truncate}
\usepackage{hyperref}
\usepackage{ucs}
\usepackage{xcolor}
\usepackage{stmaryrd}
\usepackage[toc,page]{appendix}
\usepackage{pgfplots}
\usepackage{slashed}
\usepackage{cancel}
\usepackage{mathtools}
\usepackage{bbold}
\usepackage{kpfonts}
\usepackage[all]{xy}
\usepackage{harmony}
\usepackage{upgreek}
\usepackage{accents}
\usepackage{dsfont}
\usepackage{graphicx}
\newcommand{\cl}{C \kern -0.1em \ell}
\makeatother

\newcommand{\m}{\mu}
\newcommand{\n}{\nu}
\renewcommand{\r}{\rho}
\newcommand{\s}{\sigma}
\renewcommand{\d}{\delta}
\renewcommand{\a}{\alpha}
\newcommand{\Db}{\bullet}
\newcommand{\p}{\partial}
\newcommand{\vx}{\vec{x}}
\newcommand{\vy}{\vec{y}}
\newcommand{\ot}{\otimes}

\renewcommand{\b}{\beta}
\newcommand{\g}{\gamma}
\newcommand{\be}{\begin{equation}}
\newcommand{\ee}{\end{equation}}
\newcommand{\ba}{\begin{array}}
\newcommand{\ea}{\end{array}}
\newcommand{\f}{\frac}
\newcommand{\beq}{\begin{eqnarray}}
\newcommand{\eeq}{\end{eqnarray}}
\newcommand{\om}{\Omega}
\renewcommand{\r}{{\rightarrow}}
\newcommand{\la}{\label}
\newcommand{\vp}{\varphi}
\renewcommand{\D}{\Delta}
\renewcommand{\m}{\mu}
\renewcommand{\n}{\nu}
\newcommand{\e}{\epsilon}
\renewcommand{\l}{\lambda}
\newcommand{\ci}{\circ}
\renewcommand{\G}{\Gamma}
\newcommand{\rf}{\lrcorner}
\newcommand{\w}{\wedge}

\begin{document}
\begin{center}
Kelvyn P\'aterson Sousa de Brito
\par\end{center}

\vspace{3cm}

\begin{center}
{\huge{} Espinores sobre o {\it bulk} e 
em dimensões compactificadas}
\par\end{center}{\huge \par}

\begin{center}
\vspace{4cm}
\par\end{center}

\begin{center}
\begin{minipage}[t]{0.6\columnwidth}%
\begin{center}
Tese apresentada ao Programa de Pós-Graduação em Física
da Universidade Federal do ABC (UFABC), como requisito parcial à obtenção
do título de Doutor em Física.
\par
\end{center}%
\end{minipage}
\par\end{center}

\begin{center}
\vspace{2cm}
Orientador: Prof. Dr. ROLDÃO DA ROCHA
\par\end{center}


\begin{center}
\vspace{3cm}
Santo André - SP
\par
\end{center}

\begin{center}
2017
\par
\end{center}

\begin{center}
{\large{}\thispagestyle{empty}}
\par\end{center}{\large \par}

\pagebreak{}
\par

\vfill{}

\begin{center}
\fbox{\begin{minipage}[t]{0.8\columnwidth}%
BRITO, Kelvyn P. S. de

\hspace{1cm}ESPINORES NO ESPA\c CO DE MINKOWSKI E EM DIMENS\~OES COMPACTIFICADAS/ Kelvyn Páterson Sousa de Brito - Santo André,
Universidade Federal do ABC, 2017.

\vspace{0.5cm}

\hspace{1cm}56 fls. 28 cm\vspace{0.5cm}

\hspace{1cm}Orientador: ROLDÃO DA ROCHA\vspace{0.5cm}

\hspace{1cm} Tese (doutorado) - Universidade
Federal do ABC, Programa de Pós-Graduação em Física, 2017\vspace{0.5cm}

\hspace{1cm}1. Álgebra de Clifford 2. Espinores 3. $S^7$ 4. $AdS_5$ 5.  {\it bulk}.
I. BRITO, Kelvyn Páterson Sousa de. II. Programa de Pós-Graduação em Física,
2017. III. Espinores no Espaço de Minkowski e em Dimensões compactificadas.%
\end{minipage}}
\par\end{center}

\thispagestyle{empty}
\par

\pagebreak{}
\par

\begin{minipage}[t]{0.9\columnwidth}%
\begin{center}
\par
\textbf{\large{Agradecimentos}\rule[0.5ex]{1\columnwidth}{1pt}}
\par\end{center}{\large \par}
Sou grato pelo sopro de vida e fluido vital que dá vida a tudo no Universo! Sou extremamente grato a todos as pessoas que passaram pela minha vida e me influenciaram positivamente. Principalmente, todos os professores inspiradores que tive e que me ajudaram muito, entre eles não posso esquecer a Suely Monteiro, Geane, Adelaide, Adriana Paula, Adriana Carvalho, Newton Santos, Roger Moura e Bruto Pimentel!

Agradeço a meu pai e minha mãe por estarem presentes em todas as situações, mesmo estando tão distantes fisicamente. 
Por terem aceitado cada vez mais o que sonho para minha vida.

Muito grato, em especial, à Andréia (Déia) pelo amor, por todos os momentos e pelos sonhos que compartilhamos, todos foram especiais e únicos! Também, pelos momentos que estudamos juntos e por tudo o que aprendemos um com o outro, em muitos aspectos. 

Agradeço muitíssimo ao Roldão, por ter me aceito como seu orientando, por ter me motivado sempre a lutar pelo que sonho, por sua paciência, por sua amizade e companheirismo, e por sempre ter oferecido ajuda em todas as situações que precisei. 

Sou também muito grato a todos os meus amigos do GEB! Agradeço principalmente ao Paulo, Giovane, Lucas da Lê, Kaio, Vini e Asaph por suas amizades! Que também foram muito importantes no meu crescimento pessoal! Foram os amigos, com quem pude realmente contar! Muito obrigado! 

Grato a Capes e a UFABC pelo apoio financeiro, sem isso esse trabalho não teria sido possível!

Gratos também a todos que me ajudaram de alguma forma para a realização desse trabalho!

\vspace{2cm}

\centering{AOS MEUS PAIS E MEUS IRMÃOS.}
\begin{doublespace}
\end{doublespace}

\rule[0.5ex]{1\columnwidth}{1pt}%
\end{minipage}

\thispagestyle{empty}
\par

\myclearpage
\par

\rule[0.5ex]{1\columnwidth}{1pt}
\par \par

\begin{center}
\textbf{\large{Resumo}}
\par
\end{center}{\par}
\vspace{0.5cm}

\noindent Este trabalho tem como principal objetivo explorar a natureza de campos fermiônicos, por meio de uma classificação de campos espinoriais sobre espaços físicos de interesse, tais como o {\it bulk} e o espaço compactificado $S^7$ das teorias de supergravidade. Essa classificação é de extrema utilidade na exploração e na busca por novos tipos de partículas, cujos observáveis correspondem a bilineares covariantes construídos com seus respectivos campos espinoriais. Dessa forma, esses campos são classificados em classes, através de vínculos obtidos por meio das identidades de Fierz-Pauli-Kofink e simetrias dos bilineares que constituem os mensuráveis dos campos fermiônicos. Para que esse estudo pudesse ser feito, revisamos as \'algebras de Clifford, as no\c c\~oes de geometria Riemanniana e a classifica\c c\~ao de espinores sobre o espa\c co de Minkowski segundo Lounesto. Dessa maneira, encontramos os termos da Lagrangiana para campos espinoriais livres sobre espaço de sete dimensões compactificadas em supergravidade e, como aplicação de nossa classificação de campos espinoriais sobre o {\it bulk}, associamos os invariantes sobre buracos negros axissimétrico com as componentes dos bilineares covariantes do campo fermiônico correspondente. Além disso, encontramos que um tipo de espinor denominado {\it flag-dipole} não gera brana, mas degenera-se em outro tipo de espinor chamado de {\it dipole} e, em teorias de compactificação, obtivemos a expressão para o campo quântico sobre espaços compactificados em baixas energias em termos de operadores de criação e aniquilação. 

\textbf{\Large{}\vspace{0.5cm}
}{\Large \par}

\textbf{Palavras chave: Álgebra de Clifford, Espinores, ${\bf S^7}$, $\bf {AdS_5}$, {\it bulk}.}

\thispagestyle{empty}
\par \par


\myclearpage
\par

\begin{center}
\textbf{\large{}\rule[0.5ex]{1\columnwidth}{1pt}}
\par\end{center}{\large \par}

\selectlanguage{english}

\begin{center}
\textbf{\Large{}Abstract}
\par\end{center}\par

\vspace{0.5cm}

\noindent This work has as the main aim to explore the nature of the fermionic fields, through a classification of spinor fields about physical space of interest, such as the bulk and the compactified space $S^7$ from the supergravity theories. This classification has an extreme utility in the exploration and in the search for new type of particles, whose observables correspond to bilinear covariants constructed upon their respective spinor fields. In this way, these fields are classified into classes, through constraints obtained by means of the Fierz-Pauli-Kofink identities and the bilinear symmetries that constitutes the mensurables of the fermionic fields. In order that study could be done, we revised the Clifford algebras, the notions of Riemannian geometry and the classification of spinors on the Minkowski space according to Lounesto. We found the Lagrangian terms for the free spinor fields on spaces of seven compactified dimension in supergravity and, as application of our spinor fields classification on the bulk, we associate the invariants on axisymmetric black holes to the components of the bilinear covariants  of the correspondent fermionic field. Besides, we found that a type of spinor denominated {\it flag-dipole} does not generate the brane, but degenerate itself in other type of spinor called {\it dipole} and, in compactification theories, we obtained the expression to the quantum field on compactified spaces for low energies in terms of creation and annihilation operators. 


\vspace{0.5cm}


\selectlanguage{brazil}

\thispagestyle{empty}
\par \par

\myclearpage
\par
\clearpage
\thispagestyle{empty}
\begin{titlepage}
\thispagestyle{empty}
\pagenumbering{gobble}
\tableofcontents
\end{titlepage}
\par\par
\thispagestyle{empty}
\par \par

\myclearpage
\par

\pagenumbering{arabic}
\setcounter{page}{15}
\chapter{Introdução}

O cerne deste trabalho é a classificação de espinores sobre espaços Riemannianos de sete dimensões, como por exemplo  a 7-esfera ($S^7$), e sobre espaços Lorentzianos de cinco dimensões, como por exemplo o espaço de anti-de Sitter de cinco dimensões ($AdS_5$). Dessa forma, essa classificação desempenhará um papel importante no estudo e exploração da dinâmica de campos de matéria em teorias de supergravidade. Por essa razão, iniciamos com uma contextualização histórica dessas teorias. 

No final do século XIX e início do século XX, entendia-se, de acordo com uma afirmação de William Thomson (Lord Kelvin), que toda a teoria física da época estava bem fundamentada, exceto por alguns poucos experimentos não explicados. 
Ele chegou a afirmar que a tarefa da física no século seguinte seria apenas
completar as casas decimais do que já havia sido feito. Com isso, sugeriu que os estudantes não se dedicassem à física, pois só alguns detalhes pouco interessantes faltavam,
como, por exemplo, medições mais precisas \cite{Kelvin}. No entanto, ele afirmou também que ainda haviam dois pequenos problemas a serem resolvidos: o resultado
do experimento de Michelson-Morley e o porquê da distribuição de energia
de um corpo negro \cite{Mich,Body}. Desses dois experimentos nasceriam duas áreas
que seriam amplamente estudadas no século seguinte, respectivamente a Relatividade
Restrita e a Mecânica Quântica, as quais mudariam completamente a nossa forma de ver o mundo.

As equações de Maxwell prevêem, a partir das constantes eletromagnéticas de vácuo, a natureza ondulatória da luz e sua velocidade de $\simeq 2,998\cdot 10^8 m/s$. Mas, a mecânica Newtoniana
nos diz que há um referencial privilegiado, onde a velocidade da luz seria
isotrópica. Assim, da mesma forma que as ondas mecânicas e o som necessitam de um meio elástico para se propagar, imaginava-se que o mesmo ocorreria com a luz. Esse meio,  denominado
éter, teria propriedades muito estranhas e permearia todo o espaço. Foi então proposto um experimento com interferômetro que o detectaria, caso ele existisse. Através desse experimento, Michelson e Morley mostraram que não é necessária a sua existência e que a velocidade da luz é constante no vácuo e independente do referencial. Dessa forma, diante do resultado do experimento, entre a mecânica
de Newton e o eletromagnetismo de Maxwell, foi necessário que Einstein escolhesse a segunda opção, apesar de a mecânica clássica de Newton já estar bastante enraizada, principalmente por recorrer bastante ao senso comum. Então, Einstein postulou que as leis físicas são as mesmas para todos os referenciais inerciais, entre elas as equações de Maxwell, e que a velocidade da luz é constante e aproximadamente $2,998\cdot 10^8 m/s$ em todos os referenciais inerciais. A partir desses dois postulados, ele deduziu toda a sua teoria da relatividade restrita, que discorda da nossa compreensão cotidiana de espaço e tempo \cite{Einstein}.

O outro problema também requeria uma nova forma de pensar. Foi o que fez Max Planck ao admitir, para explicar o espectro de energia de radiação de corpo negro, que a luz se propaga em {\it quanta} de energia, denominados fótons, os quais possuem energia proporcional à sua frequência $E=2\pi \hbar\nu$, onde $\hbar\approx 1,05\cdot 10^{-34} J\cdot s$ é a constante de Planck reduzida, também denominada constante de Dirac \cite{Body}. Da quantização da energia, nasceu a mecânica quântica, uma nova área de pesquisa que viria a ser bastante explorada no século XX. 
Outro fenômeno que só foi explicado pela mecânica quântica é a estabilidade das órbitas eletrônicas no átomo, pois de acordo com o eletromagnetismo clássico, elétrons acelerados emitiriam radiação, perdendo energia e isso colapsaria o átomo. Para explicar por que os elétrons descreveriam órbitas estáveis, em 1913, Niels Bohr 
postulou que elétrons localizam-se em níveis atômicos discretos bem definidos \cite{Bohr}. Dessa forma, a radiação que o elétron pode absorver ou emitir também assume valores discretos de energia. Essas ideias foram o início do que se tornaria a mecânica quântica, ramo da física que descreve fenômenos em escalas microscópicas, onde todas as noções de fenômeno clássico e cotidiano se perdem e emergem experimentos que contrariam o senso comum do ponto de vista macroscópico. Através dessa nova área do conhecimento, os fenômenos que antes eram obscuros foram muito bem explicados, como o efeito fotoelétrico, as linhas de Fraunhofer, o calor específico de sólidos, o espectro atômico descontínuo e o experimento de Stern-Gerlach, dentre outros.

 Essas duas novas teorias abriram novos caminhos para uma  gama de experimentos que surgiam, vindo novos desafios pela frente. Como descrever fenômenos microscópicos de altas energias, se isso implica em velocidades da ordem da velocidade da luz? A mecânica quântica usual já não bastaria para descrever esses experimentos? Do processo de quantização de energia das equações da teoria da relatividade, análogos ao seguido para obter a equação de Schrödinger, surgiram a equação de Klein-Gordon, que descreve campos escalares, e a equação de Dirac, que é uma linearização da equação de Klein-Gordon e descreve campos de matéria, consolidando a utilidade do formalismo de espinores. Desses desdobramentos, surgiu outra área da física que se desenvolveu bastante na segunda metade do século XX, denominada teoria quântica de campos, que aparece da tentativa de mesclar a relatividade restrita com a mecânica quântica \cite{Pesk}.

\begin{tabular}{|c|c|c|} \hline
\ &$\ell\approx 1m$& $\ell << 1m$\\ \hline
$v<<c$ &Mecânica Clássica & Mecânica Quântica\\ \hline
$v\approx c$&Relatividade& Teoria Quântica de Campos
\\ \hline
\end{tabular}
\\ {\hspace{4cm} \small Tabela 1. Àreas da física dentro dos seus limites de validade.}
\medbreak

 Como subárea da teoria quântica de campos, a eletrodinâmica quântica (QED-{\it Quantum Eletrodynamics}) foi uma das teorias de mais sucesso pelo seu grau de precisão nas previsões. Através de teorias de renormalização, pode-se fazer, então, uma unificação das forças eletromagnética e fraca, surgindo uma teoria bem sucedida, denominada teoria eletrofraca. Da mesma forma, a força forte pode ser bem estudada em altas energias, também através de processos de renormalização, o que deu origem à cromodinâmica quântica (QCD-{\it quantum cromodynamics}) \cite{Pesk}. Dessa maneira, todas as forças, exceto a força da gravidade, foram bem explicadas pelo modelo padrão com alta precisão para fenômenos de alta energia, através de renormalização. No entanto, uma teoria que surgiu no início para explicar melhor a força forte, conhecida como teoria de cordas, parece ser uma candidata a uma teoria quântica da gravidade. 
 
Kaluza e Klein conseguiram expressar as equações de relatividade geral de Einstein e do eletromagnetismo de Maxwell em uma só teoria. Foi uma das primeiras tentativas em se construir uma teoria unificada envolvendo a gravidade. A façanha foi conseguida adicionando-se uma dimensão compacta ao espaço-tempo ($\mathbb{R}^{1,3}\times S^1$). Provou-se que os modos excitados de uma partícula sem massa em cinco dimensões possuem massa quando a partícula é restrita ao espaço-tempo quadridimensional. Essa ideia pode ser facilmente estendida para outros tipos de compactificações. Dessa forma, uma teoria sem massa em $M^4\times M^n$, onde $M^4$ é o espaço-tempo e $M^n$ é um espaço compacto, adquire massa quando restrita ao espaço-tempo $M^4$, através dos modos excitados \cite{Gimb}. 
 
Em uma teoria quântica da gravidade, para que os valores divergentes devido a singularidade de um buraco negro seja contornado, as fontes de campo não são consideradas pontuais, como em teoria de campos usual, mas unidimensionais como uma corda. A Lagrangiana de uma corda bosônica é escrita em termos das coordenadas do cone de luz para que seja invariante por transformação de Poincaré e os modos normais de vibração sejam quantizados através do formalismo de operadores, os quais são denotados por $\a^\m_n$. 
O primeiro modo de vibração da corda bosônica, o qual é o modo taquiônico, admite no mínimo 26 dimensões \footnote{Quando a corda é quantizada no cone de luz, ela não preserva a simetria de Lorentz. Assim, para se retirar essa anomalia, ou seja, para que os comutadores dos geradores de Lorentz $J^{\m\n}=\int d\s M^{\m\n}_\tau =x^\m p^\n-x^\n p^\m-i\sum_{n=1}^\infty\f{1}{n}(\a^\m_{-n}\a^\n_n-\a^\n_{-n}\a^\m_n)$ sejam preservados são necessárias 26 dimensões. De forma que, em uma etapa dos cálculos, os quais não entraremos em detalhes, pode ser utilizado o somatório de Ramanujan $\sum_{n=1}^\infty n=-\f{1}{12}$, no sentido do limite da função zeta de Riemann.} e somente com esse número de dimensões a teoria é conformalmente invariante \footnote{Invariância conforme é um difeomorfismo $f:M\rightarrow M$ que preserva a métrica a menos de um fator escalar $e^{-2\omega(x)}: M\rightarrow \mathbb{R}$, ou seja, para todo $x\in M,\;f^*g_{f(x)}=e^{-2\omega(x)}g_x$, onde $g_x$ denota a métrica no ponto $x\in M$ . Sobre o espaço-tempo, essa invariância  é uma extensão da invariância de Poincaré, a qual está relacionada às transformações de Weyl, que são transformações locais no tensor métrico, dadas por: $g_{\m\n}\mapsto e^{-2\omega(x)}g_{\m\n}$, as quais produzem classes de métricas, denominadas classes conformes \cite{Naka}.} e livre de anomalias. Então, para que a teoria também incorpore férmions, é necessário introduzir supersimetria \footnote{Supersimetria é dada pela álgebra de supersimetria sobre o superespaço, que é um espaço de representação dado pelos parâmetros $x^\m$ do espaço-tempo e pelos parâmetros $\theta^\a,\bar{\theta}^{\dot{\a}}$ que anticomutam entre si. Os geradores de supersimetria são operadores de criação [aniquilação] que adicionam [subtraem] $\f{1}{2}$ ao número de spin do campo, transformando férmions em bósons e vice-versa. Para mais detalhes, veja\cite{Wess-Bagger}.}, e é denominada teoria de supercordas \cite{Polch1}. 
    Dessa forma, o número de dimensões para que esta teoria não tenha carga central \footnote{Cargas centrais são operadores de simetria que comutam com todos os outros e formam, assim, uma sub-álgebra abeliana da álgebra dos operadores de simetria.} e seja admissível deve ser igual a dez, cuja compactificação mais conhecida é $AdS_5\times S^5$. Há um total de cinco teorias de supercordas, que estão relacionadas por dualidades e que são conectadas, tornando-se uma só teoria se levamos em conta naturalmente a constante de acoplamento ao ser adicionada mais uma dimensão ao espaço da teoria de supercordas. Além disso, em outro contexto, $AdS_5$ também é um dos possíveis {\it bulks} da teoria de branas e onde está a 3-brana que representa nosso universo
. 
Existem ainda objetos de ordem mais elevadas, que possuem $p$ dimensões, conhecidos como $p$-branas. Uma compactificação bastante conhecida para espaços de onze dimensões é $AdS_4\times S^7$, em que o espaço-tempo é um espaço de anti-de Sitter e as outras sete dimensões são compactificadas 
 \cite{Polch2}. No Cap. 2, resultados já começaram a ser apresentados. Escrevemos o campo quântico sobre espaços compactificados mais gerais em baixas energias, segundo uma extensão dos modos de Kaluza-Klein. Mostramos que {\it flag-dipoles} em branas espessas, por não gerar brana, degenera-se em {\it dipoles}, os quais foram localizados na brana.
 


\subsubsection*{De forma resumida, o que faremos \cite{Kragh}:}

\qquad Iniciaremos com uma revis\~ao sobre as \'algebras de Clifford, 
as quais s\~ao extens\~oes das \'algebras dos reais ($\mathbb{R}$), dos complexos ($\mathbb{C}$) e dos quat\'ernions ($\mathbb{H}$). 
Alguns conceitos e defini\c c\~oes de geometria semi-Riemanniana, que ser\~ao utilizados ao longo do texto, s\~ao introduzidos, tais como por exemplo: tor\c c\~ao, curvatura e fibrados e, em seguida, apresentamos a teoria física acerca das compactificações em supergravidade, sobre a qual está fundamentada nosso trabalho, para uma melhor compreensão e motivações deste. Como passo intermedi\'ario, no terceiro cap\'itulo, discorremos sobre a classifica\c c\~ao de espinores no espa\c co de Minkowski segundo os observ\'aveis definidos pelos bilineares covariantes, cuja motiva\c c\~ao \'e dada a seguir:

Os geradores da \'algebra de Clifford ${\cal C}\ell_{1,3}$ aparecem naturalmente de uma lineariza\c c\~ao da equa\c c\~ao de Klein-Gordon, feita por Dirac, em 1928,
\be
\left(-\f{1}{c^2}\p^2_t+\p_i\p^i\right)\psi=\f{m^2c^2}{\hbar^2}\psi
.\ee
 Normalizando as constantes ($c=1,\hbar=1$),
$$
(\square+m^2)\phi(x)=0
,$$
 de onde segue-se covari\^ancia da equa\c c\~ao resultante, a qual \'e denominada, por raz\~oes \'obvias, equa\c c\~ao de Dirac \cite{Ald,lou2}. Essa equa\c c\~ao \'e escrita em termos da \'algebra de Clifford ${\cal C}\ell_{1,3}$ e descreve a din\^amica do el\'etron, que \'e representado pelo espinor $\psi$:
 $$(i\hbar\g^\m\p_\m-mc)\psi(x)=0,$$
 com o qual podemos expressar alguns observáveis, dentre eles, a densidade de corrente
$$
{\bf J(x)}=\g_\m\underbrace{(\psi^\dagger(x)\g_0\g^\m\psi(x))}_{J^\m}
.$$
Logo em seguida, vínculos entre esses bilineares puderam ser encontrados. Tais v\'inculos s\~ao denominados identidades de Fierz e são \'uteis na resolu\c c\~ao de diversos problemas envolvendo a din\^amica de espinores. As identidades de Fierz também foram usadas por Lounesto para classificar espinores sobre o espa\c co-tempo de Minkowski. 
Além disso, Dirac calculou as equações de onda relativísticas em notação espinorial para partículas de spin $s>\f{1}{2}$, como consta em \cite{Dirac}.

No Cap. 4, previamente listamos todos os bilineares poss\'iveis em dimens\~oes arbitr\'arias. A seguir, construímos e analisamos uma classificação para espinores sobre espaços Riemannianos de sete dimensões de acordo com seus bilineares covariantes \cite{BBR}, a qual \'e uma generaliza\c c\~ao da classifica\c c\~ao de espinores em $1+3$ dimens\~oes espa\c co-temporais. Como caso particular, em supergravidade, campos espinoriais sobre a esfera $S^7$ podem ser classificados em classes e analisadas a dinâmica de cada uma dessas classes. No entanto, isto foge do escopo deste trabalho. 

Dessa forma, as simetrias dos bilineares covariantes e suas identidades de Fierz nos dão vínculos locais para os campos espinoriais. Tais vínculos nos dão as classes obtidas aqui para os campos espinoriais sobre espaços Riemannianos de sete dimensões e Lorentzianos de cinco dimensões. Além disso, integrais  construídas com os bilineares covariantes vindas de invariantes topológicos podem nos dar vínculos globais com os campos. Assim, esses vínculos locais e globais, juntamente com o teorema de Stiefel-Whitney, nos dão uma classificação completa de campos espinoriais sobre espaços arbitrários. No entanto, os cálculos envolvendo os vínculos topológicos estão fora do escopo deste trabalho. Todavia, são satisfeitas todas as condições necessárias para a existência de estruturas espinoriais sobre os espaços utilizados aqui \cite{Lawson}.

Na seção seguinte, trabalhamos uma classifica\c c\~ao de campos espinoriais para cinco dimens\~oes e assinatura $(1,4)$ \cite{BR}. Temos boas motiva\c c\~oes f\'isicas para construir essa classificação sobre o {\it bulk} de teoria de branas, pois nos permite propor novos campos, que podem ser candidatos a matéria escura e analisar suas dinâmicas por meio de uma análise de seus bilineares covariantes \cite{BR}. Fizemos isso, ao fim do Cap. 4, com buracos negro axissimétricos de cinco dimensões ao calcularmos os bilineares covariantes de vetores de Killing e mostrarmos que perfazem todos os invariantes possíveis que possam ser construídos com as componentes desse vetor.

No \'ultimo cap\'itulo, abordamos algumas perspectivas de temas de trabalho mais gerais a serem desenvolvidos posteriormente ao curso de doutorado. Dessa forma, s\~ao abordados ideias e questionamentos bastante gerais, que intentam serem motivadores para anos de pesquisa e uma alavanca para uma pesquisa frut\'ifera. 

%
Dessa forma, nas perspectivas, iniciamos uma tentativa de classifica\c c\~ao segundo bilineares covariantes para espinores escritos em segunda quantiza\c c\~ao sobre o espa\c co de Minkowski. Fizemos, assim, uma caracteriza\c c\~ao dos espinores de Dirac e Weyl em termos dos tais bilineares. Também almejamos estudar extensões octoniônicas sobre $S^7$, explorar uma classificação de espinores nessa abordagem, através do grupo de Lie $G_2$ e suas dinâmicas para cada classe a ser encontrada.

\myclearpage
\par

\chapter{Preliminares}

\section{\'Algebras de Clifford e Grupo Spin}

\qquad A partir do processo de contagem, surgiu o conceito de número. Esse conceito foi estendido cada vez mais  para satisfazer a novas necessidades e foram obtidas novas classes cada vez maiores de conjuntos numéricos. 
Partindo-se dos n\'umeros naturais ($\mathbb{N}$), passando pelos inteiros ($\mathbb{Z}$), racionais 
 ($\mathbb{Q}$) e chegando aos n\'umeros reais ($\mathbb{R}$), complexos ($\mathbb{C}$) e quat\'ernions ($\mathbb{H}$). Estes dois \'ultimos simplificaram bastantes c\'alculos sobre os espa\c co vetoriais $\mathbb{R}^2,\mathbb{R}^3$ e $\mathbb{R}^4$. De uma forma geral, as álgebras de Clifford, que definiremos mais adiante e que incluem as álgebras $\mathbb{R,C}$ e $\mathbb{H}$, permitiram explorar bastante a geometria de espaços bastante gerais.

Para que equa\c c\~oes c\'ubicas fossem resolvidas, foi preciso lidar com a raiz quadrada de n\'umeros negativos. Da\'i, surgiu a necessidade de introduzir um novo conjunto de n\'umeros mais geral, que cont\'em a reta real. Tal conjunto \'e denominado conjunto dos n\'umeros complexos $z\in \mathbb{C}$, onde $z$ pode ser escrito como $z=a+bi$;\quad $a,b\in \mathbb{R}$ e $i=\sqrt{-1}$. Surge, ent\~ao, a quest\~ao: há algum polinômio  com coeficientes complexos que possua alguma raiz que não seja um número complexo? Gauss, em 1799, respondeu-a ao propor o conhecido teorema fundamental da \'algebra, o qual tem como corolário que todas as raízes de todos os polinômios de coeficientes complexos são complexas \cite{Gauss}.

Em 1835, Hamilton, aos 30 anos,  descobriu como tratar os n\'umeros complexos como um par de n\'umeros reais e ficou intrigado com a rela\c c\~ao existente entre $\mathbb{C}$ e a geometria de $\mathbb{R}^2$, pois essa álgebra ($\mathbb{C}$) isomorfa como espaço vetorial a $\mathbb{R}^2$ satisfaz a seguinte regra de multiplicação
$$
(a,b)\cdot(c,d)=(ac-bd,ad+bc);\quad a,b,c,d\in \mathbb{R}
$$
e cont\'em todas as solu\c c\~oes de todas as equa\c c\~oes polinomiais. Há ainda uma operação, denominada conjuga\c c\~ao, que é definida como $\bar{z}=a-bi$, com a qual a norma em $\mathbb{R}^2$ pode ser indicada por um  simples produto:
$$
|z|^2=\bar{z}z=(a-ib)(a+ib)=a^2+b^2.
$$ A partir da relação de Euler $e^{i\theta}=\cos(\theta)+i {\rm sen}(\theta)$, a multiplicação complexa escrita em coordenadas polares $z=re^{i\theta}$ torna-se óbvia:
 $$
 (r_1 e^{i\theta_1})(r_2e^{i\theta_2})=(r_1r_2)(e^{i(\theta_1+\theta_2)}).
 $$
Além disso, no contexto das funções complexas, pode-se
mostrar que se $f(z)$ é uma função analítica, então ela também é conforme, ou seja, é uma transformação em $\mathbb{R}^2$ que preserva ângulos \cite{Churchill}.

 
 Hamilton explorou essa quest\~ao por muitos anos na busca por espaços de três dimensões com propriedades semelhantes a essas, que são válidas para a álgebra dos complexos. Buscou, assim, por uma \'algebra quadr\'atica com unidade e isomorfa a $\mathbb{R}^3$ como espa\c co vetorial. Por fim, mostrou que a mesma n\~ao existe. Como ilustração, considere a extensão mais direta dos complexos, suponha que exista uma tal álgebra com elementos do tipo $\a+i\b+j\g $ e com unidade, onde $i^2=j^2=-1$ e $\a,\b,\g\in \mathbb{R}$, então para que a álgebra seja fechada, devem existir $\a,\b,\g\in \mathbb{R}$, tal que
\beq
ij=\a+\b i+\g j\quad\stackrel{{\cdot i}}{\Rightarrow}\quad -j=\a i-\b+\g ij\quad\Rightarrow\quad ij=\f{\b}{\g} -\f{\a}{\g} i-\f{1}{\g}j\\ \to
\a=\f{\b}{\g},\quad\b=-\f{\a}{\g},\quad \g=-\f{1}{\g}\quad\Rightarrow \b=\a\g,\quad \a=-\b\g,\quad \g^2=-1,
\eeq
o que é inconsistente. Mas, se adicionarmos 
 $k:=ij$ à álgebra, ela torna-se uma álgebra fechada, a qual é denotada por $\mathbb{H}$ e denominada álgebra dos quatérnions, cujos elementos são escritos como $\mathbb{H}\ni q=a+bi+cj+dk$, onde $a,b,c,d\in\mathbb{R}$ \cite{lou2,Lawson}. 

  Conta-se que, em 16 de outubro de 1843, quando passava pelo Canal Royal a caminho de um encontro na {\it Royal Irish Academy}, em Dublin, ele teve uma compreensão para o seu problema e em um ato de vandalismo matem\'atico, escreveu na ponte de Brougham a seguinte express\~ao \cite{Voight}:
\be 
i^2=j^2=k^2=ijk=-1, 
\ee 
que s\~ao exatamente as rela\c c\~oes que determinam o produto da \'algebra dos quat\'ernions. Esse é o primeiro resultado importante que \'e construído em quatro dimens\~oes e \'e fortemente relacionada à estrutura do espa\c co-tempo, a qual tem assinatura 1+3. Os quat\'ernions foram bastante explorados por f\'isicos te\'oricos e diversos modelos puderam ser simplificados com o uso dos mesmos \cite{Baez}. Por exemplo, pode se notar que o produto quaterniônico de dois vetores ${\bf a, b}\in \mathbb{R}^3$ incorporam os produtos interno e vetorial dos mesmos, conforme é dado na seguinte expressão \cite{lou2}:
\be 
{\bf ab}={\bf a}\cdot{\bf b}+{\bf a}\times {\bf b}.
\ee
Além disso, tomando-se $a\in \mathbb{H}$, rotações em $\mathbb{R}^3$ podem ser escritas em termos de uma representação quaterniônica \cite{lou2}:
\beq
\nonumber\mathbb{R}^3&\rightarrow&\mathbb{R}^3 \\
\nonumber{\bf r}&\mapsto & a{\bf r}a^{-1}
\eeq

Aparece, então, uma questão mais geral: seria poss\'ivel construir uma \'algebra quadr\'atica de dimens\~ao arbitr\'aria com unidade, ou seja, dada uma base vetorial seria poss\'ivel construir uma \'algebra que induzisse um produto interno e, consequentemente uma geometria, como os n\'umeros complexos? Seria poss\'ivel construir uma tal \'algebra que tivesse 3,4,5,... dimens\~oes? Essa busca foi feita em dimens\~oes superiores, e as \'algebras poss\'iveis encontradas foram denominadas \'algebras de Clifford. Este ramo de pesquisa foi bastante explorado e se desenvolveu bastante no \'ultimo s\'eculo. 

Desses desenvolvimentos, alguns são relatados a seguir:

 ``Em 1929, a equa\c c\~ao de Dirac foi arranjada em termos de pares de quat\'ernions por Lanczos \cite{gsponer}. Mais tarde, Juvet e Sauter, em 1930, substitu\'iram essa representa\c c\~ao de espinores como colunas por matrizes quadradas, onde apenas a primeira coluna \'e diferente de zero \cite{lou2}. Apenas em 1947, Marcel Riesz considerou os espinores como elementos de um ideal m\'inimo a esquerda em uma \'algebra de Clifford. Depois disso, uma representa\c c\~ao quaterni\^onica da equa\c c\~oes de Dirac em termos de matrizes $2\times 2$ foi dada, em 1956, por G\"ursey'' (sic) \cite{bookroldao}. Além disso, mais recentemente, foi proposta uma formulação quaterniônica da mecânica quântica \cite{deMelo:2008pv}.
\subsection{
Defini\c c\~ao}
\qquad Surgiu assim a ideia de se colocar uma geometria sobre uma \'algebra, atrav\'es de uma rela\c c\~ao de simetriza\c c\~ao do produto da \'algebra, para se obter um produto interno, que define um tipo de geometria sobre o espaço $\mathbb{R}^n$. Como dissemos acima, essa busca por uma interpreta\c c\~ao geom\'etrica de vetores e tensores deu origem às t\~ao conhecidas \'algebras de Clifford. Assim, dados $u$ e $v$ em um espa\c co vetorial $V$, a \'algebra de Clifford ${\cal C}\ell_V$ correspondente \'e uma \'algebra com unidade e definida pelas rela\c c\~oes \cite{Levi}:
$$
u\circ v+v\circ u=-2 u\cdot v,
$$
onde $\circ$ indica o produto de Clifford, que será denotado apenas pela justaposição daqui em diante, e $\cdot$, o produto interno padr\~ao. De forma equivalente, dada uma base ortonormal $\{e_1,\ldots,e_n\}$ em $V$ com assinatura $(p,q);\;\;p+q=n$, segue-se que
\be
\left(\sum_{i=1}^n \a_ie_i\right)^2=\sum_{i=1}^p(\a_i)^2-\sum_{i=p+1}^m(\a_i)^2.
\ee
 Ou seja, a \'algebra de Clifford pode ser determinada pela seguinte regra de multiplicação \cite{Porteous}:
\beq
e_i^2&=&+1;\; \qquad i=1,\ldots ,p,\\
e_j^2&=&-1;\;\qquad j=p+1,\ldots ,n=p+q,\\
e_ie_j& =&-e_je_i;\qquad \text{para }i\neq j.
\eeq
 Dessa forma, as primeiras \'algebras de Clifford podem ser encontradas e s\~ao bastante \'uteis \cite{Porteous,Porteous1}:
\beq
{\cal C}\ell_{0,0}=\mathbb{R},&{\cal C}\ell_{0,1}=\mathbb{C},&{\cal C}\ell_{0,2}=\mathbb{H},\\
{\cal C}\ell_{1,0}=\mathbb{R}\oplus\mathbb{R},&{\cal C}\ell_{1,1}=\mathbb{R}(2),&{\cal C}\ell_{1,2}=\mathbb{C}(2),\\
{\cal C}\ell_{2,0}=\mathbb{R}(2),&{\cal C}\ell_{2,1}=\mathbb{R}(2)\oplus \mathbb{R}(2),&{\cal C}\ell_{2,2}=\mathbb{R}(4),
\eeq
onde $\mathbb{K}(n)$ denota o espaço vetorial das matrizes de ordem $n$, cujas entradas são elementos do corpo $\mathbb{K}$. Em seguida, a partir dos seguintes isomorfismos, todas as \'algebras de Clifford finitas são obtidas \cite{Lawson}:
\beq\label{iso1}
{\cal C}\ell_{n,0}\otimes {\cal C}\ell_{0,2}&\cong & {\cal C}\ell_{0,n+2}\\
\label{iso2}{\cal C}\ell_{0,n}\otimes {\cal C}\ell_{2,0}&\cong &{\cal C}\ell_{n+2,0}\\
\label{iso3}{\cal C}\ell_{p,q}\otimes {\cal C}\ell_{1,1}&\cong &{\cal C}\ell_{p+1,q+1}.
\eeq
{\bf Prova}: Provemos o isomorfismo (\ref{iso1}). Seja o espaço vetorial $\mathbb{R}^{n+2}$ com uma base ortonormal $\{e_1,\ldots,e_{n+2}\}$ e considere sua álgebra de Clifford associada ${\cal C}\ell_{0,n+2}$. Considere também a base de geradores  $\{e'_1,\ldots, e'_n\}$ da álgebra de Clifford ${\cal C}\ell_{n,0}$ e os geradores $\{e''_1,e''_2\}$ de ${\cal C}\ell_{0,2}$. Então a aplicação 
\beq
&e'_i\otimes e''_1e''_2\mapsto e_i\qquad &\text{ para }1\leq i\leq n,\\
&1\otimes e''_1\mapsto e_{n+1},&\\
&1\otimes e''_2\mapsto e_{n+2},&
\eeq
quando estendida linearmente às suas respectivas álgebras de Clifford, determina o isomorfismo desejado, pois 
\beq
(e'_i\otimes e''_1e''_2)^2&=&(e'_i)^2\otimes (e''_1e''_2)^2=(+1)\otimes (-1)=-1=(e_i)^2;\qquad 1\leq i\leq n,\\
(1\otimes e''_j)^2&=&1\otimes(e''_j)^2=-1=(e_{n+j})^2;\qquad\qquad\qquad\qquad\; j=1,2.
\eeq
O isomorfismo (\ref{iso2}) é construído de forma análoga. 
Provemos o terceiro isomorfismo. Para isso, considere os geradores $\{e_1,\ldots,e_{p+1}, \e_1,\ldots,\e_{q+1}\}$ da álgebra de Clifford ${\cal C}\ell_{p+1,q+1}$, os geradores $\{e'_1,\ldots,e'_p,\e'_1,\ldots,\e'_q\}$ da álgebra ${\cal C}\ell_{p,q}$ e os geradores $\{e''_1,\e''_1\}$ de ${\cal C}\ell_{1,1}$. Se a seguinte aplicação é definida:
\beq
e'_i\otimes e''_1\e''_1 &\mapsto & e_i\qquad \text{ para }1\leq i \leq p,\\
1\otimes e''_1&\mapsto & e_{p+1}\\
\e'_j\otimes e''_1\e''_1 &\mapsto &\e_j\qquad \text{ para }1\leq j\leq q,\\ 1\otimes \e''_1 &\mapsto & \e_{q+1},
\eeq e estendida à sua álgebra de Clifford, o isomorfismo $(\ref{iso3})$ pode ser mostrado.

\newpage
A partir desses resultados, todas as álgebras de Clifford podem ser organizadas na seguinte tabela:

\begin{center}
\begin{tabular}{||c||c|c|c|c||} \hline \hline
$\begin{matrix} p-q \\ \mod \, 8 \end{matrix}$ & 0 & 1 & 2
& 3 \\ \hline
\fbox{$ \begin{matrix}   {\cal C}\ell_{p,q}   \end{matrix}$}  &
$\mathbb{R}(2^{[n/2]})$ &
$\begin{matrix} \mathbb{R}(2^{[n/2]}) \\ \oplus \\
\mathbb{R}(2^{[n/2]}) \end{matrix}$&
$\mathbb{R}(2^{[n/2]})$&
$\mathbb{C}(2^{[n/2]})$ \\ \hline
$\begin{matrix}p-q \\ \mod \, 8\end{matrix}$ & 4
& 5 & 6 & 7 \\ \hline
 \fbox{$ \begin{matrix}   {\cal C}\ell_{p,q}   \end{matrix}$}   &
$\mathbb{H}(2^{[n/2]-1})$&
$\begin{matrix} \mathbb{H}(2^{[n/2]-1}) \\ \oplus \\
\mathbb{H}(2^{[n/2]-1})\end{matrix}$ &
$\mathbb{H}(2^{[n/2]-1})$&
$\mathbb{C}(2^{[n/2]})$ \\ \hline\hline
\end{tabular}\\
\medskip
\captionof{table}{Representa\c c\~oes das Álgebras de Clifford -- Caso Real
($p + q = n$),}
\end{center}
onde $\mathbb{H}$ denota o anel dos quat\'ernions.
As representa\c c\~oes para o caso complexo s\~ao mais simples:
\medbreak
 \begin{center}
\begin{tabular}{||c|c||} \hline \hline  \fbox{$ \begin{matrix}   n = 2k
\end{matrix}$}  &$\mathbb{C}({2^k})$
\\ \hline
 \fbox{$ \begin{matrix}   n = 2k +1  \end{matrix}$ }   &
$\mathbb{C}(2^{k}) \oplus \mathbb{C}(2^{k})$
\\ \hline\hline
\end{tabular}\\
\medskip
\captionof{table}{Representa\c c\~oes das Álgebras de Clifford -- Caso Complexo}
\end{center}

Em termos mais formais, a álgebra de Clifford é construída a partir da álgebra tensorial, como uma álgebra quociente:
\be
{\cal C}\ell_V=\f{\bigotimes V}{I},
\ee
onde $I=\{x\otimes x-\langle x,x\rangle \in\bigotimes V; \;x\in V\}$ é um ideal. É interessante notar que há uma graduação sobre a álgebra de Clifford, que a faz isomorfa a álgebra exterior (${\cal C}\ell_V\approx \Lambda^*(V)$)  \cite{Lawson} e que nos permite definir dois anti-automorfismos: a reversão ($x^t$), dada diretamente pela reversão da ordem de todos os fatores
$$
(\;)^t:x_1x_2\ldots x_k\;\mapsto\; x_k\ldots x_2x_1,\;\text{ onde } x_1,\ldots, x_k\in V
$$
e a conjugação ($\bar{x}$), que é a composição de um automorfismo $\a$ que satisfaz $\a(v)=-v$ para todo $v\in V$, e da reversão $(\; )^t$, como se segue: $\bar{x}=\a(x^t)=\a(x)^t$ \cite{lou2}. A definição de norma pode ser estendida e dada por $N(x)=\langle x\bar{x}\rangle_0,$ onde $\langle\;\;\rangle_0$ denota a parte escalar.

 Como veremos, as álgebras de Clifford são bastante importantes para visualizações e compreensões no contexto das transformações ortogonais.
\subsection{Transformações Ortogonais}

\qquad As álgebras de Clifford nos permitem implementar reflexões e rotações em $\mathbb{R}^n$
, pois dado o conjunto de vetores ortonormais $\{e_i\in \mathbb{R}^n$
$;\quad (e_i)^2=-1\}$ e o hiperplano $H_i$, que por definição é o subespaço de dimensão $n-1$ ortogonal ao vetor $e_i$, temos que
\be
\varphi_{e_i}(e_j)=e_ie_je_i=\left\{\begin{tabular}{c}
$-(e_i)^2e_j= e_j;\qquad {\rm se}\; j\neq i$, \\
$(e_i)^2e_j= -e_j;\qquad {\rm se}\; j=i$
\end{tabular}\right.
\ee
descreve a reflexão do vetor $e_j$ em relação a $H_i$. Se estendida linearmente a $\mathbb{R}^n$, ela pode ser escrita como:
\beq\label{refl}
\varphi: \mathbb{R}^n\times \mathbb{R}^n&\r& \mathbb{R}^n\\
        (u,x)&\mapsto& \varphi_u(x)=-uxu^{-1}=x-2\f{\langle u, x\rangle}{u^2}u .
\eeq
 O teorema de Dieudonné-Cartan nos diz que qualquer transformação ortogonal em $\mathbb{R}^{p,q}$ pode ser escrita como uma composição de no máximo $p+q$ reflexões. Isso nos permite escrever qualquer transformação ortogonal a partir de reflexões expressas na eq. (\ref{refl}).
 Ao trabalharmos com as álgebras de Clifford, esse tratamento a respeito das rotações e transformações pode ser formalizado e estendido, como é mostrado a seguir. 
 
 Dessa forma, seja o automorfismo 
 $
 \a:{\cal C}\ell_V\r{\cal C}\ell_V
 $, tal que $\a(\a(s))=s $ e $\a(s)=-s$, onde se $s\in V$, o subgrupo $\Gamma_{p,q}$ de elementos invertíveis da álgebra de Clifford é dado por \cite{Lawson}
 $$
 \Gamma_{p,q}=\{s\in {\cal C}\ell_{p,q} |\; \a(s)xs^{-1}\in\mathbb{R}^{p,q},\;\forall x\in \mathbb{R}^{p,q}\},
$$ denominado grupo de Clifford. Uma representação adjunta contorcida de elementos da álgebra de Clifford pode ser construída através desse grupo e é definida como:
\beq
 \rho: \Gamma_{\mathbb{R}^{p,q}}&\r & Aut(\mathbb{R}^{p,q})\\ s&\mapsto & \rho_s:\mathbb{R}^{p,q}\r \mathbb{R}^{p,q}\\ 
 &&\qquad x\mapsto \rho_s(x)=\a(s)xs^{-1},
\eeq
com a qual podemos obter transformações lineares em $\mathbb{R}^{p,q}$ \cite{Lawson}. Em particular, o subgrupo de $\Gamma_{p,q}$, cuja representação adjunta contorcida é o grupo das transformações ortogonais $O(p,q)$, isto é, transformações que preservam a norma, é denominado grupo Pin e é definido, equivalentemente, por
$$
{\rm Pin}_{p,q}=\{x\in \Gamma_{p,q}| \quad N(x)=\bar{x}x=\pm 1\}=\{v_{1}v_{2}\dots v_{r}|\,\,\forall i\,\|v_{i}\|=\pm 1\}.
$$
Como pré-imagem do grupo $SO(p,q)$, obtemos o grupo Spin \cite{lou2,Porteous1}
\be 
\xymatrix{
{\rm Pin}_{p,q}\ar@/^/[r]^\rho
&O(p,q)\ar@/^/@{_>}[l]^{\rho^{-1}}\\
{\rm Spin}_{p,q}\ar[u]^i\ar[r]^\rho &{\rm SO}(p,q)\ar[u]^i}
,\ee
 ou seja, o grupo Spin é o subgrupo do grupo Pin, cujos elementos têm grau par:
 $$
{\rm Spin}_{p,q}={\rm Pin}_{p,q}\cap {\cal C}\ell^+_{p,q} \footnote{${\cal C}\ell^+_{p,q}$ denota a subálgebra de ${\cal C}\ell_{p,q}$, com elementos de grau par.}\,,
 $$ o qual também tem o subgrupo Spin$^+_{p,q}$, que é conexo para $p+q\geq 2$, com exceção de $p=q=1$:
$$ 
 {\rm Spin}_{p,q}^+=\{s\in {\rm Spin}_{p,q}\;|\;s\tilde{s}=1\}.
 $$
  O grupo Pin possui produtos de reflexões arbitrárias, enquanto que o grupo Spin engloba todos os produtos de rotações. Nota-se que os grupos Pin$_{p,q},{\rm Spin}_{p,q}$ e ${\rm Spin}^+_{p,q}$ são espaços de recobrimento duplo de ${\rm O}(p,q),{\rm SO}(p,q)$ e ${\rm SO}^+(p,q)$, respectivamente \cite{lou2}. Vejamos alguns exemplos.



\subsection{Primeiros exemplos de grupos ortogonais e espinoriais:}
\begin{enumerate}
\item {\bf $U(1)\simeq SO(2)$}

\qquad O caso mais simples e conhecido de transformações que preservam a norma é uma transformação unitária no plano complexo, representada através de uma multiplicação por um número complexo unitário 
\beq
\nonumber R_{\varphi}:\mathbb{C}&\r&\mathbb{C}\\
z=x+iy&\mapsto &e^{i\varphi}z=(\cos \varphi+i{\rm sen}\varphi)(x+iy),
\eeq onde $e^{-i\varphi}=\f{1-it}{1+it},\;\;t=\tan(\f{\varphi }{2})$. Isso é equivalente a uma rotação no plano $\mathbb{R}^2$, dada pela transformação: 
\be
\left(\begin{tabular}{c}
 $x$\\$y$
 \end{tabular}\right)\mapsto\left(\begin{tabular}{cc} $\cos(\varphi)$&-sen$(\varphi)$\\
 sen$(\varphi)$&$\cos(\varphi)$\end{tabular}\right)\left(\begin{tabular}{c}$x$\\$y$
 \end{tabular}\right),\qquad\left(\begin{tabular}{cc} $\cos(\varphi)$&-sen$(\varphi)$\\
 sen$(\varphi)$&$\cos(\varphi)$\end{tabular}\right)\equiv e^{ \mathbf{i}\varphi}\in SO(2),
 \ee
  onde usamos a equivalência:
\be 
\mathbf{1}\equiv\left(\begin{tabular}{cc}
$1$&$0$\\$0$&$1$
\end{tabular}\right),\qquad
\mathbf{i}\equiv\left(\begin{tabular}{cc}
$0$&-$1$\\
$1$&$0$
\end{tabular}\right)
.\ee Assim, nota-se o isomorfismo $U(1)\simeq SO(2)$. Observe que uma multiplicação pelo imaginário $i$ equivale a uma rotação de $\f{\pi}{2}$ no plano complexo. 

\qquad Seja o espaço vetorial $\mathbb{R}^2$, cuja norma é dada pela forma quadrática \cite{lou2}
 $(x{\bf e}_1+y{\bf e}_2)^2=x^2+y^2 $. Então,
 ${\bf e}_1^2={\bf e}_2^2=1$ e ${\bf e}_1{\bf e}_2=-{\bf e}_2{\bf e}_1$, tal que o elemento ${\bf e}_1{\bf e}_2$ não é nem escalar, nem vetor; mas um bivetor, pois satisfaz $({\bf e}_1{\bf e}_2)^2=-1$. Como resultado, temos a álgebra de Clifford ${\cal C}\ell_{2,0}$, cuja multiplicação de elementos de $\mathbb{R}^2$ é dada por:
\be\label{123}
(a_1{\bf e}_1+a_2{\bf e}_2)\,(b_1{\bf e}_1+b_2{\bf e}_2)=(a_1b_1+a_2b_2)+(a_1b_2-a_2b_1)e_{12},
\ee 
o que oferece uma abordagem geométrica bem interessante e ilustrativa, pois do lado direito da equação (\ref{123}), temos os produtos interno e vetorial, respectivamente. Extensões para dimensões arbitrárias surgem naturalmente. Se ${\bf x,y}\in \mathbb{R}^n$ e $u,v\in {\cal C}\ell_{n,0}$, então \cite{lou2}
\be
{\bf x}u={\bf x}\,\rfloor\,u +{\bf x}\wedge u, 
\ee
onde ${\bf x}\,\rfloor\,u$ é definido por:
\beq
{\bf x}\,\rfloor\,{\bf y}&=&\langle {\bf x},{\bf y}\rangle\;\;\;\qquad\qquad\qquad\quad\qquad\text{-produto interno},\\
{\bf x}\,\rfloor\,(u\wedge v)&=&({\bf x}\,\rfloor\,u)\wedge v+(-1)^{|u|}u\wedge({\bf x}\,\rfloor\,v)\; \;\text{-regra de Leibniz},
\eeq
e denominado produto interior, enquanto que ${\bf x}\wedge u$ é o produto exterior, que é associativo e distributivo, dado a partir de ${\bf x}\wedge {\bf x}=0,\;\;\forall {\bf x}\in \mathbb{R}^n$.

Desta forma, uma rotação no plano pode ser escrita ainda através da álgebra de Clifford ${\cal C}\ell_{2,0}$ \cite{lou2}
\be 
x{\bf e}_1+y{\bf e}_2\mapsto\left(\cos\f{\varphi}{2}+{\bf e}_{12}{\rm sen}\f{\varphi}{2}\right)^{-1}\left(x{\bf e}_1+y{\bf e}_2\right)\left(\cos\f{\varphi}{2}+{\bf e}_{12}{\rm sen}\f{\varphi}{2}\right),
\ee
a qual pode ser convenientemente escrita para abranger generalizações para espaços de dimensões maiores. Segue diretamente o isomorfismo $SO(2)/\mathbb{Z}_2\approx S^1$.

\item{$\bf SU(2)= Spin_{3,0}$} 
Elementos do grupo de rotações $SO(3)$ agem como $x'=R_{\theta,\phi,\psi}x$ e têm a forma \cite{lou2,Goldstein}:
{\small{\be 
R_{\theta,\phi, \psi}=\left(\begin{tabular}{ccc}$\cos\psi \,\cos \phi-\cos \theta\, {\rm sen} \phi \,{\rm sen} \psi$&$\cos\psi\, {\rm sen} \phi\,-\cos \theta\, \cos \phi \,{\rm sen} \psi$ & ${\rm sen} \psi\, {\rm sen} \theta$\\$
-{\rm sen} \psi \,\cos \phi-\cos \theta \,{\rm sen} \phi\, \cos \psi$ &$-{\rm sen} \psi\, {\rm sen} \phi +\cos \theta \,\cos \phi \,\cos \psi$& $\cos \psi \,{\rm sen} \theta$\\
${\rm sen} \theta \,{\rm sen} \phi$&$ - {\rm sen} \theta \,\cos \phi$ & $\cos \theta $
\end{tabular}\right),
\ee}}
onde $\theta, \phi$ e $\psi$ são os ângulos de Euler.
Essa transformação pode ser escrita ainda segundo a decomposição de Cartan 
\be 
R_{\theta,\phi,\psi}=e^{I_3\psi}\,e^{I_1\theta}\,e^{I_3\phi},
\ee 
onde 
\be 
e^{I_3\psi}=\left(\begin{tabular}{ccc}
$\cos(\psi)$&$\sin(\psi)$&$0$\\$-\sin(\psi)$&$\cos(\psi)$&$0$\\$0$&$0$&$1$\end{tabular}\right),\quad e^{I_1\theta}=\left(\begin{tabular}{ccc}
$1$&$0$&$0$\\$0$&$\cos(\theta)$&$\sin(\theta)$\\
$0$&$-\sin(\theta)$&$\cos(\theta)$\end{tabular}\right),
\ee
\begin{center}
\be
e^{I_3\phi}=\left(\begin{tabular}{ccc}
$\cos(\phi)$&$\sin(\phi)$&$0$\\$-\sin(\phi)$&$\cos(\phi)$&$0$\\
$0$&$0$&$1$\end{tabular}\right).\ee
\end{center}
Se representamos $p\in \mathbb{R}^3$ como matrizes complexas de ordem 2
\be 
p=\left(\begin{tabular}{cc}
$z$&$x-iy$\\$x+iy$&$z$
\end{tabular}\right),
\ee
 então pode ser encontrada uma transformação equivalente a $R_{\theta,\phi,\psi}$ nesse espaço, a qual é escrita como:
 \be 
Q_{\theta,\phi,\psi}=\exp\left(i\f{\psi}{2}\s_3\right)\,\exp\left(-\f{\theta}{2}\s_1\right)\,\exp\left(i\f{\phi}{2}\s_3\right),
 \ee
 onde $\s_i$ são as matrizes de Pauli:
\be 
\s_1=\left(\begin{tabular}{cc}
$0$&$1$\\$1$&$0$
\end{tabular}\right),\quad\s_2=\left(\begin{tabular}{cc}
$0$&$-i$\\$i$&$0$
\end{tabular}\right),\quad\s_3=\left(\begin{tabular}{cc}
$1$&$0$\\$0$&-$1$
\end{tabular}\right).
\ee
As transformações $Q_{\theta,\phi,\psi}$ são elementos de ${\rm SL}(2,\mathbb{C})$ que atuam segundo a expressão $p'=Q_{\theta,\phi,\psi}\,p\,Q_{\theta,\phi,\psi}^\dagger$. Dessa forma, é notável o seguinte isomorfismo SO$(3)\simeq {\rm SU}(2)/\mathbb{Z}_2$, onde 
$$
{\rm SU}(2)=\{s\in \mathbb{C}(2)\;|\; s^\dagger s=I, \,\det\, s=1\}.
$$ Outro isomorfismo segue imediato deste: ${\rm SU}(2)={\rm Spin}_{3,0}$. Além disso, se denotamos uma bola fechada no $\mathbb{R}^n$ centrada na origem de raio $r$ por $B_n[0,r]$, uma rotação em torno de um eixo $u\in B_3[0,\pi]$, cuja norma é usada para medir o ângulo de rotação, pode ser expressa em termos da álgebra ${\cal C}\ell_{3,0}$, como:
\be\label{rotRn}
r\mapsto \exp\left(\f{u}{2}\,e_{123}\right)\,r\,\exp\left(-\f{u}{2}\,e_{123}\right),
\ee
onde $\exp(ue_{123})=\cos(|u|)+e_{123}\f{u}{|u|}\text{sen}(|u|)$, \;$(ae_{123})^2=-|a|^2,\;u\in \mathbb{R}^3\subset \mathbb{H}$. 
Temos os seguintes isomorfismos ${\rm Spin}_{3,0}/{\rm Spin}_{2,0}={\rm SO}(3)/{\rm SO}(2)=S^2$ \cite{lou2}, onde
$$
{\rm Spin}_{3,0}=\{u\in{\cal C}\ell_3^+\;|\;u\tilde{u}=1\}.
$$ 

%

Podemos notar aqui, que se esse processo da eq. (\ref{rotRn}) é estendido para $u\in B_n[0,\pi]$, o seguinte isomorfismo mais geral  é obtido:
\be
 {\rm SO}(n+1)/{\rm SO}(n)=S^n.
\ee
 
\item ${\bf SO(1,3)}$ 
Transformações de Lorentz são descritas por elementos do grupo de Lorentz $${\rm O}(1,3)=\{L\in \mathbb{R}(4)\;|\; LgL^\intercal=g\}
,$$ os quais preservam a norma de vetores em $\mathbb{R}^{1,3}$, ou seja, têm determinante $\pm 1$. Este grupo possui quatro componentes desconexas, intercambiadas através da inversão espacial (paridade) e da reversão temporal. Os elementos com determinante $+1$ formam o grupo especial de Lorentz $SO(1,3)$, que por sua vez apresenta duas componentes intercambiáveis através da operação de paridade. Uma dessas componentes é o grupo de Lorentz ortócrono especial denotado por $SO_+(1,3)$, o qual preserva as orientações espaço-temporais \cite{lou2}.  

\qquad Quando a álgebra de Clifford é construída para o espaço-tempo de Minkowski, ela possui subgrupos que herdam uma hierarquia de componentes análogas às transformações ortogonais descritas acima. 
Se representadas na álgebra ${\cal C}\ell_{3,0}\simeq \mathbb{C}(2)$,
 a pré-imagem de $\rho$ restrita a $SO_+(3,1)$ é denotada por $\$pin_+(1,3)$
\be 
\$ pin^+_{1,3}= \{s\in {\cal C}\ell_{1,3}\;|\; s\tilde{s}= 1\}.
\ee
No entanto, se escrevemos um subgrupo com definição equivalente na álgebra ${\cal C}\ell_{3,1}\simeq \mathbb{R}(4)$, ele é denotado por ${\rm Spin}^+_{3,1}$ \cite{lou2}.
%
\end{enumerate}
\subsection{Espinores Clássicos de Dimensão Finita}
\qquad Denotamos por $S_{p,q}$ o espa\c co das representa\c c\~oes irredut\'iveis do grupo ${\rm Spin}^+_{p,q}=\{\xi\in {\cal C}\ell^+_{p,q}|\bar{\xi}\xi=1\; e\;\forall x\in \mathbb{R}^{p,q}, \a(\xi)x\xi^{-1}\in \mathbb{R}^{p,q} \} $ 
 conectadas com a unidade, o qual age sobre cada fibra do espaço de espinores cl\'assicos (fibrado espinorial) relacionados ao fibrado de Clifford sobre $M$. Através de algumas identidades muito importantes ${\cal C}\ell^+_{p,q}\simeq {\cal C}\ell^+_{q,p}\simeq{\cal C}\ell_{p,q-1}\simeq{\cal C}\ell_{q,p-1}$, obtemos a tabela abaixo: 
\begin{center}
\begin{tabular}{||c||c|c|c|c||} \hline \hline
$\begin{matrix} p-q \\ \mod \, 8 \end{matrix}$ & 0 & 1 & 2
& 3 \\ \hline
\fbox{$ \begin{matrix}   S_{p,q}   \end{matrix}$}  &
$\begin{matrix} \mathbb{R}^{2^{[(n-1)/2]}} \\ \oplus \\
\mathbb{R}^{2^{[(n-1)/2]}} \end{matrix}$&
$\mathbb{R}^{2^{[(n-1)/2]}}$&
$\mathbb{C}^{2^{[(n-1)/2]}}$&
$\mathbb{H}^{2^{[(n-1)/2]-1}}$ \\ \hline
$\begin{matrix}p-q \\ \mod \, 8\end{matrix}$ & 4
& 5 & 6 & 7 \\ \hline
 \fbox{$ \begin{matrix}   S_{p,q}   \end{matrix}$}   &
$\begin{matrix} \mathbb{H}^{2^{[(n-1)/2]-1}} \\ \oplus \\
\mathbb{H}^{2^{[(n-1)/2]-1}}\end{matrix}$ &
$\mathbb{H}^{2^{[(n-1)/2]-1}}$&
$\mathbb{C}^{2^{[(n-1)/2]}}$ &
$\mathbb{R}^{2^{[(n-1)/2]}}$ \\ \hline\hline
\end{tabular}\\
\medskip
\captionof{table}{Representa\c c\~oes Irredut\'iveis dos Espinores Cl\'assicos -- Caso Real
($p + q = n$),}
\end{center}
onde $\mathbb{H}$ denota o anel dos quat\'ernions.
As representa\c c\~oes para o caso complexo s\~ao mais simples:
\medbreak
 \begin{center}
\label{eq.4.40.1}
\begin{tabular}{||c|c||} \hline \hline  \fbox{$ \begin{matrix}   n = 2k
\end{matrix}$}  &
$\mathbb{C}^{2^{k-1}} \oplus \mathbb{C}^{2^{k-1}}$\\ \hline
 \fbox{$ \begin{matrix}   n = 2k +1  \end{matrix}$ }   &
$\mathbb{C}^{2^k}$
\\ \hline\hline
\end{tabular}\\
\medskip
\captionof{table}{Representa\c c\~oes Irredut\'iveis dos Espinores Cl\'assicos-- Caso Complexo} 
\end{center}\bigskip

Além do aspecto algébrico dos campos espinoriais, é necessário compreendermos também sua natureza geométrica, para que possa ser feita uma melhor exploração a respeito de sua classificação em espaços de dimensões arbitrárias. 
\section{Geometria Semi-Riemanniana}

\qquad As estruturas b\'asicas em uma teoria f\'isica s\~ao o espa\c co de fundo \footnote{Utilizamos a palavra ``fundo'' para se denotar {\it background}.} e os observ\'aveis medidos sobre este, respectivamente dados por uma `estrutura global' e uma `estrutura interna'. Uma estrutura desse tipo mais simples que podemos imaginar \'e um campo de temperaturas dinâmico dentro de uma sala, a qual corresponde ao espa\c co Euclidiano $\mathbb{R}^3$. Essas intuições a respeito do espaço físico são uma forma de incorporar o fenômeno dentro da geometria, a qual foi sendo construída ao longo da história.

Na Gr\'ecia antiga, para se construir as no\c c\~oes geom\'etricas faziam-se uso somente de r\'egua e compasso. Por volta de 300 a.C., por meio dos seus cinco postulados em sua obra {\it Os Elementos}, Euclides criou o que mais tarde veio a ser conhecido por geometria Euclidiana. Essa obra foi o principal livro de geometria por bastante tempo, durante o qual por mais de 2000 anos o quinto axioma de Euclides foi questionado \cite{Eucl}: {\it ``Se em um ponto de uma linha reta qualquer concorrerem
de partes opostas duas retas, fazendo com a primeira reta os
ângulos adjacentes iguais a dois retos, as retas, que concorrem
para o dito ponto, estarão em direitura uma da outra''} (p.14) \cite{Eucl1}, que pode ser parafraseado como: Duas retas não-coincidentes n\~ao podem se interceptar em mais de um ponto.
Desse questionamento, por volta do s\'eculo XVIII, surgiram, de forma mais s\'olida, geometrias mais gerais que a geometria plana de Euclides, denominadas geometrias n\~ao-Euclidianas, as quais s\~ao subdivididas ainda em
: geometria (semi-)Riemanniana, geometria de Cartan, geometria de Finsler, dentre outras. Estudaremos em mais detalhes apenas a geometria semi-Riemanniana, o qual é iniciado na subseção seguinte, pois nos será útil na construção dos fibrados espinoriais, necessários para o entendimento deste trabalho.

\subsection{Variedades Diferenci\'aveis}

\qquad Uma variedade $M$ \'e um espa\c co topol\'ogico $(M,\mathfrak{T})$ de Hausdorff que possui uma base $\{\mathfrak{T}_i\}$ cont\'avel, a qual \'e formada por abertos de um espa\c co Euclidiano de uma dada dimens\~ao $n$, ou seja, $M$ pode ser reconstru\'ido  por meio de uma união  de abertos \footnote{Em outras palavras, uma variedade (topol\'ogica) $n$-dimensional \'e um espa\c co topol\'ogico de Hausdorff segundo cont\'avel que \'e localmente Euclidiano e de $n$ dimens\~oes \cite{Lee}.} \cite{Levi}. Dado $p\in M$, h\'a um homeomorfismo $\phi$ entre uma vizinhan\c ca $V$ de $p\in M$ e um aberto do $\mathbb{R}^n$. Consequentemente, dadas duas cartas $\phi_1: U_1\rightarrow \mathbb{R}^n$ e $\phi_2: U_2\rightarrow \mathbb{R}^n$, h\'a um homeomorfismo (fun\c c\~ao $C^0$) entre os abertos $\phi_1(U_1\cap U_2)$ e $\phi_2(U_1\cap U_2)$ do $\mathbb{R}^n$. Isto nos diz que as cartas $\phi_1$ e $\phi_2$ est\~ao $C^0$-relacionadas \cite{Bishop}. 

Precisamos de uma cole\c c\~ao de cartas para cobrir toda a variedade $M$, a essa cole\c c\~ao chamamos de atlas. Para que uma dada carta $\phi$ seja admiss\'ivel em um $C^0$-atlas, \'e necess\'ario que ela esteja $C^0$-relacionada a todas as outras cartas do atlas. Este procedimento transfere propriedades da variedade $M$ para rela\c c\~oes entre abertos do $R^n$. Assim, uma dada variedade $M$ recebe o mesmo adjetivo que seus homeomorfismos em $\mathbb{R}^n$ correspondentes. O caso fundamental acima, em que os homeomorfismos entre os abertos do $\mathbb{R}^n$ caracterizam a variedade como topol\'ogica, pode ser estendido para outras caracteriza\c c\~oes de $M$ \cite{Bishop}. Dessa maneira, $C^k$ [$C^\infty$, anal\'itica, ...]-variedades  s\~ao definidas por atlas, cujas cartas admiss\'iveis s\~ao $C^k$ [$C^\infty$, anal\'itica, ...]-relacionadas entre si. Em outras palavras,
se queremos adicionar uma propriedade $P$ a uma variedade $M$,  substitu\'imos $C^0$, nas proposi\c c\~oes acima, por $P$ e a chamamos de $P$-variedade $M$. Enfim, uma variedade $M$ \'e diferenci\'avel de classe $C^k$ [$C^\infty$, anal\'itica, ...], se suas cartas admiss\'iveis s\~ao $C^k$ [$C^\infty$, anal\'itica, ...]-relacionadas entre si \cite{Bishop}.

Seja $M$ uma $C^\infty$-variedade (variedade suave). 
 A partir de curvas suaves sobre $M$, denotadas pela aplicação $c:(-1,1)\;\r\; M$, tais que dire\c c\~ao e velocidade s\~ao bem definidas em cada ponto $p\in c$, uma estrutura local linear pode ser adicionada. Para  isso, considere fun\c c\~oes suaves sobre $U\subset M$, 
\be 
f:U\rightarrow \mathbb{R},
\ee 
que podem ser usadas para se formalizar o conceito emp\'irico de campo de velocidades. Considere derivadas totais de $f$ ao longo de $c$ \footnote{H\'a um abuso de linguagem aqui, $\f{\p f }{\p x^\m }$ denota na verdade $\f{ \p(f\circ \phi^{-1}(x))}{\p x^\m}$.}:
\be 
\left.\f{df(c(t))}{dt}\right|_{t=0}=\underbrace{\left.\f{dx^\m(c(t))}{dt}\right|_{t=0}}_{X^\m}\f{\p f}{\p x^\m}=\left(X^\m\f{\p}{\p x^\m}\right)f\equiv X[f] ,
\ee 
Essa estrutura linear local $X=X^\m\p_\m$ (onde $\p_\m$ denota $\f{\p}{\p x^\m}$) \'e denominada vetor tangente a $M$ em $p=c(0)$. A partir de todas as curvas que passam por $p\in M$, obtemos o espa\c co tangente $T_pM$, que \'e tangente a $M$ em $p$. Com essa estrutura, $M$ pode ser analisado localmente nesse ponto $p\in M$. Outras estruturas lineares podem ser adicionadas ao espa\c co tangente $T_pM$
, tais como por exemplo: o espa\c co de 1-formas, a  \'algebra tensorial, a \'algebra exterior e a \'algebra de Clifford. Vejamos alguns casos.

O espa\c co vetorial dual a $T_pM$ \'e denominado espa\c co cotangente $T_p^*M$. Quando tratamos com formas diferenciais, covetores $\omega:T_pM\;\r\;\mathbb{R}$ pertencentes ao espa\c co dual $T_p^*M$  s\~ao decompostos na base natural $\{dx^\m\}$ como $\omega=\omega_\m dx^\m$ e denominados 1-formas. As duas bases, duais entre si, se relacionam por:  
\be 
\langle dx^\m ,\p_\n\rangle =\f{\p x^\m}{\p x^\n}=\d^\m_\n .
\ee
 Um produto interno pode ent\~ao ser definido trivialmente, como
\beq
\langle\;,\;\rangle:T^*_pM\times T_pM&\r& \mathbb{R}\\
(\omega, V)&\mapsto& \langle\omega,V\rangle =\omega_\m V^\m
.\eeq

Objetos multilineares constru\'idos a partir de $T_pM$ e $T_pM^*$ sobre cada ponto $p$ de $M$ s\~ao denominados $tensores$. Um tal elemento $T$, com $r$ entradas vetoriais e $s$ covetoriais, \'e dito ser um tensor do tipo $(r,s)$, expresso em componentes como
\be
T=T^{\m_1\ldots\m_r}_{\n_i\ldots\n_s}\p_{\m_1}\cdots\p_{\m_r}dx^{\n_i}\cdots dx^{\n_s},
\ee e mapeia elementos de $\otimes^rT^*_pM\otimes^sT_pM$ em $\mathbb{R}$
 \cite{Bishop, Naka, Rodrigues:2005yz}.

 Além disso, podemos munir  uma variedade diferenci\'avel $M$ de  uma estrutura de produto interno local (m\'etrica), a qual \'e uma esp\'ecie de generaliza\c c\~ao do produto interno de $R^n$. Essa variedade \'e dita ser semi-Riemanniana, se em cada ponto $p$ pertencente a $M$, temos um espa\c co tangente $T_pM$ munido de uma m\'etrica semi-Riemanniana. Essa m\'etrica \'e definida como uma forma  bilinear $g_p(\;,\;) $ sim\'etrica, n\~ao-degenerada, com assinatura $\e^\n$ 
constante sobre M, onde $\e^\n=+1$ para $1\leq \n\leq s$ e $\e^\n=-1$ para $s+1\leq \n\leq n$. Vetores tangentes $u\in T_pM$ s\~ao classificados, por meio dessa m\'etrica, como:
 
 \beq
&\text{tipo-espa\c co, se}  &g({u},{u})\,>\,0 \text{\; ou\; }u=0\\
&\text{nulo, se} &g({u},{u})\,=\,0 \text{ \;e\; }u\neq 0\\
&\text{tipo-tempo, se }& g(u,u)\,<\,0.
\eeq
Numa outra abordagem, a m\'etrica $g$ \'e um campo tensorial $g_p =g_{\m\n}(p)dx^\m\otimes dx^\n $ sobre $M$ do tipo $(0,2)$, denominado tensor m\'etrico. Considere um deslocamento infinitesimal $dx^\m\f{\p}{\p x^\m} \in T_pM,$ a quantidade que mede uma dist\^ancia infinitesimal sobre $M$,
\be
ds^2= g\left(dx^\m\p_\m,dx^\n\p_\n\right)=g_{\m\n}dx^\m dx^\n,
\ee \'e conhecida por m\'etrica. Se $t=0$ ou $n$, a variedade \'e dita ser Riemanniana. Se $t=1$ ou $n-1$, ela \'e Lorentziana
 \cite{Naka,Manf,Neil}.

Uma variedade diferenci\'avel $M$ e um espa\c co tangente em cada ponto $p\in M$ foram definidos e munidos de uma m\'etrica $g$. Com essas estruturas, somente vetores no mesmo espa\c co tangente podem ser comparados. Para que vetores em pontos diferentes de $M$ o possam tamb\'em, \'e necess\'aria a adi\c c\~ao de uma nova estrutura denominada conex\~ao.  Dados $p,q\in M$, \'e por meio dessa estrutura que dizemos como vetores em um dado $T_pM$ deve ser transladado para $T_qM$ ao longo de um caminho que une $p$ a $q$. Seja o conjunto $\Xi(M)$ de todos os campos vetoriais sobre $M$, uma conex\~ao \'e uma aplica\c c\~ao diferenciável $\nabla:\Xi(M)\times \Xi(M)\;\r\; \Xi(M)$. Para garantir a diferenciabilidade de campos vetoriais sobre $M$, precisamos garantir que a conex\~ao seja linear em suas entradas e satisfa\c ca a regra de Leibniz:
\beq
\nabla_{fX+Y}Z=f\nabla_XZ+\nabla_YZ\\
\nabla_X(Y+Z)=\nabla_XY+\nabla_XZ\\
\nabla_X(fY)=X[f]Y+f\nabla_XY\,.
\eeq
A essa conex\~ao denominamos conex\~ao afim \cite{Naka}.
Dados os campos vetoriais $X,Y,Z$, uma conex\~ao afim $\nabla$ tem as seguintes caracter\'isticas:

$\bullet$ n\~ao-metricidade: $\nabla_X g$ (compatibilidade com a m\'etrica),

$\bullet$ tor\c c\~ao: $T(X,Y)=\nabla_XY-\nabla_YX-[X,Y]$ e

$\bullet$ curvatura: $R(X,Y)Z=\nabla_X\nabla_YZ-\nabla_Y\nabla_XZ-\nabla_{[X,Y]}Z.$

É interessante escrevermos essas grandezas em termos de suas componentes, para isso considere uma carta $(U,\varphi)$ sobre $M$, cujas coordenadas s\~ao $x^\m=\varphi^\m(p)$. Os coeficientes da conex\~ao $\G^\rho_{\m\n}$ determinam o quanto uma base em $T_pM$ \'e alterada por um transporte paralelo infinitesimal:
\be
 \nabla_\m e_\n\equiv\nabla_{e_\m}e_\n=e_\rho\G^\rho_{\m\n}.
\ee

Seja  $(M,g)$ uma variedade semi-Riemanniana. Para que normas e produtos internos, sobre o espa\c co tangente $T_pM$ de cada ponto $p$ arbitr\'ario de $M$, fiquem invariantes sob transporte paralelo, \'e necess\'ario que a m\'etrica seja covariantemente constante, ou seja, que sua n\~ao-metricidade seja nula:
$$
(\nabla_\rho g)_{\m\n}=\p_\rho g_{\m\n}-\G^\l_{\rho\m}g_{\l\n}-\G^\l_{\rho\n}g_{\l\m}=0
.$$ 
Ent\~ao, essa conex\~ao afim $\nabla$, que \'e compat\'ivel com a m\'etrica, \'e tamb\'em chamada de conex\~ao m\'etrica. Se além disso, essa conexão, por motivos emp\'iricos, compara  vetores em espa\c cos tangentes infinitesimalmente separados, simplesmente projetando  um diretamente no outro, ela \'e conhecida também por conex\~ao de Levi-Civita. Em outras palavras, uma conex\~ao m\'etrica \'e de Levi-Civita, se dados $p,q\in M$, o paralelismo por meio dessa conexão entre $V_q$ em $q$ e $V_p$ em $p$ é dada através da comparação entre a proje\c c\~ao usual de $V_q$ em $T_pM$ e $V_p$.

Vejamos agora o conceito de torção. Tome dois pontos $q$ e $s$ infinitesimalmente pr\'oximos a $p\in M$, identifique $\hat{pq}$ e $\hat{ps}$ com vetores infinitesimais $\vec{pq}$ e $\vec{ps}$, respectivamente, em $T_pM$. Translade $\vec{pq}$ at\'e $s$ e $\vec{ps}$ at\'e $q$, o vetor deslocamento entre os pontos $q$ e $s$ deslocados medir\'a a tor\c c\~ao de $M$ em $p$.  
Se o paralelismo de vetores \'e  dada pela conex\~ao de Levi-Civita, obtem-se uma  tor\c c\~ao nula, isto \'e, os deslocamentos paralelos da defini\c c\~ao da tor\c c\~ao formam um paralelogramo \cite{Naka}. 
Nesse caso, a conex\~ao tem seus coeficientes sim\'etricos $\G^\rho_{\m\n}=\G^\rho_{\n\m}$ e pode-se mostrar que ela \'e \'unica, podendo ser escrita em termos do tensor m\'etrico $g_{\m\n}$ como:
\be
\G^\a_{\m\n}=\f{1}{2}g^{\a\b}\left( \p_\n g_{\b\m}+\p_\m g_{\b\n}-\p_\b g_{\m\n}\right).
\ee
Então, a curvatura sobre $M$ pode ser determinada também somente em termos do tensor m\'etrico $g$,
\be
R^\a_{\b\g\d}=\p_\g\G^\a_{\b\d}-\p_\d\G^\a_{\b\g}+\G^\a_{\g\m}\G^\m_{\b\d}-\G^\a_{\d\m}\G^\m_{\b\g}.
\ee
A partir da qual, definem-se o tensor de Ricci
\be 
R_{\m\n}:={R^\a}_{\m\a\n},
\ee
e a curvatura escalar
\be
R:=g^{\m\n}R_{\m\n}.
\ee

\subsubsection*{Espaços maximalmente simétricos}

\qquad Utilizando os conceitos revisados acima, podemos compreender melhor as simetrias de um dado espaço. Um espaço é denominado maximalmente simétrico se ele possui o mesmo número de simetrias que o espaço Euclidiano de mesmo número de dimensões, ou seja, $\f{1}{2}n(n+1)$ simetrias. Além disso, dizer que um espaço é maximalmente simétrico é equivalente a dizer que ele é homogêneo e isotrópico. O tensor de curvatura de Riemann sobre esse tipo de espaço pode ser simplificado e escrito como \cite{Naka} 
$$
R_{\m\n\rho\s}=\f{R}{n(n-1)}(g_{\m\rho}g_{\n\s}-g_{\m\s}g_{\n\rho}).
$$
Sendo assim, esses espaços são esquematizados na tabela abaixo:

{\centering
\label{my-label}{\centering{\begin{tabular}{|c||c|}\hline
&\begin{tabular}{c|c} \qquad Espaço\;\; \qquad{} & Grupo de Simetria\end{tabular}\\
\hline\hline
Lorentziano &\begin{tabular}{c|c}
\hline\qquad\qquad$dS_n$\qquad \qquad{} &$Spin_{n-1,1}$\\ \hline
Minkowski&Poincaré\\ \hline
$AdS_n$ & $Spin_{n-2,2}$\\ \hline
\end{tabular}\\ \hline 
Riemanniano&\begin{tabular}{c|c}
{}Esfera& $Spin_{n}$\\ \hline
$\mathbb{R}^n$&$SO(n)\rtimes T(n)$\\ \hline
Esp. Hiperbólico& $Spin_{n-1,1}$,
\end{tabular}\\ \hline \end{tabular}}}

\hspace{4cm}\small Tabela 4. Espaços maximalmente simétricos}

onde $\rtimes$ denota o produto semidireto e $T(n)$, o grupo de translações no $\mathbb{R}^n$.
Sobre uma variedade $M$ podem ser constru\'idos espa\c cos tangentes e opera\c c\~oes podem ser feitas com eles. Surgem ent\~ao novos objetos mais gerais, denominados fibrados, que s\~ao descritos na se\c c\~ao seguinte, como exemplo temos: o fibrado vetorial, o fibrado exterior e o fibrado de Clifford.

\subsection{Fibrados}

\qquad Considerando todos os espa\c cos tangentes $T_pM$ em cada ponto $p$ de $M$, temos o fibrado tangente $TM$, que \'e o fibrado mais intuitivo que conhecemos. Formalmente, dada uma ${\it (p+q)}$-variedade semi-Riemanniana $(M,g)$,
$$
TM:=\bigsqcup_{p\in M}T_pM=\bigcup_{p\in M}\{p\}\times T_p M
$$
 \'e chamado de fibrado tangente. A defini\c c\~ao de fibrado cotangente $T^*M$ \'e similar. A esses fibrados podemos adicionar uma estrutura que d\'a um nome mais espec\'ifico ao fibrado, como por exemplo: o fibrado exterior, o fibrado de Clifford e o fibrado espinorial.

Um fibrado de fibras $(E,\pi,M,F,G)$, que \'e o conceito mais geral de fibrados, \'e descrito por:
\begin{itemize}
\item $E$, uma variedade diferenci\'avel denominada espa\c co total,

\item $M$, uma variedade diferenci\'avel chamada de espa\c co base,

\item $F$, a variedade diferenci\'avel imagem inversa $\pi^{-1}(p)\cong F$ e chamada de fibra,

\item $\pi: E\;\r\; M$, uma aplica\c c\~ao sobrejetora, denominada proje\c c\~ao, e

\item $G$, um grupo de Lie que age sobre a fibra $F$, denominado grupo estrutural e que satisfaz:
\begin{itemize}
\item o que chamamos de trivializa\c c\~ao local $\pi\circ \phi_i(p,f)=p$, atrav\'es do difeomorfismo $\phi_i: U_i\times F\r\pi^{-1}(U_i)$, tal que $p\in M$, $f\in F$ e $\{U_i\}$ \'e uma cobertura de $M$,

\item e os difeomorfismos $\phi_i$ e $\phi_j$ s\~ao relacionados em $U_i\cap U_j \neq\emptyset$ pelas fun\c c\~oes de transi\c c\~ao $t_{ij}(p)\equiv \phi^{-1}_i(p)\circ\phi_j(p)$, como se segue \cite{Naka}:
\be
\phi_j(p,f)=\phi_i(p,t_{ij}(p)f). 
\ee
\end{itemize}
\end{itemize}
Se o fibrado $E$ decompõe-se no produto direto $M\times F$, dizemos que o fibrado \'e trivial.
Vejamos alguns casos bastante gerais de fibrado:
\subsubsection*{Fibrado Vetorial}

\qquad Se a fibra $F$ do fibrado de fibras $\pi:E\;\r\; M$ \'e um espa\c co vetorial real [complexo] $V$ de dimens\~ao $k$, ele \'e dito ser um fibrado vetorial e as fun\c c\~oes de transi\c c\~ao $\phi_i$ pertencem a $GL(k,\mathbb{R})$[$GL(k,\mathbb{C})$]. Como exemplo de fibrados vetoriais triviais temos o cilindro $S^1\times \mathbb{R}$ e a esfera com feixes de linhas normais $S^2\times \mathbb{R}$. A faixa de M\"obius \'e um caso n\~ao-trivial de fibrado, onde $M=S^1,\; F=[-1,1]$ e $G\cong \mathbb{Z}_2$.

\subsubsection*{Fibrado Principal}
\qquad Se a fibra $F$ \'e id\^entica ao grupo estrutural $G$, obtemos o fibrado principal $\pi: P\;\r\; M$, tamb\'em denotado por $P(M,G)$ e chamado de $G$-fibrado sobre $M$. Como exemplo, podemos exibir os fibrados de Hopf $\pi_1: S^3\;\r\; S^2$, onde $F\approx G= U(1)\approx S^1$, e $\pi_2:S^7\;\r\; S^4$, onde $F\approx G= SU(2)\approx S^3$.

\subsubsection*{Fibrado Associado}
\qquad O fibrado associado ao fibrado principal $P(M,G)$ \'e o fibrado $(E,\pi, M,G,F,P)$ determinado pela seguinte rela\c c\~ao de equival\^encia sobre $P\times F$:
\be
(u,f)\approx (ug,g^{-1}f).
\ee
Dessa forma, o fibrado associado \'e identificado com o quociente $\f{(P\times F)}{G}$. 

Considerando que $F$ \'e um espa\c co vetorial $V$,  inserimos uma rela\c c\~ao de equival\^encia an\'aloga no produto $P\times V$:
\be
 (u,v)\approx (ug,\rho(g)^{-1}v),
\ee e obtemos o fibrado vetorial associado $P\times_\rho V.$
\subsubsection*{Fibrado de Referenciais}

\qquad Um fibrado de referenciais \'e um fibrado principal sobre $M$ associado ao fibrado vetorial tangente $TM$, cujo grupo estrutural \'e $GL(n,\mathbb{R})$.  Um elemento ${g^i}_j\in GL(n,\mathbb{R})$ age sobre um referencial $\{X_i \}$ em $p\in M$ como $Y_j=X_i{g^i}_j.$

\vspace{1cm}
Sejam os fibrados $\pi: E\;\r\; M$ e $\pi': E'\;\r\; M$ e suas respectivas fibras $F$ e $F'$. O fibrado produto tensorial $E\otimes E'$ tem suas fibras dada pelo produto $F\otimes F'$, em cada ponto $p\in M$, e o fibrado soma de Whitney $E\oplus E'$ tem como fibras $F\oplus F'$. Atrav\'es desses conceitos de fibrado podem ser constru\'idos sobre um fibrado vetorial $ P\times_\rho V$ as mesmas estruturas constru\'idas sobre um espa\c co linear $V$, como por exemplo: o fibrado tensorial, o fibrado exterior e o fibrado de Clifford. Como as defini\c c\~oes desses fibrados s\~ao bastante an\'alogas, basta definirmos o que vem a ser um fibrado de Clifford.
\subsubsection*{Fibrado de Clifford}
\qquad Seja um fibrado vetorial $\pi: E\;\r\; M$, tal que $M$ \'e uma variedade semi-Riemanniana. Em cada fibra vetorial $E_p=\pi^{-1}(p)=V$, uma forma quadr\'atica $||v||^2=\langle v,v\rangle $ est\'a bem definida, o que permite a constru\c c\~ao da \'algebra de Clifford ${\cal C}\ell(E_p)$, e da\'i segue o fibrado de Clifford  $\pi:{\cal C}\ell(E)\;\r\; M$, que é dado pela seguinte uni\~ao disjunta n\~ao-trivial \cite{Lawson}
\be 
{\cal C}\ell(E)= \bigsqcup_{p\in M}\left(p, {\cal C}\ell(E_p)\right)
.\ee
Esse fibrado tem uma estrutura $\mathbb{Z}_2$-graduada que vem da \'algebra de Clifford ${\cal C}\ell(V)$.

\subsubsection*{Fibrado Espinorial}

\qquad Considere novamente o fibrado vetorial $\pi: E\; \r\; M$ com m\'etrica $g$ semi-Riemanniana e assinatura $p+q$, e o fibrado de refer\^enciais orientados e ortonormais $P_{SO}(E)$. O homomorfismo $s: {\rm Spin}_{p,q}\;\r\; SO_n$, para $n\geqq 3$, cujo n\'ucleo \'e isomorfo a $\mathbb{Z}_2$, induz  uma estrutura de spin sobre $E$ \cite{Lawson}
$$
s': P_{\rm Spin}(E) \;\r\; P_{\rm SO}(E),
$$
onde  $\forall \; p\in P_{\rm Spin}(E)$ e $g\in {\rm Spin}_{p,q}$, $s'(pg)=s'(p)s(g)$. Surge assim o conceito de fibrado espinorial $P_{\rm Spin}(E)$ a partir de uma estrutura de spin, o que dá origem à noção de spinor. As estruturas de spin tem aplicações na física-matemática, em particular na teoria quântica de campo, desempenhando um papel fundamental em teorias contendo férmions. Condições necessárias e suficientes para a existência de tais estruturas em espaços Riemannianos foram encontrados por A. Haefliger, as quais nos dizem que esses espaços admitem estrutura espinorial se, e somente se a segunda classe de Stiefel-Whitney $w_2(M)$ \footnote{$w_2(M)\in H_2(M,\partial)$ é um invariante topológico do fibrado vetorial.}  é trivial \cite{Hae, Hae1}. Se além disso, os grupos de cohomologia $\tilde{H}^0(M,\mathbb{Z}_2)$ e $H^1(M,\mathbb{Z}_2)$ forem triviais, pode-se afirmar que há apenas uma estrutura espinorial a menos de isomorfismos sobre M, onde $\tilde{H}$ denota a cohomologia reduzida  \cite{Hae1}. 
\subsubsection*{Fibrado de K\"ahler-Atiyah} 
\qquad Uma decomposição do produto de Clifford em partes simétrica e antissimétrica resulta nos produtos escalar e exterior, respectivamente. Dessa forma, o fibrado exterior $\wedge(M)$ munido do produto de Clifford é denominado fibrado de K\"ahler-Atiyah $(\sec \wedge(TM),\circ),$ tal que $\circ$ denota o produto de Clifford e obviamente satisfaz as álgebras de Grassmann e de Clifford. De forma que dadas as 1-formas $\phi$ e $\psi$, temos a seguinte expressão:
\be 
\phi\circ\psi= \phi\cdot\psi+\phi\wedge \psi
,\ee
onde $\cdot$ é o produto escalar.

Abordados os conceitos matemáticos, vejamos o arcabouço da teoria física necessária a compreensão deste trabalho e suas motivações.
\section{Formalismo de Compactificação em Supergravidade}
\qquad Abordaremos, nesta seção, como o formalismo de compactificação surgiu em meio às teorias de unificação. Para isso, na subseção a seguir, revisaremos uma das primeiras tentativas de unificação, buscando escrever a teoria da gravitação de Einstein e a teoria eletromagnética de Maxwell em termos de uma teoria para campos não-massivos em cinco dimensões. Isso resulta em uma quantização na massa através de modos harmônicos nas dimensões extras compactificadas. Depois disso, na seção seguinte, esse formalismo de compactificação é estendido para dimensões arbitrárias através da inserção de um campo de $s$-forma $F$, o qual foi feito por Freund e Rubin \cite{Freundrubin}. Finalmente, nas duas últimas subseções, são estudados os casos particulares, mas muito importantes, sobre os espaços $AdS_5\times S^5$, o {\it bulk} $AdS_5$ e $AdS_4\times S^7$.

\subsection{Modelo de Kaluza-Klein}
\qquad Um dos primeiros a realizar cálculos com dimensões extras na física foi Kaluza, cujo modelo foi um precursor das teorias de cordas. Posteriormente, Klein deu uma interpretação em termos da teoria de campos ao modelo de Kaluza, tornando-se conhecido como modelo de Kaluza-Klein \cite{duff86}.

Em um nível clássico, a unificação das equações da relatividade com as equações do eletromagnetismo é obtida através de uma teoria pura de gravidade em cinco dimensões. Dessa forma, as equações de Einstein escritas no espaço de cinco dimensões $M^{1,3}\times S^1$ se decompõem nas equações de Einstein clássicas e nas equações de Maxwell \footnote{Nesta seção, $M^{1,3}$ denota o espaço-tempo de quatro de dimensões, o qual é uma variedade Lorentziana, não necessariamente o espaço de Minskowski $\mathbb{R}^{1,3}$.}. Isso é feito ao se considerar uma métrica sobre $M^{1,3}\times S^1$  dada por \footnote{Note que esta j\'a era uma ideia desenvolvida exclusivamente em geometria, pois o grupo de rotaç\~oes $SO(1,3)$ (no espaço de Minkowski) pode ser visto como o grupo de rotações e {\it boosts} no espaço tridimensional Euclidiano.} \cite{duff86}:
\be 
g_{ab}\equiv \left[\begin{tabular}{cc}
$g_{\m\n}+\phi^2A_\m A_\n$ &$\phi^2A_\m$ \\ $\phi^2A_\n$ &$\phi^2$
\end{tabular}\right],
\ee
e sua matriz inversa associada
\be
g^{ab}\equiv \left[\begin{tabular}{cc}
$g^{\m\n}$ &$-A^\m$ \\ $-\phi^2A^\n$ &$g_{\a\b}A^\a A^\b +\f{1}{\phi^2}$
\end{tabular}\right]. 
\ee
Considere a velocidade da luz $c=1$. A partir da ação para gravidade em cinco dimensões
\be
S_5=\f{1}{16\pi G_5}\int\sqrt{-g_5}R_5d^4xdy,
\ee
a Lagrangiana se decompõe em
\be
\sqrt{-g_5}{R}_5=\sqrt{-g_4}\left(R-\f{1}{4}F_{\m\n}F^{\m\n}\right),
\ee
onde $R_5$ e $R$ são a curvatura escalar em cinco e quatro dimensões, respectivamente. Assim, obtemos as seguintes equações de movimento \cite{duff86}:
\beq
R_{\m\n}-\f{1}{2}g_{\m\n}R&=&-8\pi G_4\phi^2\left[{F_\m}^\g F_{\n\g}-\f{1}{4}g_{\m\n}F_{\a\b}F^{\a\b}\right]-\f{1}{\phi}\left[\nabla_\m(\p_\n\phi)- g_{\m\n}\square\phi\right],\\
\nabla^\m F_{\m\n}&=&-3\f{\p^\m\phi}{\phi}F_{\m\n},\\
\square\phi &=&4\pi G_4\phi^3 F_{\m\n}F^{\m\n}
.\eeq
Se escolhermos o campo escalar, denominado {\it radion}, constante e igual a 1, as equações tornam-se
\beq
R_{\m\n}-\f{1}{2}g_{\m\n}R&=&-8\pi G_4\phi^2\left[{F_\m}^\g F_{\n\g}-\f{1}{4}g_{\m\n}F_{\a\b}F^{\a\b}\right],\\
\nabla^\m F_{\m\n}&=&0,\\
 F_{\m\n}F^{\m\n}&=&0.
\eeq
As primeiras equações são as equações de Einstein (com o tensor de energia-momento para matéria) na presença de campo eletromagnético. A segunda equação codifica as equações de Maxwell na ausência de carga e corrente. No entanto, $\phi=1$ só é consistente, se a terceira equação é um vínculo para o campo $F^{\m\n}$, que pode ser expresso também como $E_iE^i-B_iB^i=0$ \cite{Gimb}. 
\subsubsection*{Compactificação no Modelo de Kaluza-Klein}

\qquad Em 1926, Oscar Klein propôs uma compactificação dessa dimensão extra, o que resultou em uma teoria massiva no espaço-tempo de Minkowski a partir de uma teoria não-massiva em cinco dimensões. Dessa forma, o campo escalar pode ser decomposto em uma série de Fourier 
$$
\phi(x^\m,y)=\phi(x^\m,y+2\pi {\rm R})=\f{1}{\sqrt{2\pi {\rm R}}}\sum_{n=-\infty}^\infty\phi_n(x^\m)\cdot e^{i\frac{n}{\rm R}y},$$
o que nos permite escrever a seguinte ação \cite{Gimb}
$$
S=\f{1}{2}\int d^5x\p_M\phi(x^\m,y)\p^M\phi(x^\m,y)=\int d^4 x\left[\f{1}{2}\p_\m\phi_0\p^\m\phi_0+\sum_{n=1}^\infty\left(\p_\m\phi_n^\dagger\p^\m\phi^n-\frac{n^2}{{\rm R}^2}\phi_n^{\dagger}\phi_n\right)\right].
$$
Note que o primeiro termo é o termo cinético de uma campo escalar sem massa, enquanto que os outros termos da somatória são campos escalares massivos com massa $m_n^2=\f{n^2}{{\rm R}^2}$. Esta ideia pode ser estendida para teoria de compactificação de dimensões superiores
. Em geral, para campos massivos de massa $m_0$ sobre um espaço $M^{1,3}\times T^q$ que possui $q$ dimensões extras, com raios de compactificação ${\rm R}_1,\,{\rm R}_2,\ldots,{\rm R}_q$, é válida a seguinte expressão para as massas de Kaluza-Klein:
\be
m_n^2=m_0^2+\sum_{i=1}^q\f{j_i^2}{{\rm R}_i^2}, 
\ee
onde $j$ refere-se ao $j$-ésimo modo de Kaluza-Klein.\\
\indent Até aqui, tratamos apenas compactificações sobre $S^1$ e toros $T^n=\overbrace{S^1\times\cdots\times S^1}^{n{\rm \; vezes}}$. No entanto, essas mesmas ideias podem ser estendidas para espaços compactos de dimensões superiores, como por exemplo a esfera de $q$ dimensões $S^q$. Portanto, da mesma forma que sobre $S^1$ e sobre $S^2$, os campos são decompostos em uma série de Fourier e em uma série de harmônicos esféricos, \cite{nast}:
\beq
\phi (\vec{x},y)&=&\sum_n \phi_n(\vec{x})e^{\f{iny}{\rm R}},\\
\phi (\vec{x},\theta,\phi)&=&\sum_{lm}\phi_{lm}(\vec{x})Y_{lm}(\theta,\phi),
\eeq
respectivamente, os quais são auto-funções do Laplaciano \cite{Gimb}:
\beq \p_y^2e^{\f{iny}{\rm R}}&=&-\left(\f{n}{\rm R}\right)^2e^{\f{iny}{\rm R}},\\
\nabla_2Y_{lm}(\theta,\phi)&=&-\f{l(l+1)}{{\rm R}^2}Y_{lm}(\theta,\phi).
\eeq
Similarmente, sobre o espaço compactificado $M_4\times K_q$, onde $K_q$ é um produto de esferas, podemos decompor os campos nos harmônicos: 
\beq
\phi(\vx,\vy)=\sum_{n,I_n}\phi^{I_n}_n(\vx)Y_n^{I_n}(\vy),
\eeq
os quais obtemos, através de um Laplaciano escrito neste espaço:
\beq 
\nabla_q Y_n^{I_n}(\vy)=-m^2_n Y_n^{I_n}(\vy).
\eeq
Como resultado, podemos interpretar uma teoria sem massa em $4+q$ dimensões, para cada modo $\phi_n(\vx,\vy)=\phi_n^{I_n}(\vx)Y_n^{I_n}(\vy)$, como uma teoria massiva de massa $m_n$ em $4$ dimensões, como esquematizamos a seguir \cite{nast}
\beq
 \square_{4+q}\phi_n(\vx,\vy)=(\square+\nabla_q)\phi_n(\vx,\vy)=(\square+m_n^2)\phi_n(\vx,\vy).
\eeq

Sabemos que um campo quântico $\psi(x)$ sobre o espaço de Minkowski pode ser escrito como  
 \cite{Pesk}:
\be
 \psi(x)=\int \f{d^4p}{(2\pi)^4}
 \sum_s\left(a^s_{\bf p}u^s(p)e^{-ip\cdot x}+b^{s\dagger}_{\bf p}v^s(p)e^{ip\cdot x}\right).
 \ee
No entanto, um campo quântico $\psi'(y)$ sobre $K_q$, quando escrito em termos de operadores de criação e aniquilação
, resulta em uma série de harmônicos sobre esse espaço $K_q$, cujos respectivos momentos são quantizados. Dessa forma, se $M^4$ é o espaço-tempo e $K_q$ é o espaço compactificado, o campo quântico $\Psi$ sobre $M^4\times K_q$ pode ser escrito como $\Psi=\Psi_4(\vx)\Psi_q(\vy)$, onde aos harmônicos do espaço compacto $K_q$ estão associados operadores de criação e aniquilação. Então, o campo quântico $\Psi_q$ é decomposto como se segue \cite{Gimb,nast} 
\be
\Psi_q(\vy)=\sum_{I_n}
\left[Y_n^{I_n}(\vy)u_{I_n}a_{I_n} +\left(Y_n^{I_n}(\vy)\right)^*u^*_{I_n}b_{I^*_n}\right],
\ee
onde $I^*_{l,m}=I_{l,-m}$ e $a_{I_n}$, $b_{I^*_n}$ são coeficientes a valores de operadores. Em particular, para $q=7$, $Y_n^{I_n}(\vy)$ são harmônicos esféricos sobre $S^7$. %
Esse procedimento é válido para teoria de campos sobre espaços compactificados em baixas energias e em um formalismo de operadores de criação e aniquilação.

\subsection{Formalismo de Compactificação de Freund e Rubin}

\qquad Toda essa construção feita acima pode ser bastante explorada no contexto de supergravidade, pois para que esta fosse consistente foi necessário admitir a existência de um número de dimensões maior que quatro. Isso só foi possível através de compactificações das dimensões extras ou admitindo que fazemos parte de um universo com mais que três dimensões espaciais, onde vivemos em 3-branas. Como resultado, admitindo supergravidade em $d$ dimensões, surgiram modelos maximalmente simétricos de vários tipos. Entre esses modelos, há uma classe que pode ser expressa como $[A]dS_{d-q}\times S^q$. Posteriormente, Freund e Rubin obtiveram essa teoria de compactificação geral a partir de um campo totalmente antissimétrico $F^{\a_1\ldots\a_s}$, ou seja, a partir de uma $s$-forma dada pelo {\it ansatz} 
\cite{Freundrubin},
\be\label{ansatz}
 F^{\a_1\ldots\a_s}=\left\{\begin{tabular}{cc}
 $\f{f}{\sqrt{|g|}}\e^{\a_1\ldots\a_s},$&$1\leq \a_i \leq s$\\ 
 $0,$& caso contrário.
 \end{tabular}\right.
\ee 
Há extensões do tensor eletromagnético na equação de Einstein para ordens superiores, que herdam antissimetria total:
\be
R^{\m\n}-\f{1}{2}g^{\m\n}R=-8\pi G\theta^{\m\n},
\ee
onde 
\be
\theta^{\m\n}={F_{\a_1\ldots\a_{s-1}}}^\m F^{\a_1\ldots\a_{s-1}\n}-\f{1}{2s}F_{\a_1\ldots\a_s}F^{\a_1\ldots\a_s}g^{\m\n}
.\ee
Dessa forma, admitindo uma decomposição do espaço $M$ em $M^s\times M^{d-s}$, os escalares de curvatura em cada espaço são obtidos a partir do {\it ansatz} (\ref{ansatz}) (Apêndice [\ref{apen1}])
\be\label{curvescalar}
R_{d-s}=\f{(s-1)(d-s)}{d-2}\l,\qquad R_s=-\f{s(d-s-1)}{d-2}\l , 
\ee 
onde $\l=8\pi G\cdot sgn(g_s)$. Nesse caso, um espaço maximalmente simétrico tem métrica dada pela seguinte expressão:
\be 
ds^2=dx^2_{AdS_s}+\rho d\Omega_{d-s}^2, 
\ee
onde $dx^2_{AdS_s}$ denota a métrica sobre o espaço $AdS_s$, $\rho$ é o raio de compactificação do espaço $S^{d-s}$ e $d\Omega_{d-s}$ são ângulos infinitesimais sobre $S^{d-s}$. No caso particular, em que $d=11$ e $s=4$, o espaço $M$ maximalmente simétrico tem  curvaturas escalares dadas por:
\be 
R_7=\f{7\l}{3},\qquad R=-\f{8\l}{3}.
\ee
Concluímos então, que $M^4$ é um espaço Lorentziano de curvatura escalar negativa (espaço de anti-de Sitter) e $M^7$, um espaço Riemanniano de curvatura escalar positiva (espaço esférico), ou seja, $M$ é escrito como $AdS_4\times S^7$. Como vimos, aos modos de Kaluza-Klein sobre essa compactificação estão 
associados os harmônicos esféricos sobre $S^7$. 
Através desses, uma teoria sem massa em $d=11$ torna-se uma teoria efetiva sobre $AdS_4$ com massa. Esse é um aspecto que pode ser analisado: investigar os modos de Kaluza-Klein para cada tipo de espinor que encontramos sobre $S^7$.

No outro caso, em que $d=10$ e $s=5$, o espaço de Einstein se decompõe com simetria maximal como $AdS_5\times S^5$ e possui curvaturas escalares dadas por \cite{Freundrubin}: 
\be
^{S^5}R= \f{5}{2}\l\quad \text{e}\quad ^{AdS_5}R=-\f{5}{2}\l,
\ee 
resultando em uma decomposição nos seguintes espaços de Einstein:
\beq
 ^{S^5}R_{mn}&=& \f{1}{2}\l g_{mn},\\
  ^{AdS_5}R_{\bar{m}\bar{n}}&=&-\f{1}{2}\l g_{\bar{m}\bar{n}}.
\eeq

%

\subsection{{\it Bulk} $AdS_5$}


\qquad A partir de supergravidade em $d=10$, admitindo a decomposição $M=M_5\times M'_5$, podemos obter a compactificação $AdS_5\times S^5$ através da $5$-forma $F_{\a_1\ldots\a_5}=4L^4(\e_{\a_1\ldots\a_5}+*\e_{\a_1\ldots\a_5})$, que é equivalente a abordagem acima, se identificamos $\l=\f{8}{L^2}$. Dessa maneira, inserindo essa $5$-forma nas equações de Einstein, obtemos os tensores de Ricci \cite{Freundrubin,Duff}:
\beq
R_{\m\n}=-\f{4}{L^2}g_{\m\n},\\
R_{mn}= \f{4}{L^2}g_{mn},
\eeq
que determinam, no caso maximalmente simétrico, os espaços $AdS_5$ e $S^5$, respectivamente, como veremos adiante. Isso resulta em uma curvatura escalar nula sobre $AdS_5\times S^5$. Para analisar esse caso, considere a seguinte solução
\beq 
ds^2=\f{r^2}{L^2}\eta_{\m\n}dx^\m dx^\n+\f{L^2}{r^2}\d_{ij}dx^i dx^j.
\eeq
Levando em conta a mudança de coordenadas $\d_{ij}dx^i dx^j=dr^2+r^2ds^2_{S^5}$ e pondo  $z=L^2/r$, a métrica é simplificada,
\beq
ds^2=\f{L^2}{z^2}\left(\eta_{\m\n}dx^\m dx^\n+dz^2\right)+L^2ds^2_{S^5},
\eeq
que é exatamente a métrica do espaço $AdS_5\times S^5$.
Considere então, cordas fechadas e cordas abertas em supergravidade do tipo IIB sobre um espaço assintoticamente plano $\mathbb{R}^{9,1}$ \cite{Kanno,Horava,Horava1}. Observado próximo ao horizonte de eventos, através do formalismo de cordas fechadas, ele se comporta como supergravidade do tipo IIB compactificada sobre o espaço $AdS_5\times S^5$, a qual pode ser explorada no contexto de dimensões infinitas, por meio espaços quocientes \cite{Mikhailov:2012uh}. 
Vale frisar que nossa motivação aqui é apenas compactificação, não sendo nosso intuito explorar cordas do tipo IIB.

Além disso, em outro contexto o espaço $AdS_5$ também é utilizado
na resolução do problema de hierarquia. Para isso, foi proposto o modelo de Randall-Sundrum, que descreve um {\it bulk} de cinco dimensões contido entre duas 3-branas, sendo que uma é chamada de brana de Planck, onde os parâmetros de massa e a massa de Planck se identificam, e a outra é a brana fraca, onde ocorrem as interações e os parâmetros diferem entre si. Esse espaçamento entre as duas branas é uma fatia de $AdS_5$ que determina a quantização (discretização) dos modos de Kaluza-Klein. Dessa forma, a métrica sobre esse espaço $\mathbb{R}^{1,3}\times (S^1/\mathbb{Z}_2)$ é dada pelo {\it ansatz} \cite{Gimb}
\be
ds^2=e^{-2A(y)}\eta_{\m\n}dx^\m dx^\n +dy^2. 
\ee
Considere um campo escalar sem massa $\phi$ sobre o fundo de Randall-Sundrum em cinco dimensões, sua Lagrangiana é ${\cal L}=\f{1}{2}(g^{MN}\p_M\phi \p_N\phi)=\f{1}{2}e^{3A(z)}(\eta^{MN}\p_M\phi \p_N\phi)$. Então, através das equações de Lagrange, obtemos:
\be\label{111}
e^{3A(y)}[\p_\m\p^\m+\p_y\p^y-3(\p_y A(y))\p_y]\phi(x^\m,y)=0,
\ee
a partir da qual podemos calcular o potencial efetivo, por meio da substituição $\phi(x^\m,y)=:e^{-ip\cdot x}\phi(y)=:e^{-ip\cdot x}e^{f(y)}\psi(y)$:
\be 
 -\p_y^2\psi-(2\p_y f+3\p_y A)\p_y\psi-(\p_y^2+(\p_y f)^2+3(\p_y A)\p_y f)\psi =m^2\psi,
\ee
que simplifica a eq. (\ref{111}) e assume a forma da equação de Schrödinger, se escolhemos $2\p_y f+3\p_y A=0$. Se uma segunda brana é adicionada como uma função delta em $y=L$, o seguinte potencial, denominado potencial vulcão, é obtido \cite{Gimb}
\be
V(y)=\f{15}{4}\f{k^2}{(k|y|+1)^2}-\f{3k(\d(y)-\d(y-L))}{k|y|+1}.
\ee 
No espaço entre as branas, a equação de onda \cite{Gimb}
\be 
-\p_y^2\psi_n+\left(\f{15k^2}{4(k|y|+1)^2}\right)\psi_n=m^2\psi_n
\ee
nos dá os modos do gráviton de Kaluza-Klein, em termos das funções de Bessel de primeiro e segundo tipo:
\be
 \psi_n=\sqrt{|y|+1/k}[a_nJ_2(m_n(|y|+1/k))+b_nY_2(m_n(|y|+1/k))],
\ee
cujas massas de Kaluza-Klein são quantizadas pelas condições de fronteira, o que resulta nas seguintes massas $m_n\simeq j_0^n/L$. Outras soluções deste conjunto de soluções foram propostas em diferentes contextos em \cite{HoffdaSilva:2012em,German:2012rv,Bernardini:2014vba,Bazeia:2013usa,Bazeia:2012qh}. \\ 
\qquad Se além do campo escalar, considerarmos um campo espinorial fermiônico $\xi$ sem massa, obtemos a seguinte Lagrangiana total
\be
{\cal L}=R_5+\underbrace{\f{1}{2}\p_\m\phi \p^\m\phi+V(\phi)}_{{\cal L}_{\text{escalar }\phi}}+\underbrace{\left[\f{i}{2}\left(\bar{\xi}\slashed{\nabla}\xi-\bar{\xi}\stackrel{\leftarrow}{\slashed{\nabla}}\xi\right)-m\bar{\xi}\xi+\f{\l}{2}(\bar{\xi}\xi)^2\right]}_{{\cal L}_{\text{espinor }\xi}}+\overbrace{\phi\bar{\xi}\xi}^\text{termo de Yukawa}.
\ee
A partir da qual, a dinâmica é calculada para esses campos \cite{brane}. Dessa forma, se escolhemos o campo espinorial representado por $\xi=\left(a(y),0,b(y),0\right)^\intercal$ e implementado sobre a brana, as equações de Einstein e de Dirac são escritas, para esse caso, como:
\beq\label{Einstein}
{R_a}^A-\f{1}{2}{e_a}^AR=\chi{T_a}^A+{e_a}^A\Lambda,\\
\left[i\Gamma^a{e_a}^A D_A-m+\lambda(\bar{\psi}\psi )\right]\psi =0
,\eeq
a densidade de energia pode ser calculada e é dada por ${T_0}^0=-2\l a^2b^2$ \cite{brane}. As componentes do espinor $\xi$ também podem ser espressas em termos de constantes de acoplamento e da dimensão extra \cite{brane}:
\beq
a(y)=a_0\exp\left\{my-\sqrt{\f{3\l}{\chi}}\arctan\left(2\sqrt{\f{\chi}{3\l}}my\right)-\f{1}{2}\ln\left(1+\f{4}{3}\f{\chi}{\l}m^2y^2\right)\right\},\\
b(y)=b_0\exp\left\{-my+\sqrt{\f{3\l}{\chi}}\arctan\left(2\sqrt{\f{\chi}{3\l}}my\right)-\f{1}{2}\ln\left(1+\f{4}{3}\f{\chi}{\l}m^2y^2\right)\right\}.
\eeq 

Se em vez disso, considerarmos um campo espinorial do tipo {\it flag-dipole} de duas componentes, representado por $\xi=\left(a(y),0,0,b(y)\right)^\intercal$ e uma métrica dada por $ds^2=\phi^2(y)\left(dx_0^2-dx_1^2-dx_2^2-dx_3^2\right)-dy^2$, encontramos que as componentes diagonais do tensor energia-momento $T_{\m\n}$ são identicamente nulas. Além disso, através da eq. (\ref{Einstein}) e do fato de o tensor de Ricci e de {\it vielbein} ${e_a}^A$ serem diagonais, resulta que o tensor $T_{\m\n}$ também é nulo fora da diagonal \cite{brane,Cavalcanti:2014wia,esk}. Dessa forma, como o tensor energia-momento é nulo, esse {\it flag-dipole} não gera brana e não tem existência. Pois, a partir de um dos vínculos obtido da componente ${T_{\bar{0}}}^1\,=\,0\,=\,\phi'(y)a(y)b(y),$ obtemos que pelo menos uma das componentes do espinor deve ser nula. Por essa razão, o espinor $\xi$ deve ser, na verdade, do tipo {\it dipole}, dado por: $\xi=\left(a(y),0,0,0\right)^\intercal$. Calculamos o fator de dobra $\phi$ da métrica e a componente do {\it dipole} e encontramos:
\beq
\phi(y)&=&\phi_0 e^{\pm i\sqrt{\f{\Lambda}{6}}y}\\
a(y)&=&a_0e^{\mp i\sqrt{\f{2\Lambda}{3}}y}.
\eeq

 Esses espinores sobre o {\it bulk} são classificados no capítulo seguinte e sua dinâmica pode ser explorada para cada uma das classes de espinores listadas \cite{BR}. Campos espinoriais de ordem mais elevada também podem ser explorados nesse contexto \cite{Mikhailov:2002bp}
\subsection{$AdS_4\times S^7$}


%
\qquad A principal motivação para se classificar campos de espinores sobre espaços Riemannianos de sete dimensões, em particular $S^7$, o que é feito no capítulo central deste trabalho, reside na importância, em supergravidade, de se explorar o espaço compactificado $M_4\times M^7,$ onde $M_4$ é o espaço-tempo de quatro dimensões e $M^7$, um espaço de Einstein-Riemann arbitrário de sete dimensões com uma estrutura spin compatível\cite{BBR}.

Em supergravidade, espaços de onze dimensões com uma 4-forma são compactificados através do {\it ansatz} de Freund-Rubin, resultando na compactificação $M_4\times M^7$
, podendo este ser, principalmente: $S^7,S^5\times S^2, S^4\times S^3, S^2\times S^2\times S^3, T^7$ e outros \cite{Lawson,Hae,Hae1,Freundrubin}. No entanto, se adicionamos a isso que $M^7$ é maximalmente simétrico, resta apenas a escolha $M^7=S^7$, que é invariante sob o grupo de simetria $SO(8)$ e tem uma métrica induzida do espaço Euclidiano, dada por \cite{duff86}:
\be
ds^2=\left(\d_{\m\n}+\f{y^\m y^\n}{m^{-2}-y^\a y^\a}\right) dy^\m dy^\n, 
\ee
a partir da qual, o tensor de curvatura de Riemann pode ser obtido:
\be
R_{\m\n\rho\s}=m^2(g_{\m\rho}g_{\n\s}-g_{\m\s}g_{\n\rho}) 
.\ee

Em outras palavras, uma compactifica\c c\~ao espont\^anea sobre $S^7$ de supergravidade em onze dimens\~oes 
resulta em uma teoria de supergravidade em quatro dimens\~oes com simetria maximal
. \la{AdS4} Espa\c cos-tempos com simetria maximal podem ser apenas espa\c cos de de Sitter, Minkowski ou anti-de Sitter, isto \'e, cujo v\'acuo \'e invariante sob a a\c c\~ao do grupo $SO(4,1)$, Poincar\'{e} ou $SO(3,2)$, respectivamente. Esses têm como solução o vácuo e são produzidos por meio de uma constante cosmol\'ogica positiva, zero ou negativa, respectivamente. Dessa forma, um possível espaço de $11$ dimensões maximalmente simétrico é $AdS_4\times S^7$. É interessante notar que um campo fermiônico sobre um espaço maximalmente simétrico pode ter valor esperado de vácuo nulo.
 
 Essa compactificação -- $AdS_4\times S^7$ -- em supergravidade em $d=11$ é uma solu\c c\~ao  maximalmente sim\'etrica para o espa\c co-tempo, a qual é compatível com o produto direto $M_{4} \times M_{7}$ e é obtida através do {\it ansatz} de Freund e Rubin \cite{Freundrubin}:
\be
F_{\mu\nu\rho\sigma}=\frac{3a}{2}\epsilon_{\mu\nu\rho\sigma}
\la{FRansatz2}
,\ee
onde todas as outras componentes s\~ao nulas e $a$ \'e um escalar. 
Isso pode ser verificado substituindo esse {\it ansatz} nas equa\c c\~oes de campo. Sobre o espa\c co-tempo de quatro dimensões, obtemos
\be
R_{\mu\nu}=-3a^{2}g_{\mu\nu}=-\frac{12}{L^{2}}g_{\mu\nu}
\la{einstein12}
\ee
e sobre o espa\c co de Einstein-Riemann de sete dimens\~oes,
\be
R_{mn}=\frac{3a^{2}}{2}g_{mn} =\frac{6}{L^{2}}g_{mn}
\la{einstein22}.
\ee
\\ \indent Entre os campos espinoriais sobre $S^7$, os quais classificamos neste trabalho e encontramos duas novas classes não-triviais, há um tipo que é muito útil na exploração de simetrias em supergravidade, os denominados espinores de Killing \footnote{De uma forma geral espinores de Killing $\xi$ são definidos como $\nabla_X \xi=\l X\xi$, onde a justaposição é a multiplicação de Clifford e $X$ são vetores tangentes. Em particular, sobre o espaço $AdS_4$, ele é dado através da seguinte equação $\nabla_\m \xi= im\g\G_\m\xi$, onde $\l$ é constante e $\l^2=1$. No outro caso, em que o espaço é a esfera $S^7$, ele é dado pela equação $\nabla_i \psi=\f{i}{2}m\G_i\psi$ \cite{Eug, Luy}.}, e satisfazem à seguinte equação sobre $S^7$:
\be
\tilde{D}_m\xi=(\overbrace{\p_m-\f{1}{4}{\omega_m}^{ab}\Gamma_{ab}}^{D_m}-\f{1}{2}m{e_m}^a\Gamma_a)\xi,
\ee
onde $\tilde{D}_M$ denota derivadas supercovariantes calculadas em 11 dimens\~oes. De fato, as matrizes gama são decompostas através da compactificação de Freund-Rubin
\be
 \Gamma_{A}=(\gamma_{\alpha} \otimes 1, \gamma_{5} \otimes
 \Gamma_{a}),
\ee
onde
\beq
\{\gamma_{\alpha},\gamma_{\beta}\}= -2\eta_{\alpha\beta} ;\qquad \a,\b=0,1,2,3,\\
\{\Gamma_{a},\Gamma_{b}\}=-2\delta_{ab};\qquad a,b=1,2,\ldots,7,
\eeq
o que resulta também em uma decomposição das derivadas 
\beq
\tilde D_{\mu} =D_{\mu}+\frac{1}{L}\gamma_{\mu}\gamma_{5},
\\ \tilde D_{m} = D_{m}-\frac{1}{2L} \Gamma_{m}
\la{internalderiv},
\eeq
tal que $\gamma_{\mu}=e_{\mu}^{\alpha}\gamma_{\alpha}$ e
$\Gamma_{m}=e_{m}^{a}\Gamma_{a}$.

  Tendo em mãos essas definições, a Lagrangiana pode ser escrita para supergravidade em $d=11$ e é dada por \cite{Freundrubin,Duff,eng4,logi09,SUSYBr}:
\beq
{\cal L}&=&-\f{V}{4K^2}R(\omega)\\
&-&\f{iV}{2}\bar{\psi}_\m\G^{\m\n\rho}D_\n
\left(\f{\omega+\bar{\omega}
}{2}\right)\psi_\rho-\f{V}{48}F_{\m\n\rho\s}F^{\m\n\rho\s}\\
&+&\f{KV}{192}\left(\bar{\psi}_\m
\G^{\m\n\a\b\g\d}\psi_\n
+12\bar{\psi}^\a\G^{\g\d}\psi^\b
\right)\left(F_{\a\b\g\d}+\hat{F}_{\a
\b\g\d}\right)\\
&+&\f{2K}{(144)^2}e^{\a_1\a_2
\a_3\a_4\b_1\b_2\b_3\b_4\m\n\rho} F_{\a_1\a_2
\a_3\a_4}F_{\b_1\b_2 \b_3\b_4}A_{\m\n\rho}
,\eeq
onde $[\a_1\dots\a_n]$ indica que os \'indices nos colchetes s\~ao antissimetrizados e $A_{\nu\rho\sigma}$ \'e um campo de calibre que pode ser escrito como uma forma bilinear $A_{npq}=c\bar{\xi}\G_{npq}\xi$, onde $\xi$ é um espinor de Killing sobre $S^7$ \cite{BBR}. Dessa forma, o campo $4$-forma $F_{mnpq}$ sobre $S^7$ é um invariante de calibre, pois pode ser escrito como $F_{mnpq}\equiv 4\p_{[m}A_{npq]}$ e tem sua dinâmica regida pela equação:
 \be
 \nabla_m F^{mpqr}= -\f{1}{6}m\e^{rstupqr}F_{rstu}.
 \ee
A import\^ancia de nossos resultados para supergravidade em $d=11$ reside no fato que a identifica\c c\~ao da 3-forma bilinear covariante $\bar{\psi}\g_{\m\n\rho}\psi$ com o tensor de torção $A_{\m\n\rho}$ nos d\'a uma forma mais natural e formal de expressar, por meio desse {\it ansatz}, que a torção sobre $S^7$, possivelmente, deve ser n\~ao-nula quando consideramos campos fermiônicos de Majorana sobre a mesma:
\be
A_{jkl}=\f{\l}{4!}S_{jkl}= \f{\l}{4!}m_7\bar{\xi}\g_j\g_k\g_l\xi
,\ee onde 
 $m_7$ é o parâmetro de massa sobre $S^7$. Este {\it ansatz} \'e motivado pela equa\c c\~ao de estrutura entre os geradores:
\begin{equation}
\left[e_\alpha ,e_\beta \right]={\alpha_{\alpha\beta}}^\gamma e_\gamma
.\end{equation}
H\'a tamb\'em uma relação com as constantes de estrutura octoni\^onicas \cite{duff86,Duff}
\be
e_\a e_\b=-\d_{\a\b}+\sum_{\g}c_{\a\b\g} e_\g
.\ee
Ent\~ao, um caso que pode ser considerado \'e $S_{jkl}=-m_7c_{jkl}$. Nesse caso, a 4-forma intensidade de campo $F_{jklm}=4!\partial_{[j }A_{klm]}$ (curvatura) \'e zero, mas a 4-forma dual de Hodge $\star A$ \'e n\~ao-nula \cite{SUSYBr,BBR}. 
Dessa forma, a tor\c c\~ao $A_{jkl}$, sobre $AdS_4\times S^7$, 
comporta-se como um campo de calibre. É interessante notar, que já em estudos da dinâmica de campos espinoriais sobre o espaço-tempo de Minkowski surge a proposta de se incluir uma torção diferente de zero, no contexto da geometria de Riemann-Cartan, através de uma extensão da geometria da relatividade geral, denominada geometria de Lyra \cite{Casana:2005de,casana}.

Sendo assim, uma das principais motiva\c c\~oes para se classificar campos espinoriais fermiônicos sobre $S^7$ \'e que a n\~ao-trivialidade desses espinores sobre $AdS_4\times S^7$ surge ao escolhermos somente espinores de Majorana, quando restritos a $S^7$. Pois, a forma $A=A_{jkl}\g^j\g^k\g^l$ pode ser identificada com a 3-forma $\hat{\varphi}_3$, que \'e n\~ao-nula. Além disso, a 4-forma dual tamb\'em é n\~ao-nula, diferentemente do campo intensidade $F_{jklm}=4!\partial_{[j }A_{klm]}$, que pode ser zero \cite{BBR,SUSYBr}. Atrav\'es de uma extens\~ao complexa dos espinores de Majorana sobre $M^7$, em particular $S^7$, encontramos duas novas classes não-triviais: uma singular e uma regular. Abordaremos em detalhes, no Cap. 4, tal formulação. Al\'em disso, essas duas novas classes encontradas sobre $S^7$ nos sugerem novos espinores, que podem ser novos candidatos às soluções das equações de campos em supergravidade 
\cite{BBR}.

Um modelo muito estudado em onze dimensões, como vimos acima, é $AdS_4\times S^7$. É interessante notar, que sobre seu espaço compacto $S^7$ há mais duas geometrias como soluções das equações de movimento,  além da geometria Riemanniana trivial \cite{cart26,cart26a}. No entanto, foi mostrado posteriomente haver uma família de geometrias a um parâmetro sobre $S^7$ \cite{eng4}. Encarando isso de uma forma mais abrangente, 
temos como perspectivas, estender essas geometrias para uma fam\'ilia de geometrias a um parâmetro octoniônico, através de um produto octoniônico bastante geral conhecido como produto-$u$, generalizando o que foi feito em \cite{eng4,logi09}, pois permite introduzir geometrias paralelizáveis sobre $S^7$, incorporando a estas aspectos das álgebras de Clifford \cite{trae,trae1,ced,akiv}. 
\myclearpage
\par
\chapter{Classifica\c c\~ao de Espinores de Acordo com Bilineares Covariantes}

\qquad Neste capítulo, apresentamos uma classifica\c c\~ao de espinores sobre o espa\c co-tempo de Minkowski segundo bilineares covariantes \cite{lou2}. Essa mesma abordagem ser\'a utilizada, no cap\'itulo seguinte, para classificarmos espinores em espaços Riemannianos de sete dimens\~oes, em particular $S^7$, e sobre espaços Lorentzianos de cinco dimensões, em especial o {\it bulk} da teoria de branas. 
Tomamos como arcabouço o que foi abordado no cap\'itulo anterior: as \'algebras de Clifford e a geometria (semi-)Riemanniana. Bilineares covariantes s\~ao escritos utilizando-se de elementos da \'algebra de Clifford correspondente. Assim, a classificação abordada neste capítulo serve de base à classificação de espinores sobre $S^7$ [sobre o {\it bulk}], por meio de bilineares covariantes, através do uso de simetrias e das identidades de Fierz. Em seguida, por meio de uma complexificação, é possível generalizar essa classificação, resultando em outras duas [seis] classes não-triviais.

\section{Bilineares Covariantes}


\qquad Apesar de já havermos definido as ferramentas necessárias para a construção dos bilineares covariantes, cumpre-nos formalizar todo seu arcabouço através da introdução sucinta de cada um desses conceitos. Considere, então, em particular, espinores no espa\c co-tempo de Minkowski 
 $(M,g)$
, o qual possui uma m\'etrica localmente Lorentziana 
$ sgn(g) = \mathrm{diag}(-1,1,1,1)$, onde $sgn(g)$ denota o sinal da métrica. O conjunto $\{x^{\mu }\}$ s\~ao as coordenadas globais adaptadas ao referencial inercial 
${e}_{0}=\partial/\partial x^{0}$ e ${e}_{i}=\partial /\partial x^{i}$, $i=1,2,3$, que s\~ao se\c c\~oes do fibrado dos referenciais $\mathbf{P}_{\mathrm{SO}_{1,3}^{e}}(M)$ sobre o espa\c co de Minkowski, cujas fibras sofrem a a\c c\~ao do grupo $\mathrm{SO}_{1,3}^{e}$.
Uma m\'etrica $g$ 
do espa\c co $\bigwedge^1(TM)= T^*M$ de 1-formas pode ser estendida para o espa\c co das formas arbitr\'arias 
$\sec\bigwedge(TM)$. Para isso, considere de uma forma geral, um funcional linear definido a partir da métrica $g$ \cite{bookroldao}:
\beq
^\#:V &\r &V^*\\
v&\mapsto &v^\# :V\r \mathbb{R}\\
&&\quad\quad u\mapsto v^\#(u)=g(u,v).
\eeq
Em seguida, considere um campo $k$-vetorial, que é uma se\c c\~ao do fibrado exterior, denotado por $a\in\sec\bigwedge^k (TM)$. Dessa forma, a seguinte extensão do funcional linear pode ser construída:
\be 
(v_1\wedge v_2\wedge \cdots \wedge v_p)^\# =v_1^\#\wedge v_2^\#\wedge \cdots \wedge v_p^\#.
\ee
Consequentemente, a métrica pode ser também estendida para elementos arbitrários da álgebra exterior $\bigwedge(V)$ \cite{bookroldao}
\beq
{\rm G}:\bigwedge_p(V)\times\bigwedge_p(V)&\r & \mathbb{R} \\
(v_1\wedge\cdots\wedge v_p,u_1\wedge \cdots\wedge u_p)&\mapsto&{\rm G}(v_1\wedge\cdots\wedge v_p,u_1\wedge \cdots\wedge u_p),
\eeq onde ${\rm G}(v_1\wedge\cdots\wedge v_p,u_1\wedge \cdots\wedge u_p)=(v_1\wedge  \cdots \wedge v_p)^\#(u_1\wedge \cdots\wedge u_p)$. Explicitamente, temos:
\be
{\rm G}(v_1\wedge\cdots\wedge v_p,u_1\wedge \cdots\wedge u_p)=\left|\begin{tabular}{cccc}
$g(v_1,u_1)$&$g(v_1,u_2)$ &$\ldots$ &$g(v_1,u_p)$\\
$g(v_2,u_1)$&$g(v_2,u_2)$&$\ldots $&$g(v_2,u_p)$\\
$\vdots $& $\vdots $&$\ddots $&$\vdots$\\
$g(v_p,u_1)$&$g(v_p,u_2)$&$\ldots $&$ g(v_p,u_p)$
\end{tabular} \right|.
\ee
Resta então, para completar a definição acima, fixar a relação bilinear entre elementos de graus diferentes:
\be
{\rm G}(v_1\wedge\cdots\wedge v_p,u_1\wedge\cdots\wedge u_q)=0, \quad {\rm se }\; p\neq q.
\ee
Além disso, o operador dual de Hodge $\star:\sec\bigwedge(TM)
\rightarrow\sec\bigwedge(TM)$ \'e definido por $a\w\star b = {\rm G}(a,b)$. É interessante notar que a \'algebra de Grassmann $(\bigwedge (TM),{\rm G})$, enquanto espaço vetorial, \'e isomorfa \`a \'algebra de Clifford correspondente $\cl_{p,q}$, pois ambas s\~ao \'algebras quadr\'aticas associadas ao espa\c co $\sec\bigwedge^1(TM)$ $\simeq\mathbb{R}^{p,q}$. 
Determinada a métrica $G$ sobre o espaço onde moram os bilineares covariantes, vamos introduzir os espinores no espaço de Minkowski e construir seus bilineares covariantes.

 Espinores cl\'assicos $\psi$ s\~ao elementos de um espa\c co associado a uma representa\c c\~ao $\rho: D^{(1/2,0)}\oplus D^{(0,1/2)}$ de $\mathrm{SL}(2,\mathbb{C})$. Dessa forma, o fibrado vetorial associado $\mathbf{P}_{\mathrm{Spin}_{1,3}^{e}}(M)\times _{\rho }\mathbb{C}^{4}$ \'e denominado fibrado espinorial, onde $\rho $ \'e estabelecido pela representa\c c\~ao $D^{(1/2,0)}\oplus D^{(0,1/2)}$ do grupo de Lorentz.  
Os bilineares covariantes obtidos a partir desses espinores s\~ao se\c c\~oes do fibrado exterior $\bigwedge(TM)$ \cite{lou2,moro,cra}. Para que esses bilineares possam ser definidos, é necessário introduzir os seguintes endomorfismos:
\beq
{\rm involu\texttt{\c{c}}\tilde{a}o\ \
graduada}&\quad\hat{a}=&(-1)^{k}a, \\ \nonumber
{\rm revers\tilde{a}o}&\quad \tilde{a}=& a_k a_{k-1}\ldots a_1=(-1)^{\f{k(k-1)}{2}}a
\\ \nonumber {\rm conjuga\texttt{\c{c}}\tilde{a}o}&\quad \bar{a}=&\tilde{\hat{a}}=\hat{\tilde{a}}=(-1)^k a_k a_{k-1}\ldots a_1=(-1)^{\f{k(k+1)}{2}}a,
\eeq
onde $a=a_1a_2\ldots a_k\in \bigwedge^k(TM)$. Usando a nota\c c\~ao padr\~ao $\bar\psi=\psi^\dagger\gamma^0$, os bilineares covariantes s\~ao listados abaixo:
\begin{eqnarray}
\sigma &=& \bar{\psi}\psi ,\nonumber\\
 \quad \mathbf{J}&=&J_{\mu }\theta ^{\mu }=\bar{\psi}\gamma _{\mu }\psi \theta ^{\mu },\nonumber\\
\quad \mathbf{S}&=&S_{\mu \nu }\theta ^{\mu \nu }=\tfrac{1}{2}i\bar{\psi}\gamma _{\mu \nu }\psi \theta ^{\mu }\wedge \theta ^{\nu }, \\
\mathbf{K} &=& K_{\mu }\theta ^{\mu }=i\bar{\psi}\gamma_{0123}\gamma _{\mu }\psi \theta ^{\mu },\nonumber\\
\quad \omega &=&-\bar{\psi}\gamma _{0123}\psi \nonumber ,
\end{eqnarray}
onde a base vetorial de matrizes gama $\{\mathbf{1}_{4},\gamma _{\mu },\gamma _{\mu }\gamma _{\nu },\gamma _{\mu }\gamma _{\nu}\gamma _{\rho },\gamma _{0}\gamma _{1}\gamma _{2}\gamma _{3}\}$ satisfaz $\gamma _{\mu }\gamma _{\nu }+\gamma _{\nu }\gamma_{\mu }=2\eta _{\mu \nu }\mathbf{1}_{4}$ e \'e uma base para $\mathcal{M}(4,\mathbb{C})$, tal que $\mu ,\nu
,\rho =0,1,2,3$ e $\mu <\nu <\rho $, onde a justaposi\c c\~ao denota o produto de Clifford \cite{rod}.

Se nos preocupamos em descrever o el\'etron como um espinor de Dirac (classes 1), cada bilinear acima descrito est\'a associado com uma propriedade do el\'etron, como se segue: ~${\bf J}$ \'e uma 1-forma tipo-tempo correspondente \`a densidade de corrente; a densidade de momento eletromagn\'etico intrínseco do elétron \'e representada pela 2-forma ~$\mathbf{S}$; 
a dire\c c\~ao do spin do el\'etron ou corrente quiral \'e dada pela 1-forma tipo-espa\c co~$\mathbf{K}$ e o pseudoescalar $\omega$ nos dá indícios do espinor sob a simetria $CPT$ \cite{lou2}. Esses observáveis podem ser vinculados entre si através de identidades envolvendo as matrizes gama, denominadas identidades de Fierz, que são descritas na seção seguinte. Tal classificação também foi implementada e estudada em regimes de gravidade quântica em \cite{Ablamowicz:2014rpa}.
\section{Identidades de Fierz-Pauli-Kofink}

\qquad Identidades de Fierz gerais s\~ao bastantes \'uteis para se escrever rela\c c\~oes de completeza e ortogonalidade  \cite{6,7}, com as quais é constru\'ida uma equival\^encia entre representa\c c\~oes espinorial e tensorial. Assim, atrav\'es do estudo dessas identidades surgiram novas classes de campos espinoriais, principalmente nessa \'ultima d\'ecada, quando esse assunto e suas aplica\c c\~oes têm sido cada vez mais explorados. Entre essas aplicações, vínculos relacionados ao termo exótico da equação de Dirac podem ser calculados \cite{exotic,daSilva:2016htz,daRocha:2016bil}, outros aspectos sobre operadores de Dirac são abordados em \cite{Vassilevich:2015soa}, uma compreensão algébrica e geométrica do Elko \footnote{Elko é uma abreviação para {\it Eigenspinoren des LadungsKonjugationsOperators}, que quer dizer autoespinores do operador de conjugação de carga} e suas propriedades, juntamente com suas relações com o espinor de Dirac podem ser encontrados
\cite{daRocha:2007pz}. Assim como, sua radiação Hawking através de tunelamento de horizonte de corda negra 
  \cite{bht,Cavalcanti:2015nna} e, através de uma Lagrangiana em supergravidade, uma ação de gravidade para cada tipo de espinor, em particular, o próprio Elko \cite{daRocha:2009gb}. Este pode ser útil em revelar simetrias escondidas no modelo Duffin-Kemmer-Petiau, mostrando que uma teoria da relatividade muito especial seria suficiente ao experimento de Michelson-Morley
\cite{Cavalcanti:2014uta, Cohen:2006ky}. Um resultado mais recente nos revela um caráter mais realístico do Elko, através de um mapeamento entre os campos de Dirac e os campos de Elko, que se dá por meio de uma extensão do modelo padrão que inclui a matéria escura \cite{HoffdaSilva:2009is}.

Conceitualmente, um produto bilinear entre dois espinores pode ser escrito como uma
 combina\c c\~ao linear de produtos de bilineares covariantes de cada um dos seus respectivos espinores, essas 
 importantes identidades recebem o nome de identidades de Fierz-Pauli-Kofink, cujas aplicações mostramos acima. No entanto, utilizaremos a denominação identidade de Fierz daqui para frente. 
Vejamos, se considerarmos o produto interno $\langle A, B\rangle:=tr(B^\dagger A)$, entre as matrizes $\G^A$ este é dado por: 
$$
\f{1}{4}tr[\Gamma^A\Gamma^B]=\eta^{AB},
$$
então as identidades de Fierz mais gerais em quatro dimens\~oes são escritas como \cite{Pesk}
\be
(\bar{u}_1\Gamma^Au_2)(\bar{u}_3\Gamma^B u_4)=\sum_{C,D}C^{AB}_{CD}(\bar{u}_1\Gamma^Cu_4)(\bar{u}_3\Gamma^D u_2), 
\ee
onde o coeficiente $C^{AB}_{CD}$ \'e obtido a partir da rela\c c\~ao de completeza das 16 matrizes $\Gamma^A$: 
\be
C^{AB}_{CD}=\f{1}{16}tr[\Gamma^C\Gamma^A\Gamma^D\Gamma^B]
.\ee
Essas identidades de Fierz podem ser construídas em termos dos bilineares covariantes estudados na se\c c\~ao anterior, seguindo uma abordagem diferente \cite{lou2,cra,hol}.
\begin{equation}\label{fifi}
-(\omega+\sigma\gamma^{5})\,\mathbf{S}=\mathbf{J}\wedge\mathbf{K},\qquad\mathbf{K}^{2}+\mathbf{J}^{2}
=0=\mathbf{J}\cdot\mathbf{K},\qquad
\mathbf{J}^{2}=\omega^{2}+\sigma^{2}\,,
\end{equation}
as quais como dissemos é bastante útil na classificação de espinores. Assim, o espinor é dito ser regular, se $\s+\omega\g_5\neq 0$. Caso contrário, ele é singular. Um campo multivetorial $Z$ pode ser construído fazendo-se uso dos bilineares covariantes associados às propriedades importantes do el\'etron, acima listados \cite{lou2}:
\begin{equation}
Z=\omega\gamma_{5}+i\mathbf{K}\gamma_{5}+i\mathbf{S}+\mathbf{J}+\sigma
\label{boomf}
.\end{equation} Esse objeto \'e denominado agregado de Fierz, se $\omega, \mathbf{S}, \mathbf{K}, \mathbf{J}$ e $\sigma$ satisfazem as identidades de Fierz (\ref{fifi}). Al\'em disso, se $Z$ satisfaz a identidade $\gamma^{0}Z\gamma^{0}=Z^{\dagger}$, ele \'e chamado de {\it boomerang} \cite{lou2}.

Quando espinores singulares s\~ao examinados, as identidades de Fierz (\ref{fifi}) podem ser substitu\'idas por express\~oes mais gerais, válidas também para espinores regulares \cite{cra}:
\beq\nonumber
 Z^{2} =4\sigma Z,\qquad Z\gamma_{\mu}Z=4J_{\mu}Z,\qquad iZ\gamma_{\mu\nu}Z=4S_{\mu\nu}Z,\nonumber\\
 iZ\gamma_{5}\gamma_{\mu}Z=4K_{\mu}Z,\qquad -Z\gamma_{5}Z=4\omega Z.
\label{boom}
\eeq
No espa\c co-tempo de Minkowski 4D, as identidades de Fierz vinculam os espinores e nos d\~ao importantes propriedades sobre esses, com algumas aplica\c c\~oes inesperadas. Se tentamos generalizar essas identidades para dimens\~oes superiores, podemos encontrar v\'inculos mais gerais com os bilineares covariantes. Isso foi feito por meio da \'algebra geom\'etrica para dimens\~oes e assinaturas arbitr\'arias \cite{1,Babalic:2013fm} e \'e baseado no que foi desenvolvido em \cite{okubo11,rand}.
 {

Uma abordagem diferente do agregado de Fierz pode ser obtida através de representações espinoriais da álgebra de Lorentz $\mathfrak{so}(1,3)$. Suas representações espinoriais fundamentais são $(\f{1}{2},0)$ e $(0,\f{1}{2})$, as quais levam em conta os espinores de Weyl levógiro $\psi_-$ e dextrógiro $\psi_+$, respectivamente. Por essa razão, ela é isomorfa à soma direta $\mathfrak{su}(2)\oplus \mathfrak{su}(2)$. De outra forma, elementos da representação $(0,\f{1}{2})\oplus (\f{1}{2},0)$ agem sobre espinores de Dirac e representações irredutíveis do grupo de Lorentz podem ser obtidas, através do seguinte produto tensorial \cite{Ticc}
\be\nonumber
\hspace{-.5cm}{\small{\begin{tabular}{cccccc}
&$\sigma$&${\bf J}$&${\bf S}$&${\bf K}$&$\omega$\\
$[(0,\f{1}{2})\oplus(\f{1}{2},0)]\otimes [(0,\f{1}{2})\oplus(\f{1}{2},0)] =$&$\overbrace{(0,0)}\oplus $&$\overbrace{\left(\f{1}{2},\f{1}{2}\right)}\oplus$ &$\overbrace{(1,0)\oplus(0,1)}\oplus$ &$\overbrace{\left(\f{1}{2},\f{1}{2}\right)}$&$\oplus\overbrace{(0,0)}$%
.\end{tabular}}}
\ee
Note que esse produto é decomposto em uma soma direta das partes homogêneas do agregado. Considerações físicas determinam um conteúdo geométrico subjacente ao agregado de Fierz, que é representado acima de outra forma ao levarmos em conta o isomorfismo ${\rm Spin}_{1,3}\simeq {\rm SL}(2,\mathbb{C})$. 
Por exemplo, o vetor densidade de corrente ${\bf J}$, que é representado por $\psi^*_+\s^\m\psi_+$ e  $\psi^*_-\bar{\s}^\m\psi_-$, é rotacionado  ou {\it boosted} por ações dos elementos de $(\f{1}{2},\f{1}{2})$. Enquanto, que o termo $(1,0)\oplus (0,1)$ da densidade de spin denota as representações à esquerda e à direita e está relacionado às matrizes antissimétricas.
Existe ainda uma rela\c c\~ao entre a exist\^encia de duas formas bilineares covariantes naturais  sobre o espa\c co de espinores e elementos da \'algebra exterior \cite{5} e, reciprocamente, h\'a uma reconstru\c c\~ao de espinores a partir das componentes multivetorias de $Z$, através das identidades de Fierz \cite{14}.

\subsection{Reconstrução de espinores}
\qquad Para espinores regulares, o agregado $Z$, expresso na eq. (\ref{boomf}), é fatorado da seguinte forma:
\be\label{fator}
Z=(\s+{\bf J}+\omega\g_5)(1+i(\s+\omega\g_5)^{-1}{\bf K}\g_5).
\ee
Se os multivetores $\s,{\bf J,S,K},\omega$ satisfazem as identidades de Fierz, cada um dos multivetores está associado a um bilinear do espinor $\psi$, que pode ser reobtido a menos de uma diferença de fase a partir do agregado de Fierz $Z$ por meio da fatorização (\ref{fator}) \cite{lou2,cra}:
 \be
 \psi=\f{Z\zeta}{2\sqrt{\zeta^\dagger\g_0 Z\zeta}},\quad\zeta\,\text{ é um espinor arbitrário não-nulo}. 
 \ee
 Reciprocamente, 
 \be
 Z=4\psi\psi^\dagger\g^0=4\psi\bar{\psi}. 
 \ee
O que torna as identidades de Fierz de grande relev\^ancia s\~ao suas recentes aplica\c c\~oes na classifica\c c\~ao de novos campos de part\'iculas espinoriais, as quais tem sido cada vez mais abrangentes, principalmente nos \'ultimos dez anos. Elas foram usadas por Lounesto para classificar espinores sobre o espa\c co de Minkowski por meio de bilineares covariantes \cite{lou2}. De fato, ele mostrou que espinores constru\'idos sobre o espa\c co-tempo podem ser distribu\'idos em seis classes disjuntas. Consequentemente, tr\^es dessas s\~ao regulares e englobam espinores de Dirac, enquanto as outras tr\^es classes s\~ao singulares, onde Weyl e Majorana são casos particulares. Além dessas, há outras três classes com ${\bf J}=0$ \cite{J0}. Na próxima seção, explicaremos e detalharemos melhor essas classes. 

\section{Classifica\c c\~ao de Espinores no Espa\c co-Tempo de Minkowski}

\qquad Fisicamente, cada bilinear de um espinor est\'a associado a um observável. Uma classifica\c c\~ao dos espinores baseada em seus bilineares covariantes imp\~oe v\'inculos sobre estes e s\~ao de extrema import\^ancia para a f\'isica de part\'iculas na busca por novos campos de part\'iculas e suas propriedades.

Lounesto classificou espinores sobre o espa\c co-tempo nas seis  seguintes classes disjuntas \cite{lou2}. Por raz\~oes f\'isicas, temos para todas as classes que a corrente $\mathbf{J}\neq 0$. Est\'a impl\'icito que as classes  (1), (2) e (3) abaixo possuem $\mathbf{K}$ e $\mathbf{S}$ n\~ao-nulos:

\begin{itemize}
\item[1)] $\sigma\neq0,\;\;\; \omega\neq0$.

\item[2)] $\sigma\neq0,\;\;\; \omega= 0$.\label{dirac1}

\item[3)] $\sigma= 0, \;\;\;\omega\neq0$.\label{dirac2}

\item[4)] $\sigma= 0 = \omega, \;\;\;\mathbf{K}\neq0,\;\;\; \mathbf{S}\neq0$.\label{tipo4}

\item[5)] $\sigma= 0 = \omega, \;\;\;\mathbf{K}= 0, \;\;\;\mathbf{S}\neq0$.\label{type-(5)1}

\item[6)] $\sigma= 0 = \omega, \;\;\; \mathbf{K}\neq0, \;\;\; \mathbf{S} = 0$.
\end{itemize}

Os tipos de espinores (1), (2) e (3) s\~ao chamados de espinores regulares para part\'iculas de {\it spin}-1/2 pelo fato de o bilinear covariante escalar ou pseudoescalar ser diferente de zero. De outra sorte, para as outras tr\^es classes de espinores (4), (5), e (6), ambos os bilineares covariantes escalar e pseudoescalar s\~ao nulos e, da\'i, esses serem denominados espinores singulares, respectivamente, {\it flag-dipoles}, {\it flagpoles} e {\it dipoles}. 


Este \'ultimo tipo de espinor -- {\it dipole} -- inclui os espinores de Weyl \cite{lou2}, os quais, sendo levógiros $\xi_-$ ou dextrógiros $\xi_+$, satisfazem $\f{1}{2}(1\pm \g_5)\xi_{\mp}=0, \;\; \f{1}{2}(1\pm\g_5)\xi_{\pm}=\xi_{\pm}$ e possuem uma rica geometria \cite{lou2,bookroldao}. Esses espinores têm uma nova din\^amica, ditada pelas seguintes equações:
\be  i\bar{\s}^\m\p_\m\psi_L=0,\qquad i\s^\m\p_\m\psi_R=0, 
\ee onde $\s^\m\equiv(1,\s^1,\s^2,\s^3)$ e $\bar{\s}^\m\equiv(1,-\s^1,-\s^2,-\s^3)$. Essas equações são resultantes do desacoplamento das equações de Dirac em duas outras para espinores de duas componentes, quando escolhemos $m=0$. 
Espinores do tipo {\it flagpole} foram considerados recentemente em cosmologia \cite{shank} e t\^em sido explorados como candidatos a mat\'eria escura em v\'arios contextos \cite{exotic,daSilva:2016htz,daRocha:2016bil,alex,lee2,lee1}. Como caso particular de {\it flagpole}, temos um tipo especial de campos fermiônicos de dimensão canônica de massa um, denominado Elko 
  \cite{daRocha:2009gb,daRocha:2007sd}, que \'e utilizado em m\'etodos de tunelamento em buracos negros, como mencionamos mais acima \cite{bht,Cavalcanti:2015nna}. No apêndice \ref{elko}, expomos de maneira mais detalhada os fundamentos sobre esse tipo de espinor. Além desse, há outro {\it flagpole}, o espinor de Majorana, que pode ser construído a partir de um espinor de Weyl:
 $$\phi_M=\xi_-+\xi_-^\dagger=\left( \begin{tabular}{c}
 $\xi$ \\ $-i\s_2\xi^*$
\end{tabular}\right),$$
o qual satisfaz a equação de Majorana $-i\slashed{\p}\psi+mC\psi=0$, onde $C=\g^2K$ é o operador de conjugação de carga 
e $K$ é a conjugação complexa 
 \cite{Frank}. É interessante notar que para partículas de spin-$\f{1}{2}$, a conjugação de carga ($C$), juntamente com a paridade ($P$) e a inversão temporal ($T$), totalizam as simetrias discretas do espaço-tempo subjacente. Disso, emerge um novo tipo de campo, denominado campo de anti-matéria \cite{DVA}. Em geral, para {\it flagpoles}, os autovalores do operador de carga podem ser quaisquer elementos do grupo $U(1)$ [$C\psi=e^{i\theta}\psi$]. Em particular, $C$ pode ter autovalores $\pm 1$, que correspondem ao Elko e ao espinor de Majorana. 
O primeiro exemplo f\'isico de espinor do tipo {\it flag-dipole} foi encontrado recentemente  no  contexto das teorias de gravidade ESK (Einstein-Sciama-Kibble), como solu\c c\~oes das equa\c c\~oes de Dirac em um fundo-$f(R)$ com tor\c c\~ao \cite{esk,Fabbri:2010pk,Fabbri:2013gza,Casana:2002fu,Fabbri:2012qr,Fabbri:2011mi}.

Os espinores de Dirac usualmente estudados estão restritos a primeira classe dos espinores regulares. Os quais, em particular, podem ser obtidos como soluções livres da equação de Dirac, para as quais são obtidas representações através de um {\it boost} na representação do referencial da partícula, como segue.

No referencial da partícula, $p=(m,\vec{0})$ e a equação de Dirac $(\g^\m p_\m-m\cdot 1_4)u(p)=0$ se torna $(m\g^0-m)u(p_0)=0$, cuja solução é $u(p_0)=\sqrt{m}\left(\begin{tabular}{c}
$\xi$\\ $\xi$ 
\end{tabular}\right)$, onde $\xi=\left(\begin{tabular}{c}
$1$\\ $0$
\end{tabular}\right)$ indica que a partícula tem spin ao longo da direção $z$. Por meio da parametrização $\left(\begin{tabular}{c}
$ E$\\ $p^3$
\end{tabular}\right)=\left(\begin{tabular}{c}
$ m\cosh \eta$\\ $m \sinh \eta$
\end{tabular}\right),$ essa solução pode ser escrita em um referencial arbitrário, através de um {\it boost}, o qual resulta no seguinte espinor \cite{Pesk}
$$
u(p)=\sqrt{m}\left(\begin{tabular}{c}
$\sqrt{p\cdot\s}\xi$\\ $\sqrt{p\cdot\bar{\s}}\xi$ 
\end{tabular}\right).
$$ 
Essa é a representação mais geral para o espinor de Dirac.

Todas essas classes de espinores foram caracterizadas com detalhes recentemente \cite{Cavalcanti:2014wia} 
e uma revis\~ao completa sobre classifica\c c\~ao de espinores foi feita, juntamente com suas aplica\c c\~oes a teoria de campos e gravita\c c\~ao \cite{daSilva:2012wp}. Essa classifica\c c\~ao tamb\'em foi bastante explorada no contexto de termodin\^amica de buracos negros \cite{bht,Cavalcanti:2015nna}. 

Além dessas seis classes que possuem espinores com densidade de corrente não-nula, foram encontradas outras três classes com ${\bf J}=0$. Por causa disso, esses espinores não descrevem partículas de Dirac, mas apenas satisfazem a equação de Klein-Gordon e, daí, sua dimensão de massa é igual a um. Essas classes são listadas abaixo \cite{J0}:
\beq\nonumber
\s=\omega=0,\;\;\left\{\begin{tabular}{cc}
${\bf S}\neq 0$,& ${\bf K}\neq 0$,\\ \nonumber
${\bf S}\neq 0$,&${\bf K}= 0$,\\ \nonumber
${\bf S}= 0$,&${\bf K}\neq 0$.
\end{tabular}\right.
\eeq

Conjectura-se que tais classes representam campos fantasmas. Tal estudo ainda está em desenvolvimento. 

Para que a din\^amica desses campos, de cada classe, possa ser explorada, a descrição dada pela Lagrangiana através dos bilineares covariantes correspondentes é suficiente. Relembremos, então, o estudo das simetrias dessa Lagrangiana e suas respectivas corrente e carga conservadas. No entanto, para facilitar essa compreensão acerca dos campos espinoriais, torna-se necessária uma abordagem para o campo escalar. Seja, então, uma transformação infinitesimal no campo escalar $\phi$ \cite{Pesk}
\be
\phi(x)\mapsto \phi(x)+\d \phi(x). 
\ee
Considere a densidade Lagrangiana ${\cal L}(\phi(x),\p_\m\phi(x))$, então o princípio da mínima ação de Hamilton : 
\beq
0=\d \int d^4 x \,{\cal L}=\int d^4 x\left\{\left[\f{\p {\cal L}}{\p \phi}-\p_\m\left(\f{\p {\cal L}}{\p(\p_\m\phi)}\right)\right]\d \phi+\p_\m\left(\f{\p {\cal L}}{\p(\p_{\m}\phi)}\d \phi\right)\right\}
\eeq
garante que a Lagrangiana é invariante a menos de uma transformação de calibre 
$$
{\cal L}(x)\mapsto {\cal L}(x)+\p_\m{\cal J}^\m_\d(x).
$$
Como resultado, obtemos \cite{Pesk}
\be
\p_\m\underbrace{\left(\f{\p{\cal L}}{\p(\p_\m \phi)}\Delta \phi-{\cal J}^\m\right)}_{j^\m(x)}=0, 
\ee
onde $j^\m(x)$ é a densidade de corrente, que está associada à simetria do sistema. 
Essa simetria nos garante uma grandeza que é constante no tempo, denominada carga $Q:=\int d^3 x\; j^0$. Note que a equação de continuidade  $\p_\m j^\m=0$ deve ser satisfeita. Isso pode ser verificado também para
campos espinoriais $\psi$, para o qual a densidade de corrente é  $J^\m=\bar{\psi}\g^\m\psi$. De fato, como $\psi$ satisfaz a equação de Dirac $\g^\m\p_\m\psi=-im\psi$ [$\p_\m\bar{\psi}\g^\m=im\bar{\psi}$], temos \cite{Pesk}
\beq
\p_\m J^\m&=&(\p_\m\bar{\psi})\g^\m\psi +\bar{\psi}\g^\m(\p_\m\psi)=0.
\eeq 
Além disso, se escrevemos o campo $\psi$ em termos dos operadores de criação e aniquilação \cite{Pesk}:
\be \psi(x)=\int \f{d^3p}{(2\pi)^3}\f{1}{\sqrt{2E_{\bf p}}}\sum_s\left(a^s_{\bf p}u^s(p)e^{-ip\cdot x}+b^{s\dagger}_{\bf p}v^s(p)e^{ip\cdot x}\right),
\ee
a carga associada à densidade de corrente $J^\m$ é dada por
\be
Q=\int d^3 x\; :\bar{\psi} \g^0\psi:=\int d^3 x\; :\psi^\dagger(x)\psi(x):=\int \f{d^3p}{(2\pi)^3}\sum_s\left(:a^{s\dagger}_{\bf p}a^s_{\bf p}:+:b^s_{-{\bf p}}b^{s\dagger}_{-{\bf p}}:\right),
\ee
onde $:\;\;:$ denota o ordenamento normal. De forma equivalente, ainda no formalismo de primeira quantização, usando a equação de Dirac, obtemos a equação
de continuidade para a densidade de corrente quiral $J^{\m 5}=\bar{\psi}\g^\m\g^5\psi=K^\m$:
\beq
\p_\m J^{\m 5}=(\p_\m\bar{\psi})\g^\m\g^5\psi +\bar{\psi}\g^\m\g^5(\p_\m\psi)=2im\bar{\psi}\g^5\psi,
\eeq
onde a mesma ($K^\m=J^{\m 5}$) se conserva apenas quando $m=0$, como nota-se diretamente da equação acima. Além da densidade de corrente $J^\m$, a simetria rotacional para espinores também pode ser obtida através do teorema de Noether. Dessa forma, a partir de uma rotação infinitesimal do espinor $\psi$ de um ângulo $\theta$ em torno do eixo $z$ \cite{Pesk}:
\be
\d \psi(x)=\Lambda_{\f{1}{2}}\psi(\Lambda^{-1}x)-\psi(x)=-\theta(x\p_y-y\p_x+\f{i}{2}\Sigma^3)\psi(x)\equiv \theta \Delta\psi,
\ee
onde $\Lambda_{\f{1}{2}}$ e $\Lambda$ são as transformações de Lorentz espinoriais e vetoriais respectivamente, é obtida a componente temporal da corrente de Noether:
\be
J^0=\f{\p {\cal L}}{\p(\p_0\psi)}\Delta\psi=-i\bar{\psi}\g^0(\overbrace{x\p_y -y\p_x}^{L_z} +\f{i}{2}\Sigma^3)\psi.
\ee
Nesta expressão, o spin intrínseco pode ser representado por $\Sigma^j:=1_2\otimes\s^j,\;j=1,2,3$.
Seguindo o mesmo raciocínio para as outras componentes da densidade de corrente para a simetria rotacional, a seguinte expressão  para o operador de momento angular pode ser obtida \cite{Pesk}:
\be
{\bf J}=\int d^3x\psi^\dagger\left({\bf x}\times\left(-i\nabla\right)+{\small \f{1}{2}}{\bf \Sigma}\right)\psi. 
\ee
Considere partículas em repouso, então apenas o termo espinorial contribui. Dessa forma, a componente $z$ do momento angular pode ser expressa por \cite{Pesk}:
\beq 
J_z&=&\int d^3x\int \f{d^3pd^3p'}{(2\pi)^6}\f{1}{\sqrt{2E_{\bf p}2E_{{\bf p}'}}}e^{-i{\bf p}'\cdot {\bf x}}e^{-i{\bf p}\cdot {\bf x}}\\ &&\times\sum_{r,r'} \left(a^{r'\dagger}_{{\bf p}'}u^{r'\dagger}({\bf p}')+b^{r'}_{-{\bf p}'}v^{r'\dagger}(-{\bf p}')\right)\f{\Sigma^3}{2}\left(a^r_{\bf p}u^r({\bf p})+b^{r\dagger}_{-{\bf p}}v^r(-{\bf p})\right).
\eeq
\indent
Depois de serem exploradas as simetrias e suas cargas associadas, podemos observar que os espinores de Dirac \footnote{Espinores de Dirac são espinores que satisfazem a equação de Dirac e de paridade [$P\psi=\pm \psi$].} (bem como Weyl e Majorana) não são o único tipo de espinores que existem. Apesar de haver um teorema em que apenas eles são permitidos, se satisfeitas algumas hipóteses. Isso pode ser contornado, através da violação de pelo menos uma dessas hipóteses, permitindo a existência de outros espinores que não são de Dirac. O teorema mencionado é o teorema de Weinberg-Witten \cite{Wein1,Wein}, do qual resulta que espinores de spin-$\f{1}{2}$, dimensão de massa $\f{3}{2}$, com grupo de simetria ${\rm SL}(2,\mathbb{C})$ e que satisfazem o grupo completo de transformações de Lorentz sobre o espaço de Minkowski só podem ser espinores de Dirac. Sobre considerações a respeito desse teorema, generalizações para partículas de spin de ordens superiores e suas respectivas equações de campo reportados e discutidos na referência \cite{Lee:2012td}, consideramos aqui apenas férmions de spin-$\f{1}{2}$ e bósons de spin $0$ e $1$ \cite{Dirac}. Além disso, Fierz e Pauli estudaram o caso de partícula de spin arbitrário na presença de um campo eletromagnético \cite{Fierz}.

}
 
 Em outras palavras, 
se os espinores deixam de satisfazer uma das condições, não será obrigatório que esse teorema seja satisfeito, consequentemente o espinor não será necessariamente de Dirac. Assim, uma das hipóteses é contornada, se a dimensão de massa é um. Nesse caso, podem ser considerados {\it flagpoles}, como por exemplo o Elko. Todos esses campos de espinores têm sido amplamente estudados na última década: foi feita uma abordagem bastante abrangente sobre álgebras de Clifford e campos de espinores \cite{bookroldao}; um novo campo quântico de spin-$\f{1}{2}$ com dimensão de massa $1$ foi construído, quando o teorema de Weinberg é contornado \cite{Wein}; foi mostrado que o Elko pode ser um bom candidato a matéria escura, pelo fato de haverem poucas interações com outras partículas do modelo padrão \cite{lee2,lee1,Cheng}, além de evitar, no contexto do período de expansão acelerada do nosso universo, o problema do ajuste fino associado às condições iniciais \cite{shank}; foram construídos possíveis cenários advindos do acoplamento exóticos entre esses campos espinoriais escuros (Elko) e campos escalares topológicos \cite{exotic,daSilva:2016htz,daRocha:2016bil,alex}. Além dessas, foram feitas outras aplicações desses campos de espinores em gravitação, cosmologia, matéria escura, branas e física de partículas. 
Não entraremos em mais detalhes a respeito.

No cap\'itulo seguinte, utilizaremos o mesmo m\'etodo utilizado aqui para classificar espinores, de acordo com seus bilineares covariantes, sobre espaços Riemannianos de sete dimens\~oes, em particular a esfera $S^7$, bastante útil em supergravidade em $AdS_4\times S^7$, e sobre espaços Lorentzianos de cinco dimensões, por exemplo o {\it bulk} ($AdS_5$), atrav\'es dos v\'inculos vindos das identidades de Fierz.

\myclearpage
\par

\chapter{Classifica\c c\~ao de Espinores em Dimens\~oes Arbitr\'arias}

Neste capítulo, exploraremos vários aspectos de espinores em espaços de dimensões e assinaturas arbitrárias e listaremos todos os bilineares covariantes possíveis. Além disso, estenderemos os espinores de Majorana através de uma complexificação do grupo spin $Spin^+_{p,q}$, obtendo, assim, elementos de $Spin_{p,q}$. Em particular, classificaremos nas seções que se seguem espinores sobre espaços Riemannianos de sete dimensões, por exemplo a esfera $S^7$, e sobre espaços Lorentzianos de cinco dimensões, como o {\it bulk} ($AdS_5$). 
Esses casos são muito importantes na exploração da dinâmica de espinores em supergravidade. É interessante notar que uma teoria sem massa em onze dimensões ganha um espectro de massas de Kaluza-Klein, quando compactificada. Esse espectro poderá ser analisado futuramente, tanto sobre o espaço $AdS_4\times S^7$, quanto sobre o espaço $M_4\times S^1/\mathbb{Z}_2$, onde $M_4$ denota o espaço-tempo de Minkowski, que é o {\it bulk} no caso do modelo de Randall-Sundrum, através do cálculo do potencial efetivo para cada uma das classes de campos espinoriais obtidas nesse trabalho. Isso também poderá ser feito futuramente. Não entraremos em mais detalhes sobre supergravidade \footnote{Verifique mais informações nessa referência \cite{BBS}.}\cite{BBS}.

Para classificarmos campos espinoriais sobre espaços Riemannianos de sete dimensões  \footnote{Motivado pela compactificação de $AdS_4\times M^7$ é usual considerar-se $M^7$ como sendo uma variedade compacta \cite{engl82}.}, serão introduzidas uma estrutura real $D$ e uma complexa $J$, que agem como uma conjugação complexa ${\psi}^*$ e como um produto com a unidade imaginária $i\psi$, respectivamente. A admissibilidade de um dado bilinear $B$ é dada por alguns sinais, que permitem que alguns vínculos sejam obtidos, os quais  anulam algumas componentes dos bilineares covariantes. Posteriormente, o uso das identidades de Fierz, que também são v\'inculos entre os bilineares covariantes, garantirá a existência de apenas uma classe de espinores de Majorana em sete dimensões. Finalmente, após complexificar esses bilineares covariantes e obter espinores e bilineares covariantes mais gerais, garantiremos a existência de duas novas classes não-triviais de espinores em espaços de sete dimensões Riemannianas, como por exemplo, sobre a esfera $S^7$: uma classe regular e outra singular.
Um procedimento bastante análogo é seguido ao classificar campos de espinores sobre o {\it bulk} $AdS_5$.

Essas classifica\c c\~oes são baseadas no mesmo crit\'erio pelo qual Lounesto classificou espinores em $\mathbb{R}^{1,3}$ em seis classes disjuntas, isto \'e, com base nos bilineares covariantes \cite{SUGRA,SUSYBr,engl}.
\section{Formas Bilineares 
em Dimens\~oes e Assinaturas Arbitr\'arias}

\qquad Nesta seção, através das estruturas real $D$ e complexa $J$ e das simetrias de uma forma bilinear covariante admissível $B$, outros bilineares covariantes serão construídos a partir desta. Por último, todos os bilineares covariantes para espinores em espaços de dimensões e assinaturas arbitrárias serão listados em termos de suas componentes. 
\subsection{Estruturas Real $D$ e Complexa $J$} 

\qquad Seja o fibrado de K\"ahler-Atiyah $(\sec \bigwedge(TM),\circ)$, onde $\circ$ denota o produto de Clifford. A partir de sua parte par $(\sec \wedge^{2k}(TM),\circ)$, o fibrado espinorial $S$ associado \`a variedade $(M,g)$, a qual possui dimensão $n$ e assinatura $p+q$, pode ser construído, atrav\'es de uma estrutura de m\'odulo\footnote{Módulo é uma extensão do conceito de espaço vetorial, tal que os escalares não são elementos de um corpo, mas pertencentes a um anel. Por isso, a multiplicação é apenas associativa e distributiva.} dada pelo morfismo\footnote{O conjunto dos morfismos é o conjunto de aplicações que possui a identidade e cuja composição é associativa.} $\g:(\wedge(TM),\circ)\;\r\; (End(S),\Ganz)$ \cite{moro,1,Babalic:2013fm,bonora,face,lazaroiu}.

Um endomorfismo\footnote{Endomorfismo é um morfismo de um espaço nele mesmo, em nosso caso o fibrado $S$.}  idempotente\footnote{$R$ é idempotente, pois satisfaz $R\circ R=1$.} $R$ sobre $S$, que em algumas dimens\~oes e assinaturas pode ser identificado com o elemento de volume $\g^{n+1}=\g^1\Ganz\cdots\Ganz\g^n$
, o qual satisfaz $(\g^{n+1})^2=(-1)^{\f{n(n-1)}{2}+q}$ e decomp\~oe este mesmo fibrado $S$ nos sub-fibrados $S^+$ e $S^-$, ou seja, $S=S^+\oplus S^-$. Esses sub-fibrados $S^+$ e $S^-$ são determinados pelos auto-valores de $R$: $+1$ e $-1$, respectivamente, 
e obtidos atrav\'es de proje\c c\~oes de spin $\Pi_\pm=\f{1}{2}(I\pm R)$, onde $I$ denota a identidade sobre $S$ e $S^\pm=\Pi_\pm(S).$ Por meio dessas decomposi\c c\~oes, podemos encontrar representa\c c\~oes irredut\'iveis para $S$. Essa decomposi\c c\~ao \'e dita ser n\~ao-trivial, se ambos $S^+$ e $S^-$ s\~ao n\~ao-nulos.

Sobre esse fibrado $S$, podem ser definidos ainda os endomorfismos {\it D} e {\it J}, de forma que o primeiro nos fornece uma estrutura real que implementa a conjuga\c c\~ao complexa, enquanto que ${\it J}$ define uma estrutura complexa \cite{1,okubo}, isto é, $D$ e $J$ são definidos por:
\be
{\xi}^*:=D(\xi);\qquad J^2=-id_S,
\ee
 as quais obedecem as seguintes propriedades: 
\be 
D\Ganz D=(-1)^{\f{1+p-q}{4}}id_S,\;\; [J,D]_{+,\Ganz}=0,\;\;[J,\g(\omega)]_{-,\Ganz}=0, \;\;\forall\omega\in\sec \wedge(TM),
\ee
onde denotamos o comutador com uma multiplica\c c\~ao $\Ganz$ restrita a $S^\pm$ por $[\ \ ,\ \ ]_{\pm,\Ganz}$. Em particular, se $p-q=3,7 \mod 8$, a estrutura complexa $J$ se identifica com o elemento de volume $\g^{n+1}$, a menos de um sinal.

Espinores reais em espaços de dimensões arbitrárias s\~ao caracterizados, juntamente com sua estrutura real $D$, conforme \'e dado na tabela abaixo
\medbreak

\begin{tabular}{||c||c|c||}
\hline\hline
$p-q$
&$D^2$&\;$R$ (espinores reais)\;\\
\;\text{mod} 8\;&&\\
\hline\hline
0&-&$\g^{n+1}$ (Majorana-Weyl)-$S^\pm$\\
\hline
1&-&-\\
\hline
2&-&-\\
\hline
3&\,$-id_S$\,&-\\
\hline
4&-&$\g^{n+1}$ (Majorana-Weyl simpl\'etico)-$S^\pm$\\
\hline
5&-&-\\
\hline
6&-&$\g^{n+1}\circ J$ (Majorana simpl\'etico)-$S^+$\\
\hline
7&$+id_S$&(Majorana duplo)-$S^+$\\
\hline
\end{tabular}
\medbreak   
 \noindent
 
{\hspace{1cm} \small Tabela 5. {Tipos de espinores reais em dimensões arbitrárias e sua estrutura real}
}
Determinadas as estruturas real e complexa sobre o fibrado $S$, resta-nos, na subseção seguinte, determinar simetrias com essas estruturas para as formas bilineares admissíveis, para que seja possível encontrar os vínculos, bastante úteis na nossa classificação dos campos espinoriais em classes.

\medbreak
\subsection{Forma Bilinear Admissível $B$ e Suas Simetrias}

\qquad Seja um referencial local $\{\e_{\a}\}_{\a=1\dots\Delta}$ sobre $S_+$, cuja dimens\~ao real \'e dada por $\D$ ( onde $\D=2^{[d/2]}$), constru\'ido a partir do co-referencial ortonormal local $\{e^a\}_{a =1,\, \ldots,\, n }\subset {\bf P}_{SO^e_{p,q}}(M)$ de $(M,g)$ \cite{alek1}. Dada uma forma bilinear arbitr\'aria $B$, para que ela seja admiss\'ivel, deve satisfazer:
\begin{itemize}
\item[a)] $B$ \'e sim\'etrica ou antissim\'etrica,
\item[b)] $S^+$ e $S^-$ s\~ao isotr\'opicos ou ortogonais com respeito a $B$, para $p-q\equiv 0,4,6,7\mod
8$,
\item[c)]Dados $\xi\in \sec\bigwedge (TM)$ e $\g(\xi)$, uma representa\c c\~ao de $\xi$, a identidade $\gamma(\xi)^\intercal =\gamma(\underaccent{\tilde}{\xi})$ é satisfeita, se e somente se $B(\gamma(\xi)\psi,\psi')= B(\psi,\gamma(\underaccent{\tilde}{\xi})
\psi')$, tal que, se $B$ \'e sim\'etrico, $\underaccent{\tilde}{\xi}$ \'e a revers\~ao usual $\tilde{\xi}$ ou a conjuga\c c\~ao de Clifford, caso contr\'ario.
\end{itemize}

 A partir dessa forma bilinear $B$ e das estruturas introduzidas $D$ e $J$, um bilinear mais geral pode ser construído por meio de uma complexificação, juntamente com outros três bilineares covariantes. Para isso, considere os campos espinoriais $\psi,\psi'\in \Gamma(S),$ onde $\Gamma(S)$ denota o espa\c co das se\c c\~oes suaves do fibrado espinorial $S$, 
então obtemos uma fibra correspondente 
 $\mathbb{C}$-bilinear $\b$ a partir de uma complexifica\c c\~ao do bilinear $B$ restrito a $S_+\otimes S_+$ ($B|_{S_+\otimes S_+}$) \cite{1}:
\be
\!\!\!\!\!\!\beta(\psi,\psi')\!=\! B\!\left(\!{}^{\rm (Re)}\psi,\!{}^{\rm (Re)}\psi'\right)\!-\!
B\!\left(\!{}^{\rm (Im)}\psi,\!{}^{\rm (Im)}\psi'\right)\!+\!i\!\left[B\!\left(\!{}^{\rm
(Re)}\psi,\!{}^{\rm (Im)}\psi'\right)\!+\! B\!\left(\!{}^{\rm (Im)}\psi,\!{}^{\rm
(Re)}\psi'\right)\right],\label{formaa}
\ee\noindent  onde as partes real e imagin\'aria de $\psi$ s\~ao, respectivamente, ${}^{\rm (Re)}\psi=\f{1}{2}(\psi+D(\psi))$ e ${}^{\rm
(Im)}\psi=\f{1}{2}(\psi-D(\psi))$. Tal bilinear $\b$ satisfaz:
 \beq
 \beta(J\xi,\xi')&=& \beta(\xi,J\xi')= i\beta(\xi,\xi'), \quad \forall \xi,\xi' \in \Gamma(M,S),\\
 \beta(\xi_+,\xi'_+)&=&B(\xi_+,\xi'_+), \quad\forall \xi,\xi' \in \Gamma(M,S_+),
 \eeq 
 onde $\G(M,S)$ denota o espa\c co de todas as se\c c\~oes ${\cal C}^\infty$ do fibrado espinorial $S$. Partindo de $\b$, reciprocamente, outros quatro bilineares  $\mathscr{B}_0,\mathscr{B}_1,\mathscr{B}_2,\mathscr{B}_3$ podem ser definidos por:
\beq
\mathscr{B}_0(\xi,\xi')&=&\beta({}^{\rm (Re)}\xi,{}^{\rm (Re)}\xi')-(-1)^{\f{d(d+1)}{2}}\beta({}^{\rm (Im)}\xi,{}^{\rm (Im)}\xi'),\\
\mathscr{B}_1(\xi,\xi')&=&-\beta({}^{\rm (Re)}\xi,{}^{\rm (Im)}\xi')-(-1)^{\f{d(d+1)}{2}}\beta({}^{\rm (Im)}\xi,{}^{\rm (Re)}\xi'),\\
\mathscr{B}_2(\xi,\xi')&=&\beta({}^{\rm (Re)}\xi,{}^{\rm (Re)}\xi')+(-1)^{\f{d(d+1)}{2}}\beta({}^{\rm (Im)}\xi,{}^{\rm (Im)}\xi'),\\
\mathscr{B}_3(\xi,\xi')&=&-\beta({}^{\rm (Re)}\xi,{}^{\rm (Im)}\xi')+(-1)^{\f{d(d+1)}{2}}\beta({}^{\rm (Im)}\xi,{}^{\rm (Re)}\xi')
.\eeq

É necessário introduzir alguns sinais, para que $B$ possa ser analisado com mais detalhes, explorado um pouco mais e uma classificação de espinores seja encontrada. 
O sinal $\s_\mathscr{B}=\pm 1$ é definido a partir da transposta de $\mathscr{B}_k$, por
\be
\mathscr{B}_k(\xi,\xi')=\s_k\mathscr{B}_k(\xi',\xi),\quad\forall\,\xi,\xi'\in \Gamma(M,S),{\rm\; onde\;}k=0,1,2,3,
\ee
 e \'e chamado de  simetria de $\mathscr{B}_k$. 
Outro sinal, que aparece de uma rela\c c\~ao entre os sub-fibrados $S^\pm$ e atrav\'es de $\mathscr{B}_k$, é $\iota_k$, denominado isotropia  de $\mathscr{B}_k$:
\beq
{\rm para}&\;p-q\equiv_8 0,4,6,7,&\iota_k:=\left\{\begin{array}{cc}+1,&\rm{se}\;\mathscr{B}_k(S^+,S^-)=0\,,\\ -1,&\rm{se}\;\mathscr{B}_k(S^\pm ,S^\pm)=0\;,
\end{array}\right.\\
{\rm para }& p-q\equiv_8 1,2,3,5,& \iota_k {\rm\ \ n\tilde{a}o\ \ \acute{e}\ \ definido}.\ \ 
\eeq 
Por fim, $\epsilon_k$ vem da transposta das matrizes $\g^A$ e \'e denominado  tipo de $\mathscr{B}_k$:
\be\label{e1}
(\g^m)^t=\e_k\g^m\;\Leftrightarrow\;\mathscr{B}_k(\g^m\xi,\xi')=\e_k\mathscr{B}_k(\xi,\g^m\xi').
\ee
No nosso caso, em que $p=7$ e $q=0$, estes sinais $\s_k,\e_k$ e $\iota_k$ são calculados e mostrados na tabela abaixo, pois serão úteis mais adiante \cite{1}:
 \beq
 \begin{array}{|c||c|c|c|c|c|}
\hline k&0&1&2&3\\ \hline
 \hline
\s_k &+1&+1&+1&-1\\
\hline
\e_k &-1&-1&+1&+1\\
\hline
\iota_k &+1&-1&+1&+1
\\
\hline
 \end{array}.
 \eeq 
 {\hspace{4cm} \small Tabela 6. {Sinais da forma bilinear $B$ para $p=7$ e $q=0$.}}
 \subsection{Bilineares covariantes em dimensões e assinaturas arbitrárias}

\qquad No cap\'itulo anterior, para que se produzisse uma quantidade invariante de Lorentz \footnote{Daqui em diante, para multi-\'indices $(\alpha_1,\ldots,\alpha_k)$, $1\leq \alpha_k\leq n$, usaremos a nota\c c\~ao $e^{\alpha_1\ldots \alpha_k}= e^{\alpha_1}\wedge\cdots\wedge e^{\alpha_k}$ e uma equivalente para termos contravariantes. Al\'em disso, o produto $\Ganz$ ser\'a denotado apenas por justaposi\c c\~ao.}
, o espinor dual de $\psi$ foi identificado com a conjuga\c c\~ao de spin $\bar\psi=\psi^\dagger\gamma^0$. Procurando uma quantidade invariante em dimens\~oes arbitr\'arias, encontramos formas bilineares mais gerais, que s\~ao produtos invariantes de spin e que podem ser escritas como 
$$\beta(\psi,\psi')=a^{-1}\tilde{\psi}\psi'= \psi^\dagger a^{-1}\psi',$$
onde $\psi, \psi'$ s\~ao campos espinoriais e $a\in\G({\rm End}(S))$ \cite{bt}. Dessa forma, denotando uma involu\c c\~ao adjunta arbitr\'aria por $\;\mathring{\mathring{}}\;$, resulta de $\tilde{a}=a$ e $\mathring{\mathring{b}}=ab^\dagger a^{-1}$ que $a^{-1}\tilde{a}=a^\dagger a^{-1}=1$. Logo, $a$ \'e hermitiano: $a^\dagger=a$ \cite{bt}. De uma forma geral, a conjuga\c c\~ao espinorial 
\'e dada por $\bar\psi=a^{-1}\tilde{\psi}=\psi^\dagger a^{-1}$, de onde segue o bilinear mais geral que pode ser escrito sobre $S$ e que \'e bastante estudado \cite{Pesk, 26}:
\be\label{formab}
\beta(\xi,\gamma_{\alpha_1\dots\alpha_k} \xi)\equiv {\overline{\xi}}{\gamma}_{\alpha_1\dots\alpha_k} {\xi}, {\rm\ \ onde\ \ }k<n.
\ee
%
Feito isso, fica claro como listar todos os bilineares covariantes em espaços de dimensões e assinaturas arbitrárias, os quais são escritos em termos de se\c c\~oes $\psi$ do fibrado espinorial $S$:
 \begin{subequations} \beq
 \Omega_1 &=&\bar{\psi}\psi\\
J_{\a_1}&=&\bar{\psi}\g_{\a_1}\psi\\
S_{\a_1\a_2}&=&\bar{\psi}\g_{\a_1\a_2}\psi\\
 &\vdots&\nonumber\\
 S_{\a_1\dots\a_{d+1}}&=&\bar{ \psi}\g_{\a_1\ldots\a_{d+1}} \psi\,,\qquad
d<k-1\\
 &\vdots&\nonumber\\
 S_{\a_1\ldots\a_k}&=&\bar{ \psi}\g_{\a_1\ldots\a_k}\psi \\
 K_{\a_1\ldots\a_{n-k-1}}&=&\bar{\psi}\g_{\a_1\ldots\a_{n-k-1}}\g_{n+1}\psi
\\
  &\vdots&\nonumber\\
 K_{\a_1\dots\a_m}&=&\bar{\psi} \g_{\a_1\ldots\a_m}\g_{n+1}\psi\,, \quad m<n-k-1\\
 &\vdots&\nonumber\\
 K_{\a_1}&=& \bar{\psi}\g_{\a_1}\g_{n+1}\psi\\
\Omega_2 &=&\bar{\psi}\gamma_{n+1}\psi\;.
 \eeq \end{subequations}

Dados multivetores arbitrários $\Omega_1,\, J_\m,\, \ldots,\, \Omega_2$ [um escalar, um vetor, \ldots, um escalar], se as identidades de Fierz são satisfeitas para esses multivetores, o seguinte agregado é denominado  agregado de Fierz 
\beq
{\rm Z}=\Omega_1 +J_{\mu}\g^{\mu}+S_{\mu_1\mu_2}\g^{\mu_1\mu_2}+\dots +
S_{\mu_1\ldots\mu_k}\g^{\mu_1\ldots\mu_k}+\nonumber\\
+K_{\a_1\ldots\a_{n-k-1}}\g^{\a_1\ldots\a_{n-k-1}}\g_{{n+1}}+\dots
+K_{\mu}\g^{\mu}\g_{{n+1}}+\Omega_2 \g_{{n+1}}\,,
\eeq \noindent que se reduz ao agregado de Fierz padr\~ao (\ref{boomf}) para $n=4$. Uma generaliza\c c\~ao do {\it boomerang} pode ser obtida de forma imediata. Assim, se as componentes $\Omega_1,\, J_\m,\, \ldots,\, \Omega_2$ são os bilineares covariantes de algum espinor $\psi\in S$ 
 e o agregado de Fierz generalizado acima satisfaz $\gamma^{0}{\rm Z}\gamma^{0}={\rm Z}^{\dagger}$, então $Z$ \'e denominado {\it boomerang} generalizado \cite{lou2,bt}.

Identidades de Fierz geom\'etricas s\~ao bastante utilizadas na busca por v\'inculos entre bilineares covariantes de espinores, os quais, como dissemos acima, nos d\~ao uma classifica\c c\~ao desses campos espinoriais de $S$ \cite{lou2}. Isso é feito na seção (\ref{42}) para espinores sobre variedades Riemannianas de sete dimensões, como por exemplo $S^7$, e na seção (\ref{43}), sobre variedades Lorentzianas de cinco dimensões, em particular o {\it bulk} ($AdS_5$).1


%

\section{Classifica\c c\~ao de Campos Espinoriais em Espaços Riemannianos de Sete Dimens\~oes }\label{42}
\qquad Nosso interesse aqui \'e determinar a natureza de campos espinoriais por meio de uma classificação em classes. 
Assim, campos espinoriais de Majorana são explorados nesses espaços, em particular, a esfera $S^7$, e o agregado de Fierz graduado pode ser definido. 
Como resultado, é obtida só uma classe de espinores de Majorana, já existente na literatura, de acordo com uma classifica\c c\~ao baseada em bilineares covariantes \cite{1}. Isso endossa os teoremas de álgebras de Clifford já conhecidos \cite{bookroldao}, mas analisamos isso por meio das identidades geométricas de  Fierz \cite{1}.
 A complexificação desses bilineares nos permitirá explorarmos campos 
pinoriais e obtermos mais duas classes n\~ao-triviais, ou seja, h\'a quatro classes de espinores para o caso complexo sobre espaços Riemannianas de sete dimensões: uma trivial, uma de Majorana em $S^+$, uma regular e uma singular.  
O endomorfismo $D$ mencionado acima, para um espaço $7$-dimensional Riemanniano, determina a conjuga\c c\~ao complexa, isto \'e, decomp\~oe um espinor $\psi$ em partes real e imagin\'aria. Além disso, quando $p-q=7\mod\, 8$, que \'e o nosso caso, $D$ \'e id\^entico ao endomorfismo $R$. Os projetores $\f{1}{2}(I\pm R)$ produzem uma decomposi\c c\~ao em fibrados espinoriais reais  $S=S^+\oplus S^-$. Reciprocamente,  o fibrado espinorial total $S$, pode ser identificado com a complexifica\c c\~ao do fibrado real $S^+$, que é formado por espinores de Majorana.

 Dado um bilinear $B$ admiss\'ivel sobre $S$, outros tr\^es bilineares podem ser obtidos. Se $n=7$, definimos os seguintes bilineares:
\be \label{b1b2} B_1:=  B \,\Ganz\,( I \otimes J), \qquad B_2:= - B \,\Ganz\, [ I \otimes (J\,\Ganz\, D)]\,,\qquad B_3:= B \,\Ganz\, ( I \otimes
D)\,\,.
\ee
Torna-se necessário também introduzir sobre $S$ o seguinte endomorfismo $A_{\psi|\psi'}$ \cite{1}:
\be
A_{\psi_1 |\psi_2}(\psi):= B (\psi,\psi_2)\psi_1\,,\quad \text{para todo}\;\;\;\;\;
\psi,\psi_1,\psi_2 \in \Gamma(S)\,,
\ee
o qual \'e bastante importante na determina\c c\~ao das identidades de Fierz geom\'etricas, que constam da seguinte express\~ao: 
\begin{equation}
A_{\psi_1 |\psi_2}\,\Ganz\, A_{\psi_3 |\psi_4}= B
(\psi_3,\psi_2)A_{\psi_1 |\psi_4}\,.
\end{equation}
Uma rela\c c\~ao de completeza, denominada agregado de Fierz graduado, construída com a forma bilinear $B$ tamb\'em pode ser considerada (\ref{boomf}):
\begin{equation}\nonumber
A_{\psi |\psi^\prime}=\frac{\ell}{2^n}\sum_k  \frac{1}{k!} (-1)^{k}  B
(\psi,\gamma_{\alpha_1\ldots \alpha_k} \psi')e^{\alpha_1\ldots
\alpha_k}\,,
\end{equation}
 tal que, se $p-q=0,1,2$, temos $\ell = 2^{\llceil\frac{n}{2}\rrceil}$ ou $\ell = 2^{\llceil\frac{n}{2}\rrceil+1}$  caso contr\'ario, onde denotamos $\llceil\frac{n}{2}\rrceil \equiv \frac{n(n-1)}{2}  \mod\,2$.
A decomposi\c c\~ao de $A_{\psi |\psi^\prime}\in\Gamma({\rm End}(S))$, através do endomorfismo $D$, \'e escrita de forma \'unica como 
$
A_{\psi |\psi'}=D\,\Ganz\, A^1_{\psi |\psi'}+A^0_{\psi |\psi'}$ \cite{1},
onde
\begin{subequations}
\begin{eqnarray}
 A  ^0_{\psi |\psi'} &=&\frac{\ell}{2^n}\sum_k  \frac{(-1)^{k}}{k!}   B (\psi,
\gamma_{\alpha_1\ldots \alpha_k}\psi')e^{\alpha_1\ldots \alpha_k}\,,\label{eoo1} \\
 A  ^1_{\psi |\psi'} &=&\frac{\ell}{2^n}\sum_{k}  \frac{1}{k!}(-1)^{\left(k+\frac{1+p-q}{4}\right)}
 B (\psi, D\,\Ganz\, \gamma_{\alpha_1\ldots
\alpha_k}\psi')e^{\alpha_1\ldots \alpha_k}\,. \label{eoo2}
\end{eqnarray}
\end{subequations}  
Cada componente pode ser decomposta ainda, de forma graduada, como $
 A^{\lambda}_{\psi |\psi'} =\frac{\ell}{2^n}\sum_k
A^{\lambda,k}_{\psi |\psi'}$, onde $\lambda = 0,1$, e as componentes  $A^{\lambda,k}_{\psi |\psi'}\in \sec\bigwedge^k(TM)$ s\~ao dadas por
\begin{subequations}\beq
 A ^{0,k}_{\psi |\psi'}&=& \frac{1}{k!} (-1)^{k}   B (\psi,
\gamma_{\alpha_1\ldots \alpha_k}\psi')e^{\alpha_1\ldots
\alpha_k}\,,\label{eok1}\\
 A ^{1,k}_{\psi |\psi'}&=& \frac{1}{k!} (-1)^{\frac{1+p-q}{4}+k}  B (\psi,
D\,\Ganz\, \gamma_{\alpha_1\ldots \alpha_k}\psi')e^{\alpha_1\ldots
\alpha_k}\,.\label{eok}
\eeq \end{subequations}
Segue então, que as identidades de Fierz (\ref{fifi}) podem ser generalizadas para dimens\~oes arbitr\'arias:
\begin{subequations}\begin{eqnarray}
 \widehat{A  ^0_{\psi_1 |\psi_2}}\circ  A  ^1_{\psi_3 |\psi_4} +   A
^1_{\psi_1 |\psi_2}\circ  A  ^0_{\psi_3 |\psi_4}
&=&   B (\psi_3,\psi_2) A  ^1_{\psi_1 |\psi_4}\,,\label{fp1}\\
  A  ^0_{\psi_1 |\psi_2}\circ  A  ^0_{\psi_3 |\psi_4} + (-1)^{\frac{1+p-q}{4}}
\widehat{A  ^1_{\psi_1 |\psi_2}}\circ  A  ^1_{\psi_3 |\psi_4}
&=&   B (\psi_3,\psi_2) A  ^0_{\psi_1 |\psi_4}\,.\label{fp2}
\end{eqnarray}\end{subequations}
Voltando à nossa discuss\~ao, dado um espinor de Majorana $\xi$ e uma forma bilinear admiss\'ivel $B$, essa pode ter seus graus de liberdade reduzidos, pois $B(\xi,\g^A\xi)$ \'e diferente de zero apenas quando $|A|=|(a_1,...,a_m)|=m$ \'e par \cite{1}. Outro vínculo resulta da simetria de $B$ e da transposta de $\g^A$. A equa\c c\~ao (\ref{e1}) implica a rela\c c\~ao
$(\g_a)^\intercal=\e_0\g_a=-\g_a$, que resulta em $(\g_{a_1\dots a_m})^\intercal=(-1)^{\f{m(m+1)}{2}}\g_{a_1\dots a_m}$ e juntamente com a simetria de $B$, nos dá
\be\label{3.19}
B_i(\xi,\g^A\xi')=\s_i B_i(\g^A\xi',\xi)=\s_i(-1)^{\f{m(m+\d_i)}{2}}B_i(\xi',\g^A\xi),\qquad i=0,1,2,3,
\ee
onde $\d_0=\d_1=1$, $\d_2=\d_3=-1$ e $B_0=B$. Então, temos que  
\be
B(\xi,\g^A\xi)=(-1)^{\f{m(m+1)}{2}}B(\xi,\g^A\xi)
\ee
 se anula, a menos que $(-1)^{\f{m(m+1)}{2}}=+1$, ou seja, $m(m+1) = 0\text{ mod } 4$, ou ainda, $m=0,3,4,7$. Em outras palavras, se, juntamente a isso, o espinor $\xi$ \'e de Majorana, ou seja, $m$ \'e par, temos que $A^{0,m}_{\psi|\psi^\prime}$ se anula, a n\~ao ser que $m=0,4$.
 
A partir dos resultados acima para o bilinear $B$, um {\it boomerang} alternado pode ser definido como
$$
 A^{(0)}_{\xi|\xi'}=\f{\Delta}{2^d}(-1)^{m}B(\xi,\g_A\xi')e^A,\quad |A|\;{\rm arbitr\acute{a}ria}.
$$
Bilineares de $\psi$ s\~ao, então, denotados de forma simplificada por 
$$
\varphi_k:=|A^{0,k}_{\psi|\psi}|=\f{1}{k!}B(\psi,\g_A\psi)e^A,\quad k=|A|={\rm constante}.
$$ 
Logo, os \'unicos  bilineares covariantes possivelmente n\~ao-nulos que podem ser escritos com a forma bilinear $B$, tal que o espinor $\xi$ é normalizado por $\varphi_0=B(\xi,\xi)=1$, s\~ao
\begin{eqnarray}
\nonumber \vspace{-1cm}\varphi_0 &\vspace{-1cm}=\bar{\xi}\xi & \equiv \qquad B (\xi,\xi)=1\\
{\bf \varphi_4}&=\bar{\xi}\g_{ijkl}\xi &
=\f{1}{24}B(\xi,\g_{ijkl}\xi)e^{ijkl}
 \label{B0},
\end{eqnarray} 
os quais s\~ao as componentes do primeiro gerador da \'algebra de Fierz $A^0_{\psi|\psi}=\f{1}{16}(1+\varphi_4)$. Isso simplifica as 
identidades de Fierz (\ref{fp1},\ref{fp2}) e as transforma na igualdade
\be\label{Fierz7}
(\varphi_4+1)\Ganz(\varphi_4+1)=8(\varphi_4+1)
.\ee
Essa identidade pode ser decomposta em componentes $\psi_k\in \wedge^k(T^*M)$ atrav\'es do produto $\D_k:\sec\bigwedge(TM)\times\sec\bigwedge(TM)\r\sec\bigwedge(TM)$, dado pela itera\c c\~ao
\be 
\chi\D_{k+1}\vartheta=\f{1}{k+1}g^{ab}(e_a\lrcorner\chi)\D_k(e_b\lrcorner\vartheta),\qquad\chi, \vartheta\in \wedge T^*M,
\ee
onde $\lrcorner$ \'e o produto interior entre duas formas de $\wedge T^*M$. O primeiro termo \'e $\D_0=\w$, que \'e o produto exterior padr\~ao. A partir dessa itera\c c\~ao surgem os seguintes termos:
\beq
\chi\D_1\vartheta&=&g^{ab}(e_a\lrcorner\chi)\w(e_b\lrcorner\vartheta)\\
\chi\D_2\vartheta&=&\f{1}{2}g^{cd}(e_c\lrcorner\chi)\D_1(e_d\lrcorner\vartheta)=\f{1}{2}g^{ab}g^{cd}[e_a\rf(e_c\rf\chi)]\w[e_b\rf(e_d\rf\vartheta)].
\eeq
Dessa forma, a partir da decomposição $\varphi_4\Ganz\varphi_4=||\varphi_4||^2-\varphi_4\D_2\varphi_4$, obtemos as identidades de Fierz:
\be
||\varphi_4||^2=7\quad{\rm e}\quad\varphi_4\D_2\varphi_4=-6\varphi_4.
\ee

Mais v\'inculos advindos dos outros tr\^es bilineares restantes s\~ao necessários para uma classifica\c c\~ao mais refinada. 
Os v\'inculos obtidos para o bilinear $B$ s\~ao equivalentes aos encontrados tamb\'em para o bilinear $B_2$, pois
$$
B_2(\xi,\g_A\xi)=B(\xi,D\circ\g_A\xi)=(-1)^{|A|}B(\xi,\g_A\xi).
$$
 Como $D$ denota uma conjugação complexa e estamos considerando espinores de Majorana, os outros dois bilineares $B_1$ e $B_3$ restantes tamb\'em s\~ao equivalentes entre si, através de uma regra de seleção
\be\label{B_1}
B_1(\xi,\g_A\xi):=B(\xi, J\circ\g_A\xi)=-B(\xi,J\circ D\circ\g_A \xi)=:B_3(\xi,\g_A\xi),\quad{\rm onde}\quad  |A|=m.
\ee
Pois, temos que a eq. (\ref{B_1}) é nula, a n\~ao ser que $m$ seja {\'\i}mpar. Disso resultam os mesmos v\'inculos. Podemos simplificar a nota\c c\~ao desses, denotando-os por
\beq
 \label{phij}
\check{\varphi}_k=\f{1}{k!}  B (\xi, J\ci\g_{\alpha_1\ldots
\alpha_k}\xi)e^{\alpha_1\ldots \alpha_k}\in\sec\bigwedge^k(TM)\,.
\eeq
 Similarmente ao c\'alculo para a forma bilinear $B$, consideramos a transposta das matrizes $\g^A$. 
$J
=\g^8=\g^1\cdots\g^7$ nos diz que $[J,\g_{a_i}]=0$ e $J^\intercal=(-1)^{\f{d(d+1)}{2}}J=+J$. Adicionado a isso a identidade $(\g^A)^\intercal=\e^{m}_{B_i}
(-1)^{\f{m(m+1)}{2}}\g^A$, temos
$$
(J\circ\g_A)^\intercal=(\g_A)^\intercal J^\intercal=(-1)^{\f{m(m+1)}{2}}J\circ\g_A
$$
e de maneira equivalente
\be
B_1(\xi',\g_A\xi)=B(\xi', J\circ\g_A\xi)=B(J\circ\g_A\xi,\xi')=(-1)^{\f{m(m+1)}{2}}B(\xi, J\circ\g_A\xi').
\ee
Ou seja, $B_1(\xi,\g_A\xi)$ se anula, exceto se ${m(m+1)}= 0\mod\; 4$, isto \'e, se $m=0,3,4,7$. Finalmente, esse v\'inculo, juntamente com o proveniente da regra de sele\c c\~ao, nos diz que $B_1(\xi,\g_A\xi)=0$, a menos que $m=3$ ou $7$. Fornecendo, assim, os \'unicos bilineares covariantes que podem ser n\~ao-nulos
\beq
\check{\varphi}_3:=\f{1}{3!}B(\xi,J\ci\g_{a_1a_2a_3}\xi)e^{a_1a_2a_3}:=(\check{\varphi}_3)_{a_1a_2a_3}e^{a_1a_2a_3},\\
\check{\varphi}_7:=\f{1}{7!}B(\xi, J\ci\g_{a_1\dots a_7}\xi)e^{a_1\dots a_7}:=(\check{\varphi}_7)_{a_1\dots a_7}e^{a_1\dots a_7}
.\eeq

Como o elemento de volume $\g^8$ define uma estrutura complexa $J$, a seguinte expressão é válida 
$$
\g_{\a_1\ldots\a_k}= \f{(-1)^{\f{k(k-1)}{2}}}{(7-k)!}{\e_{\a_1\ldots\a_k}}^{\a_{k+1}\ldots\a_7}\g_{\a_{k+1}\ldots\a_7}J.
$$
 Isso nos garante a dualidade de Hodge entre as formas dos bilineares $B$ e $B_1$, isto é, $\check{\varphi}_k= \star\varphi_{7-k}$. 
 Usando isso, o fato de que os bilineares est\~ao normalizados por $B(\xi,\xi)=1$ e as identidades de Fierz (\ref{Fierz7}), essas podem ser escritas em componentes da \'algebra exterior:
\begin{eqnarray}
\label{Fierz.}
||\varphi_4||^2=||\check{\varphi}_3||^2=7,&\;\; \check{\varphi}_3\D_1\check{\varphi}_3=\varphi_4\D_2\varphi_4=-6\varphi_4,&\\
\nonumber\check{\varphi}_3\wedge\varphi_4=7\g^8,&\;\; \check{\varphi}_3\D_2\varphi_4=-6\check{\varphi}_3,&\;\;\check{\varphi}_3\D_1\varphi_4= \check{\varphi}_3\D_3\varphi_4=0.
\end{eqnarray}
Por quest\~ao de clareza, as componentes das identidades de Fierz (\ref{Fierz.}) podem ser escritas em uma dada representa\c c\~ao
\beq
(\check{\varphi}_3)_{kij}{(\check{\varphi}_3)^k}_{lm}&=&(\varphi_4)_{pqij}{(\varphi_4)^{pq}}_{lm}=-6(\varphi_4)_{ijlm},\\
(\check{\varphi}_3)_{ijk}(\varphi_4)^{ilmn}&=&0,\\
 (\check{\varphi}_3)_{irs}{(\varphi_4)^{irs}}_j&=&0,\\
(\check{\varphi}_3)_{ijk}{(\varphi_4)^{ij}}_{lm}&=&-6(\check{\varphi}_3)_{klm}
,\eeq tal que os \'indices livres s\~ao antissim\'etricos. 
Assim, a decomposição
\be
||\varphi_4||^2=7\quad{\rm  e}\quad \varphi_4 \Delta_2\varphi_4=-6\varphi_4,
\ee
 nos diz que o bilinear $\varphi_4$ n\~ao pode ser nulo, isto \'e, temos uma parte da classifica\c c\~ao desejada,
\beq
\varphi_0=1\neq 0,\quad \varphi_4\neq 0.
\eeq
\indent Finalmente, como a dualidade de Hodge  $\check{\varphi}_k= \star\varphi_{7-k}$ \'e satisfeita, as identidades de Fierz, que se resumem às eqs. (\ref{Fierz.}), nos d\~ao somente uma classe de espinores de Majorana no fibrado espinorial advindo do fibrado de Clifford ${\cal C}\ell_{7,0}$ de acordo com uma classifica\c c\~ao  baseada em bilineares covariantes, como se segue:
\begin{eqnarray}
\!\!\!\!\!\!\!\!\!\!\!\!\!\!\varphi_0\neq 0, \quad \varphi_1=0, \quad\varphi_2= 0, \quad\varphi_3= 0,\quad
\varphi_4\neq 0,\quad\varphi_5=0, \quad\varphi_6=0, \quad
\varphi_7=0\, ,\\ 
\!\!\! \!\!\!\!\!\!\!\!\!\!\!\!\!\!\!\!\!\quad\check{\varphi}_7\neq0,  \quad\check{\varphi}_6=0,
\quad\check{\varphi}_5=0,  \quad\check{\varphi}_4=0, \quad\check{\varphi}_3\neq
0,\quad\check{\varphi}_2= 0,\quad\check{\varphi}_1=0, \quad\check{\varphi}_0 =
0\,. 
\end{eqnarray}
%
 Essa classifica\c c\~ao de campos espinoriais de Majorana padr\~ao $\xi\in\G(S^+)$ sobre variedades Riemannianas de sete dimens\~oes, em particular $S^7$, pode ser estendida para se\c c\~oes $\psi$ do fibrado $\G(S)$ atrav\'es da defini\c c\~ao de um bilinear covariante mais geral
\beq
\upvarphi_k:=
\frac{1}{k!}\upbeta_k(\psi,\gamma_{\alpha_1\ldots\alpha_k}\psi)e^{
\alpha_1\ldots\alpha_k}\,,\eeq
onde as componentes da generaliza\c c\~ao de $\varphi_k$ s\~ao constru\'idas de forma a englobar o caso complexo:
 \begin{eqnarray}
  \label{formac}
\upbeta_k(\psi,\gamma_{\alpha_1\ldots\alpha_k}\psi')&=& B\left({}^{\rm
(Re)}\psi,\gamma_{\alpha_1\ldots\alpha_k}{}^{\rm (Re)}\psi'\right)-
B\left({}^{\rm (Im)}\psi,\gamma_{\alpha_1\ldots\alpha_k}{}^{\rm
(Im)}\psi'\right)\\&&\nonumber\qquad\qquad+i\left[B\left({}^{\rm
(Re)}\psi,\gamma_{\alpha_1\ldots\alpha_k}{}^{\rm (Im)}\psi'\right)+
B\left({}^{\rm (Im)}\psi,\gamma_{\alpha_1\ldots\alpha_k}{}^{\rm
(Re)}\psi'\right)\right]\,. \nonumber
 \end{eqnarray}
 A partir dos seguintes resultados \cite{1},
\begin{eqnarray}
 B (\psi,\gamma_{\alpha_1 \ldots\alpha_k}\psi)
=\begin{cases}\label{ijk}
 B\left({}^{\rm (Re)}\psi,\gamma_{\alpha_1 \ldots\alpha_k}{}^{\rm
(Im)}\psi\right)+  B\left({}^{\rm (Im)}\psi,\gamma_{\alpha_1
\ldots\alpha_k}{}^{\rm (Re)}\psi\right)\,, \,\text{se k \'e par}\\
 B\left({}^{\rm (Re)}\psi,(J\;\Ganz\;\gamma_{\alpha_1 \ldots\alpha_k}){}^{\rm
(Re)}\psi\right)-  B\left({}^{\rm (Im)}\psi,(J\;\Ganz\;\gamma_{\alpha_1
\ldots\alpha_k}\right){}^{\rm (Im)}\psi)\,,\\\hspace{9cm} \,\text{se $k$ \'e \'impar}\,,\nonumber
\end{cases}
\end{eqnarray}
 e
\begin{eqnarray}
 B(\psi,J\,\Ganz\,\gamma_{\alpha_1 \ldots\alpha_k}\psi)
=\begin{cases}
- B\left({}^{\rm (Re)}\psi,\gamma_{\alpha_1 \ldots\alpha_k}{}^{\rm
(Im)}\psi\right)+  B\left({}^{\rm (Im)}\psi,\gamma_{\alpha_1
\ldots\alpha_k}{}^{\rm (Re)}\psi\right)\,, \,\text{se $k$ \'e par}\\
 B\left({}^{\rm (Re)}\psi,(J\;\Ganz\;\gamma_{\alpha_1 \ldots\alpha_k}){}^{\rm
(Re)}\psi\right)+  B\left({}^{\rm (Im)}\psi,(J\;\Ganz\;\gamma_{\alpha_1
\ldots\alpha_k}){}^{\rm (Im)}\psi\right)\,, \\\hspace{8.5cm}\,\text{se $k$ \'e \'impar}\,,\label{fgh}
\end{cases}
 \end{eqnarray}
 obtemos que tanto os termos da parte real como os termos da parte imagin\'aria de (\ref{formac}) se anulam. Esses cancelamentos tornam poss\'iveis o anulamento de pelo menos um dos bilineares $\varphi_0$ e $\varphi_4$, que nos garante a existência de classes de espinores adicionais na vers\~ao complexa do espinor $\psi$. Logo, s\~ao encontradas por meio dos v\'inculos acima quatro classes de campos espinoriais $\psi\in\Gamma(S)$ sobre variedades Riemannianas 7-dimensionais
\begin{subequations}
\begin{eqnarray}
 & \upvarphi_0=0,\quad\upvarphi_4=0&\;\;(\text{espinores nulos - resultado espúrio}),
 \\ & \upvarphi_0=0,\quad\upvarphi_4\neq0,&\text{(espinores singulares)}\label{c12}\\
& \upvarphi_0\neq0,\quad\upvarphi_4=0,&\text{(espinores regulares)}\label{c13}\\
& \upvarphi_0\neq0,\quad\upvarphi_4\neq0\,&\text{(espinores de Majorana)}.\label{c14}
\end{eqnarray}
\end{subequations}
Para os outros valores $k=1,2,3,5,6,7$,\; $\varphi_k=0$.
Além disso, os bilineares $\check{\varphi}_k$ s\~ao duais de Hodge dos bilineares $\varphi_{7-k}$, isto \'e, $\check{\varphi}_k=\star\varphi_{7-k}$. Dessa forma, uma classifica\c c\~ao de acordo com $\check{\varphi}_k$ \'e id\^entica à descrita acima e, portanto, n\~ao nos d\'a nenhuma classe ou resultado adicional. Esses quatro pares de desigualdades resumem bem as tr\^es classes n\~ao-triviais de campos espinoriais poss\'iveis e uma classe trivial, que obviamente é composta de um campos espinorial nulo.  Se restringimos $\psi$ a $\G(S^+)$, isto \'e, tal que $^{({\rm Im})}\psi=0$, obtemos das eq. (\ref{fgh}) somente uma classe de espinores de Majorana, como esperado.

\indent Além disso, da mesma forma que o agregado de Fierz (\ref{boomf}) foi constru\'ido por Lounesto sobre o espa\c co-tempo de Minkowski \cite{lou2}, um agregado graduado pode ser construído aqui:
\beq{\rm Z}=\f{\ell}{2^{n}}\sum_{k=0}^7(-1)^{k} B
(\psi,\g_{\alpha_1\ldots \alpha_k}\psi)e^{\alpha_1\ldots \alpha_k},\eeq onde os bilineares $B(\psi,\g_{\alpha_1\ldots\alpha_k}\psi) = 0$, exceto se $ k\;=\;0,3,4,7$. 

A classifica\c c\~ao dos espinores dada sobre espaços Riemannianos de sete dimensões, em particular $S^7$, pode restringir a Lagrangiana para os campos de mat\'eria e garantir a obten\c c\~ao de todos os seus termos. De acordo com \cite{Top}, essa %
construção depende sobre que realiza\c c\~ao os campos s\~ao escritos.
Como em uma realiza\c c\~ao real ou octoni\^onica, uma poss\'ivel Lagrangiana n\~ao teria termos cin\'eticos, mas apenas termos de massa de n\~ao-Weyl do tipo $M_\perp=tr(\bar{\xi}_+\xi_-+\bar{\xi}_-\xi_+)$, n\~ao faz sentido escrev\^e-la nessas realiza\c c\~oes. Pois n\~ao h\'a din\^amica de campo aqui. Note que a assinatura que estamos usando \'e $(p,q)=(7,0)$: [$t \mod 4=0\;{\rm e}\; t-s \mod 8=1$, onde $t$ é o número de coordenadas tipo-tempo e $s$, tipo-espaço]. Portanto, de acordo com \cite{Top}, somente uma realiza\c c\~ao quaterni\^onica torna poss\'ivel a constru\c c\~ao de tal Lagrangiana, cujos termos s\~ao:
\beq
M&=&tr(\bar{\psi}\psi)\\
M_\m&=&tr(\bar{\psi}\gamma_\m\psi)\\
K_{\n\rho}&=&tr(\bar{\psi}\gamma^\mu \g_{\n\rho}\partial_\mu \psi)\\
K_\m&=&tr(\bar{\psi}\gamma^\mu\g_\m\partial_\mu \psi)
,\eeq
onde $\m,\n,\rho=1,\ldots,7$ s\~ao tipo-espa\c co, $M, M_\m$ são os termos de massa e $K_{\n\rho},K_\m$ são os termos cinéticos. Obviamente, consideramos a associatividade dos quat\'ernions e a conjuga\c c\~ao $\bar{\psi}$ como sendo a mesma usada em (\ref{formab}). 

Resumindo, nesta seção os bilineares construídos com espinores $\pi\in S$ foram explorados através das estruturas $D$ e $J$.  Foram listados todos os bilineares covariantes poss\'iveis em dimens\~ao e assinatura arbitr\'arias e escrito o agregado de Fierz mais geral correspondente. Um bilinear admissível $B$ foi definido através de alguns sinais ($\s_k,\e_k,\iota_k;\quad k=1,2,3$), que juntamente com uma regra de seleção nos permitiram obter alguns vínculos entre seus bilineares e adicionando a isso as identidades de Fierz, encontramos uma classificação para espinores sobre espaços Riemannianos de sete dimensões, em particular $S^7$. 
Ap\'os complexificar esse caso, foram classificados espinores gerais sobre essas variedades de sete dimens\~oes por meio de v\'inculos e foram obtidas duas novas classes não-triviais de espinores e um resultado espúrio. 
Finalmente, encontramos os termos massivos e cinéticos de uma Lagrangiana de campos de mat\'eria em uma realiza\c c\~ao quaterni\^onica sobre esses espaços.

Na seção abaixo, classificaremos espinores sobre espaço Lorentziano de cinco dimensões, de uma forma análoga a realizada nesta seção. Como exemplos da classificação a seguir temos o {\it bulk} da teoria de branas, o espaço de de Sitter $dS_5$ e de anti-de Sitter $AdS_5$. Esses espinores possuem uma estrutura quaterniônica e requerem mais atenção em sua classificação.

\section{Classifica\c c\~ao de campos espinoriais no espa\c co Lorentziano de cinco dimens\~oes}\label{43}

\qquad Nesta seção, classificamos espinores sobre o {\it bulk} da teoria de branas. O {\it bulk} é um espaço de anti-de Sitter de cinco dimensões, onde está imersa a 3-brana que descreve nosso universo. 
  Se a brana é localizada em quatro dimensões, por exemplo, pela função delta de Dirac, ela é denominada brana fina. Em outro caso, onde a localização da brana ao longo da dimensão extra é mais suave, é conhecida como 
 brana espessa. Além disso, um estudo mais realístico de localização de matéria sobre branas espessas tornou-se possível ao ser construída uma Lagrangiana para campos espinoriais sobre o {\it bulk} a partir de seus bilineares covariantes, em vez de se considerar campos escalares acoplados com a gravidade \cite{brane}. Através dessa classificação de espinores sobre o {\it bulk}, que abordaremos a seguir, poderemos calcular a partir do tensor energia-momento como a matéria se distribui próximo da brana para cada tipo de espinor. No entanto, isso foge do escopo deste trabalho.

Através de nossa classificação, no final da próxima seção, verificamos de forma imediata que as componentes dos bilineares covariantes sobre buraços negros axissimétricos em cinco dimensões são todos os invariantes possíveis que podem ser construídos a partir de uma representação de um campo de férmion (espinorial) arbitrário 
\cite{MeiBH,Burk}.

A partir daqui abordaremos a classificação de espinores sobre espaços Lorentzianos de cinco dimens\~oes e assinatura $4+1$ (1 dimensão temporal e 4 espaciais) (\,+\,+\,+\,+\,-\,), introduzindo vínculos construídos com identidades de Fierz, simetrias da forma bilinear $B$ e regras de seleção, de um modo an\'alogo à feita pelo Lounesto. Considere, então, tal espaço e seu fibrado de Kähler-Atiyah $(\Omega(TM),\circ)$, que é um fibrado espinorial $S$ munido com uma estrutura de módulo dada por um morfismo $\g:\wedge T^*M\rightarrow End(S)$, que não é injetivo, nem sobrejetivo. Como o elemento de volume $\g^6:=\g^0\g^1\g^2\g^3\g^5$ satisfaz $\g^6\circ\g^6=+1$, este pode ser representado pela identidade: $\g^6={\rm id}_S$. Uma escolha para as matrizes $\gamma$'s pode ser dada por:
\be
\g^0=\left(\begin{tabular}{cc}
$0$&$1_2$\\
$1_2$&$0$
\end{tabular}\right),\;\g^i=\left(\begin{tabular}{cc}
$0$&$-\s_i$\\
$\s_i$&$0$
\end{tabular}\right)\; (i=1,2,3),\;\g^5=\left(\begin{tabular}{cc}
$i1_2$&$0$\\
$0$&$-i1_2$
\end{tabular}\right).
\ee 
De forma bastante an\'aloga à classifica\c c\~ao em sete dimens\~oes, considere tr\^es endomorfismos locais $J_i: S \r S$ que satisfazem
 $$
[J_i,\g(\omega)]_{-,\circ}=0,\quad \forall\omega\in \sec\wedge(TU),\qquad J_i\circ J_j=-\d_{ij}{\rm id}_S+\e^{ijk}J_k\;,\;
$$
onde $\e^{ijk}$ \'e o s\'imbolo de Levi-Civita 
e $\sec\wedge(TU)$, o conjunto de seções do fibrado tangente restrito ao aberto $U\subset M$.  
Por meio dessas estruturas quaterniônicas $J_\m$ é possível definir quatro bilineares admiss\'iveis n\~ao-degenerados:
\beq
B_\m=B_0\circ ({\rm id}_S\otimes J_\m)\;,\;{\; \rm onde\;}J_0 ={\rm id}_S\;{\rm e }\;B_0=B.
\eeq


Através do endomorfismo $A_{\psi|\psi'}$ sobre $S$, que é definido a partir da forma bilinear $B$, como
$$
A_{\psi|\psi'}(\xi)=B(\xi,\psi')\psi,
$$ as identidades de Fierz são escritas da seguinte forma \cite{1}: 
\be\label{fierzident}
A_{\psi_1|\psi_2}\circ A_{\psi_3|\psi_4}=B(\psi_3,\psi_2)A_{\psi_1|\psi_4}.
\ee O endomorfismo $A_{\psi|\psi'}$ admite ainda uma única decomposição através das estruturas quaterniônicas $J_\m$:
 \be
A_{\psi|\psi'}=J_\m\Ganz A_{\psi|\psi'}^{(\mu)}
,\ee cujas componentes, em termos dos bilineares covariantes sobre o {\it bulk}, são expressas como
\beq
A_{\psi|\psi'}^{(0)}&=&\f{1}{8}\sum_{A} B(\g_A^{-1}\psi,\psi')\g_A,\\
A_{\psi|\psi'}^{(i)}&=&\f{1}{8}\sum_{A} B((\g^{-1}_A\Ganz J_i^{-1})\psi,\psi')\g_A.
\eeq
Através dessa decomposição, as identidades de Fierz (\ref{fierzident}) também podem ser decompostas 
\beq\nonumber
A_{\psi_1|\psi_2}^{(0)}\circ A^{(0)}_{\psi_3|\psi_4}-\sum_{i=1}^3 A_{\psi_1|\psi_2}^{(i)}\circ A^{(i)}_{\psi_3|\psi_4}&=&B(\psi_3,\psi_2)A_{\psi_1|\psi_4}^{(0)}, \\ A_{\psi_1|\psi_2}^{(0)}\circ A^{(i)}_{\psi_3|\psi_4}+A_{\psi_1|\psi_2}^{(i)}\circ A^{(0)}_{\psi_3|\psi_4}+\sum_{j,k=1}^3\e_{ijk}A_{\psi_1|\psi_2}^{(j)}\circ A^{(k)}_{\psi_3|\psi_4}&=&B(\psi_3,\psi_2)A_{\psi_1|\psi_4}^{(i)}.\quad\label{2fierz}
\eeq
Além disso, para que os vínculos dos espinores possam ser melhor analisados precisamos levar em conta alguns sinais da forma bilinear $B$, denominados simetria $\s_B$ e tipo  $\e_B$ do bilinear covariante não-degenerado $B$, definidos sobre o fibrado espinorial $S$ e dados respectivamente por
\beq
B(\psi,\psi')&=& \s_B B(\psi',\psi)\\
B(\g^A\psi,\psi')&=& \e_B B(\psi,\g^A\psi').
\eeq
Se essas igualdades são válidas para $B$, dizemos que ele é uma forma bilinear admissível e temos
\beq
B (\psi,\gamma_{\alpha_1\ldots
\alpha_k}\psi)&=& B (\gamma_{\alpha_1\ldots
\alpha_k}\psi,\psi)\\
B (\gamma_{\alpha_1\ldots
\alpha_k}\psi,\psi)&=&(-1)^{\f{k(k-1)}{2}}B (\psi,\gamma_{\alpha_1\ldots
\alpha_k}\psi)
,\eeq
 onde $k=|A|$. É fácil ver que temos um vínculo a partir desses dois sinais e, como resultado, alguns bilineares são obrigatoriamente nulos. Segue-se que
 \beq
\label{ddd}
\varphi_k:= \vert A^{0,k}_{\psi |\psi}\vert \;\;= \frac{1}{k!}B (\psi,\gamma_{\alpha_1\ldots
\alpha_k}\psi)e^{\alpha_1\ldots \alpha_k} \eeq\noindent é nulo, exceto se $k(k-1)\equiv 0   \mod 4$, ou seja, $A^{0,k}\equiv A ^{0,k}_{\psi |\psi}$
é zero, a não ser se $k=0,1,4,5$. Portanto, a partir dos vínculos obtidos da forma bilinear $B$, apenas esses quatro bilineares podem ser não-nulos: $\varphi_0,\varphi_1,\varphi_4$ e $\varphi_5$.
 

Outros tr\^es bilineares $B(\psi',J_i\circ \psi)$ podem ser constru\'idos a partir das estruturas complexas quaterni\^onicas $J_i,\ \ i=1,2,3$. Utilizando a comutatividade $[J_i, \g_j]=0$ e a $B$-transposta $J_i^t=-J_i$, temos
 \beq
 B(\psi,J_i\circ\g_A\psi)&=&B(J_i\circ\g_A\psi,\psi)=B(\g_A\circ J_i\psi,\psi),\\
 B(\g_A\circ J_i\psi,\psi)&=&(-1)^{\f{l(l-1)}{2}+1}B(\psi,J_i\circ \g_A\psi)
 ,\eeq
onde $l=|A|$. Segue-se que, se definimos o bilinear $\tilde{\varphi}_l^i$ como
 \be 
 \tilde{\varphi}_l^i= \frac{1}{l!}B (\psi,J_i\Ganz\gamma_{\alpha_1\ldots
\alpha_l}\psi)e^{\alpha_1\ldots \alpha_l},
\ee
este é nulo, a não ser que $l(l-1)+2\equiv 0  \mod 4$, ou seja, exceto se $l=2,3$. Disso resulta seis outros bilineares possivelmente não-nulos: $\tilde{\varphi}_2^i$ e $\tilde{\varphi}_3^i
$.

No entanto, esses vínculos podem ser reduzidos ao considerarmos, através da dualidade de Hodge, apenas metade dos bilineares covariantes:
$$
\varphi_5=\star\varphi_0,\;\; \varphi_4=\star\varphi_1,\;\;\tilde{\varphi}^i_3=-\star\tilde{\varphi}^i_2\;.
$$ Para isso, são definidas algumas formas não-homogêneas que contêm os bilineares \cite{1}
\be
\om_0=\f{1}{16}(\varphi_0+\varphi_1+\varphi_4+\varphi_5)\quad{\rm e}\quad\om_i=\f{1}{16}(\tilde{\varphi}^i_2+\tilde{\varphi}^i_3),
\ee
que podem ser reduzidas, através dualidade de Hodge:
\be
\om_0=2\Pi(\om_0^{\mathring{}})\quad{\rm e}\quad\om_i=2\Pi(\om_i^{\mathring{}}),
\ee
onde $\Pi=\f{1}{2}(1+\star)$ é uma projeção,  $\om^{\mathring{}}_0=\f{1}{16}(\varphi_0+\varphi_1)$ e $\om^{\mathring{}}_i=\f{1}{16}\tilde{\varphi}^i_2$ são os geradores truncados da álgebra truncada $(\om^{\mathring{}}(M),\bullet)$, que é definida a partir da álgebra $(\om(M),\circ)$ e do homomorfismo $\Pi(\varphi\bullet\psi)=\Pi(\varphi)\circ\Pi(\psi)$. %
Além disso, as identidades de Fierz (\ref{2fierz}) também podem ser escritas na forma truncada
\beq\label{FI}
\om_0^<\bullet\om_0^<-\sum_i\om_i^<\bullet\om_i^<=\f{1}{2}\varphi_0\om_0^<\;,\\ \nonumber
\om_0^<\bullet\om_i^<+\om_i^<\bullet\om_0^<+\sum_{j,k=1}^3 \e_{ijk}\om^<_j\bullet\om^<_k=\f{1}{2}\varphi_0\om_i^<
\eeq
e são simplificadas 
\cite{1}
\begin{subequations}\label{fierz}\beq\label{fierz2}
\varphi_1\bullet\varphi_1-\sum_i\tilde{\varphi}^i_2\bullet \tilde{\varphi}^i_2=7(\varphi_0)^2+6\varphi_0\varphi_1,\\
\vp_1\Db\tilde{\vp}_2^i+\tilde{\vp}_2^i\Db\vp_0+\sum_{j,k}\e_{ijk} \tilde{\vp}_2^j\Db\tilde{\vp}_2^k=6\vp_0\tilde{\vp}_2^i
.\eeq
\end{subequations}
Para uma melhor análise dessas expressões, precisamos decompô-las em $k$-formas e isso é feito iterativamente através do produto
$\Delta _k:\Omega (TM)\times \Omega
(TM)\rightarrow \Omega (TM)$ introduzido na seção anterior \cite{1}: 
\beq
\theta\,\Delta
_{k+1}\,\vartheta=\f{1}{k+1}g^{ab}(e_a\lrcorner\theta)\,
\Delta_k\,(e_b\lrcorner\vartheta)\,,\qquad\theta, \vartheta\in \Omega (TM)\,,
\eeq\noindent onde $g^{ab}$ denota os coeficientes do tensor métrico. Algumas identidades são obtidas para esse produto $\Delta
_{k}$ \cite{1} e serão úteis mais adiante:
\beq
\label{6}\sum_i\big(\tilde{\vp}^i_2\Db\tilde{\vp}^i_2&-&\star(\tilde{\vp}^i_2 \wedge\tilde{\vp}^i_2)+||\tilde{\vp}^i_2||^2\big)=0,\\
\label{7}\vp_1\Db\tilde{\vp}^i_2+\tilde{\vp}^i_2\Db\vp_1&=&2\star(\vp_1\wedge \tilde{\vp}^i_2),\\ 
\e_{ijk}\Big[\tilde{\vp}^j_2\Db\tilde{\vp}^k_2&-& \left(\star(\tilde{\vp}^j_2\wedge\tilde{\vp}_2^k-\tilde{\vp}_2^j\;\D_1\; \tilde{\vp}_2^k-\tilde{\vp}_2^j\;\D_2\; \tilde{\vp}_2^k\right)\Big]=0.
\eeq
Note que, a partir da expressão $\vp_1\lrcorner\tilde{\vp}^i_2=0$, temos que $\vp_1\Db\tilde{\vp}_2^i$ \'e id\^entica à sua componente $\star(\vp_1\wedge\tilde{\vp}_2^i)=a\;\tilde{\vp}_2^i$. Dessa forma, as eqs. (\ref{fierz}) podem ser decompostas em elementos de $\wedge^k T^*M$
\beq\label{1}
||\vp_1||^2+\sum_i||\tilde{\vp}_2^i||^2&=&7(\vp_0)^2,\\\label{2} \sum_i\star(\tilde{\vp}_2^i\wedge\tilde{\vp}_2^i)&=&-6\vp_0\vp_1,\\
\label{3}2\star(\vp_1\wedge\tilde{\vp}_2^i)-\sum_{j,k}\e_{ijk}(\tilde{\vp}_2^j\Delta_1\tilde{\vp}_2^k)&=&
6\vp_0\tilde{\vp}_2^i,\\\label{4} \sum_{j,k}\e_{ijk}\star(\tilde{\vp}_2^j\wedge\tilde{\vp}_2^k)= \sum_{j,k}\e_{ijk}\tilde{\vp}_2^j\Delta_2\tilde{\vp}_2^k&=&0
.\eeq
Essas equações podem ser manipuladas e encontrados alguns vínculos para as formas $\vp_1$ e $\tilde{\vp}_2^i$. Com esse fim, analisamos a expressão $\delta_{ij}\vp_1\Db\tilde{\vp}_2^i\Db\tilde{\vp}_2^j$. Substitua a Eq. (\ref{2}) na Eq. (\ref{6}):
\beq\label{phi12}
\sum_i(\vp_1\Db\tilde{\vp}_2^i)\Db\tilde{\vp}_2^i=a\;\sum_i\tilde{\vp}_2^i\Db\tilde{\vp}_2^i=-6a\;\vp_0\vp_1 -a\;\sum_i||\tilde{\vp}^i_2||^2.
\eeq
De outra forma, as Eq. (\ref{2}) resultam em:
\beq\label{phi122}
\sum_i\vp_1\Db(\tilde{\vp}_2^i\;\Db\;\tilde{\vp}_2^i)=-6\;\vp_0(\vp_1\,\Db\,\vp_1)-\sum_i||\tilde{\vp}^i_2||^2\vp_1
.\eeq
A partir das Eq. (\ref{phi12}) e (\ref{phi122}), através da associatividade do produto $\bullet$ e sua graduação em $k$-formas, obtemos \cite{1}
\beq\label{varphi2}
6a\;\vp_0&=&\sum_i||\tilde{\vp}^i_2||^2\\
 a\;\sum_i||\tilde{\vp}^i_2||^2&=& 6\;\vp_0||\vp_1||^2,
\eeq
onde é considerado 
$\vp_1\Db\vp_1=||\vp_1||^2$. Um resultado imediato que vem dessas equações é $a^2=||\varphi_1||^2$. Se adicionado a isso, a eq. (\ref{varphi2}) for substituída na eq. (\ref{1}), é encontrada a expressão $a^2+6a\vp_0-7\vp_0^2=0$. Esta tem como solução $a=\vp_0$ e $a=-7\vp_0$. No entanto, apenas a solução $a=\vp_0$ é útil, pois a outra não é compatível com a eq. (\ref{varphi2}): $\sum_i||\tilde{\vp}^i_2||^2=6a\varphi_0=6(-7\varphi_0)\varphi_0=-42(\vp_0)^2$, o que \'e um absurdo. \\
\indent Finalmente, a partir das identidades de Fierz (\ref{fierz}), alguns vínculos são obtidos para as formas bilineares $\vp_1$ e $\tilde{\vp}_2^i$:
\beq
||\vp_1||^2=\vp_0^2=\frac16\delta_{ij}\tilde{\vp}^i_2\circ\tilde{\vp}^j_2=6\vp_0^2,\;\;\;\;\;\;\;\;\;\vp_1\lrcorner\tilde{\vp}^i_2=0,\;\;\;\;\;\;\;\;\;
\vp_1\lrcorner\star\tilde{\vp}^i_2=-\vp_0\tilde{\vp}^i_2,\\\delta_{ij}\tilde{\vp}_2^i\wedge\tilde{\vp}_2^j=-6\vp_0\star\vp_1,\;\;\;\;\;\;
\e_{ijk}\tilde{\vp}_2^j\Db\tilde{\vp}_2^k=-\e_{ijk}\,\tilde{\vp}_2^j\,\Delta_1\,\tilde{\vp}_2^k= 4\vp_0\tilde{\vp}_2^i.
\eeq
Se o espinor $\psi$ é normalizado: $\vp_0=B(\psi,\psi)=1$, garantimos a não-nulidade dos bilineares passíveis de serem não-nulos e exibimos, assim, uma classificação de espinores em espaços Lorentzianos de cinco dimensões de acordo com bilineares covariantes
\beq
\!\!\!\!\!\!\!\!\!\!\!\!\!\!\varphi_0\neq 0, \quad \varphi_1\neq 0, \quad\varphi_2= 0, \quad\varphi_3= 0,\quad
\varphi_4\neq 0,\quad\varphi_5\neq 0, \label{class.}\\
\!\!\!\!\!\!\!\!\!\!\!\!\!\!\tilde{\varphi}_0 =0, \quad \tilde{\varphi}_1= 0, \quad\tilde{\varphi}_2\neq 0, \quad\tilde{\varphi}_3\neq 0,\quad
\tilde{\varphi}_4= 0,\quad\tilde{\varphi}_5= 0,\label{class1}
\eeq
onde $\tilde{\varphi}_n\neq 0$ denota que $\tilde{\varphi}^k_n\neq 0$ para pelo menos um valor de $k$ e $\varphi_i,\tilde{\vp}_i^k\in\Omega^i(TM).$ Através da dualidade de Hodge $\varphi_n= \star\varphi_{5-n}$, mais uma simplificação é obtida, pois o número de bilineares covariantes necessários cai pela metade 
\beq
\varphi_0\neq 0, \quad \varphi_1\neq 0, \quad\varphi_2= 0,\quad\tilde{\varphi}_0 =0, \quad \tilde{\varphi}_1= 0 \text{\,  e } \quad\tilde{\varphi}_2\neq 0.
\eeq
Podemos fazer algo análogo à classificação dos espinores mais gerais  em espaços Riemannianos de sete dimensões, obtida na seção anterior após uma complexificação. Relembremos a expressão:
\beq
 B (\psi,\gamma_{\alpha_1 \ldots\alpha_k}\psi)
=
\begin{cases}
 B\left({}^{\rm (Re)}\psi,\gamma_{\alpha_1 \ldots\alpha_k}{}^{\rm
(Im)}\psi\right)-  B\left({}^{\rm (Im)}\psi,\gamma_{\alpha_1
\ldots\alpha_k}{}^{\rm (Re)}\psi\right)\,, \,\text{se $k=2\iota$}\\
 B\left({}^{\rm (Re)}\psi,(J\;\Ganz\;\gamma_{\alpha_1 \ldots\alpha_k}){}^{\rm
(Re)}\psi\right)+  B\left({}^{\rm (Im)}\psi,(J\;\Ganz\;\gamma_{\alpha_1
\ldots\alpha_k}\right){}^{\rm (Im)}\psi)\,,\\\hspace{9cm} \,\text{se
$k=2\iota+1$}\,,\nonumber
\end{cases}
 \eeq
 e
\beq
 B(\psi,J_i\,\Ganz\,\gamma_{\alpha_1 \ldots\alpha_k}\psi)
=\begin{cases}
- B\left({}^{\rm (Re)}\psi,\gamma_{\alpha_1 \ldots\alpha_k}{}^{\rm
(Im)}\psi\right)-  B\left({}^{\rm (Im)}\psi,\gamma_{\alpha_1
\ldots\alpha_k}{}^{\rm (Re)}\psi\right)\,, \,\text{se $k=2\iota$}\\
 B\left({}^{\rm (Re)}\psi,(J\;\Ganz\;\gamma_{\alpha_1 \ldots\alpha_k}){}^{\rm
(Re)}\psi\right)-  B\left({}^{\rm (Im)}\psi,(J\;\Ganz\;\gamma_{\alpha_1
\ldots\alpha_k}){}^{\rm (Im)}\psi\right)\,, \\\hspace{8.5cm}\,\text{se $k=2\iota+1$}\,,
\end{cases}\nonumber
 \eeq
Essas expressões nos ajudam a ter uma compreensão melhor a respeito de bilineares covariantes com espinores em espaços Lorentzianos de cinco dimensões.
Para isso, \'e importante que coloquemos os bilineares em termos dos endomorfismos $J_1, J_2$ e $J_3$ do nosso caso em cinco dimens\~oes. Ent\~ao, \'e necess\'ario trocar implicitamente as estruturas complexa e real, pelas estruturas quaterniônicas, esquematizadas como se segue:
\beq
^{(Re)}\psi=\f{1}{2}(1+D)\psi\mapsto\f{1}{2}(1+J_1)\psi&\\
^{(Im)}\psi=\f{1}{2}(1-D)\psi\mapsto\f{1}{2}(1-J_1)\psi&\\
J\mapsto J_2,\qquad -J\circ D\mapsto J_3&\\
D\mapsto J_i,\quad J\mapsto J_j,\quad -J\circ D\mapsto J_k&\quad{\rm onde\ \ } (ijk) \ \ {\rm \acute{e} \ \ ciclico}
.\eeq
Dessa forma, a eq. (\ref{formab}) que descreve bilineares covariantes de grau $k$ pode ser estendidas para o caso quaterni\^onico sobre espa\c cos de $4+1$ dimens\~oes. As expressões abaixo nos mostram que depois de uma complexificação de espinores restritos pela regra de seleção, obtemos espinores mais gerais possíveis, os quais se dividem em um maior número de classes:
\beq
 B (\psi,\gamma_{\alpha_1 \ldots\alpha_k}\psi) 
=\begin{cases}\f{1}{2}\left[
 B\left(J_1\psi,\gamma_{\alpha_1 \ldots\alpha_k}\;\Ganz\;J_1\psi\right)-  B\left(\psi,\gamma_{\alpha_1
\ldots\alpha_k}\psi\right)\right]\,, \,\text{se $k=2\iota$}\\
\f{1}{2}\left[B\left(\psi,\gamma_{\alpha_1 \ldots\alpha_k}\;\Ganz\;J_2\psi\right)-  B\left(J_1\psi,\gamma_{\alpha_1
\ldots\alpha_k}\;\Ganz\;J_3\psi\right)\right]\,, \,\text{se
$k=2\iota+1$}\,,\nonumber
\end{cases}
 \eeq
 e
\beq
 B(\psi,\gamma_{\alpha_1 \ldots\alpha_k}\,\Ganz\,J_j\psi) 
=\begin{cases}
\f{1}{2}\left[B\left(J_i\psi,\gamma_{\alpha_1 \ldots\alpha_k}\;\Ganz\;J_i\psi\right)-  B\left(\psi,\gamma_{\alpha_1
\ldots\alpha_k}\psi\right)\right]\,, \,\text{se $k=2\iota$},\\
\f{1}{2}\left[B\left(J_i\psi,\gamma_{\alpha_1 \ldots\alpha_k}\;\Ganz\;J_j\psi\right)-  B\left(\psi,\gamma_{\alpha_1
\ldots\alpha_k}\;\Ganz\;J_k\psi\right)\right]\,,\,\text{se $k=2\iota+1$}\,.
\end{cases}\nonumber
 \eeq
Isso nos permite escrever a forma bilinear $B$ de uma forma mais geral
: 
 \beq\label{formac1}
\upbeta_k(\psi,\gamma_{\alpha_1\ldots\alpha_k}\psi')= B\left(\psi,\gamma_{\alpha_1\ldots\alpha_k}\;\Ganz\;J_1\psi'\right)-
iB\left(\psi,\gamma_{\alpha_1\ldots\alpha_k}\;\Ganz\;J_3\psi'\right)\,.
\eeq\noindent 
Segue-se, então, uma extensão imediata para os bilineares covariantes
\beq
\upvarphi_k:=
\frac{1}{k!}\upbeta_k(\psi,\gamma_{\alpha_1\ldots\alpha_k}\psi)e^{
\alpha_1\ldots\alpha_k}\,.\eeq 
Na eq. (\ref{formac1}), os termos da parte real podem se cancelar e o mesmo ocorre com a parte imaginária. Uma classificação de espinores mais geral é dada através da forma bilinear $\upbeta_k$, que é a versão complexificada da forma bilinear $B$, e nos dá um número maior de classes, pois nessa classificação são possíveis valores nulos para os bilineares $\varphi_0,$  $\varphi_1$ e $\tilde{\varphi}_2$. Logo, surgem seis novas classes não-triviais de espinores: três de espinores regulares e outras três de espinores singulares. Essa classificação de espinores em espaços Lorentzianos de cinco dimensões é esquematizado em termos de seus bilineares covariantes como se segue:

\begin{subequations}
\beq
\upvarphi_0\neq 0, \quad \upvarphi_1\neq 0, \quad\tilde{\upvarphi}_2\neq 0\label{c11},\\
\upvarphi_0\neq 0, \quad \upvarphi_1\neq 0, \quad\tilde{\upvarphi}_2=0,\\
\upvarphi_0\neq 0, \quad \upvarphi_1= 0, \quad\tilde{\upvarphi}_2\neq 0,\\
\upvarphi_0\neq 0, \quad \upvarphi_1= 0, \quad\tilde{\upvarphi}_2=0,\\
\upvarphi_0= 0, \quad \upvarphi_1\neq 0, \quad\tilde{\upvarphi}_2\neq 0,\\
\upvarphi_0= 0, \quad \upvarphi_1\neq 0, \quad\tilde{\upvarphi}_2= 0,\\
\upvarphi_0= 0, \quad \upvarphi_1= 0, \quad\tilde{\upvarphi}_2\neq 0,\\
\upvarphi_0= 0, \quad \upvarphi_1= 0, \quad\tilde{\upvarphi}_2=0\label{c18},
\eeq
\end{subequations} onde $\upvarphi_n=\star \upvarphi_{5-n},\, \tilde{\upvarphi}_m=\star\tilde{\upvarphi}_{5-m}$,\,$\upvarphi_l=0$ para $l=2,3$\, e\, $\tilde{\upvarphi}_p=0$ para $p=0,1,4,5$. 

Uma expressão mais geral para o agregado de Fierz graduado é definida em termos de uma classe de equivalência:
\beq\label{boomg}{\rm Z}\sim\sum_{k}(-1)^{k} B 
(\psi,\g_{\alpha_1\ldots \alpha_k}\psi)e^{\alpha_1\ldots \alpha_k},
\eeq 
 

Os tipos de campos espinoriais poss\'iveis de acordo com essa classifica\c c\~ao limita os termos que uma Lagrangiana correspondente em $4+1$ dimens\~oes  pode ter, al\'em disso esses termos dependem de qual realiza\c c\~ao \'e levada em conta para os campos espinoriais.

Na próxima seção, exploraremos um pouco mais espinores em espaços Lorentzianos de cinco dimensões. Escreveremos os bilineares covariantes não-nulos em termos das componentes de seu espinor e concluiremos que totalizam todos os invariantes que podem ser construídos sobre o {\it bulk}, em particular sobre buracos negros axissimétricos em cinco dimensões.

\section{Espinores sobre buracos negros axissimétricos de cinco dimensões}
 
\qquad Nesta seção, consideraremos buracos negros axissimétricos em cinco dimensões, cujos análogos no espaço-tempo são os buracos negros de Kerr. Sobre esses buracos negros 5-dimensionais, construiremos campos espinoriais e calcularemos seus bilineares covariantes, os quais são invariantes sobre o {\it bulk}. Esses bilineares covariantes são os mesmos listados na seção anterior, aqui escreveremos esses bilineares em termos das componentes do espinor em uma base local de {\it fünfbein} \cite{Mei1}.
 \\ \indent
 Considere, então, buracos negros axissimétricos em um espaço-tempo de cinco dimensões. Eles são caracterizados totalmente por sua massa e seu momento angular \cite{Myers:1986un,Axi}. De acordo com Mei \cite{Mei2}, vetores de Killing\footnote{Vetores de Killing são campos de vetores $\xi$, em cujas direções as propriedades geométricas do espaço-tempo permanecem invariantes, isto é, $(\nabla_\xi g)_{\m\n}=0$, o que nos diz quantas simetrias $\xi$ esse espaço possui. De forma que, os campos de Killing são determinados equivalentemente pela equação $\nabla_\m\xi_\n+\nabla_\n\xi_\m=0$. 
É interessante notar que a simetria do espaço-tempo referente a translações temporais ($\xi=\p_t$) no horizonte de eventos nos dá vetores $\xi$ que são tipos-nulos e normais ao horizonte: $\nabla^\a(\xi_\m\xi^\m)=-2\kappa\xi^\a$ \cite{nast}.} nulos no horizonte de eventos se transformam em vetores de Killing tipo-tempo que estão a uma distância espacial infinita do buraco negro, quando interpolados através de um campo vetorial específico $\xi^\m$ \cite{Mei2}. No entanto, esses vetores de Killing são preservados quando escritos na base {\it fünfbein}, que é o referencial próprio e é, por definição, a base {\it vielbein} $e^A={e^A}_\m dx^\m$ para o espaço de cinco dimensões. Nessa seção, calcularemos os bilineares quando escritos nessa base.  Para isso, é admitida a existência de uma 1-forma densidade de corrente $(\bar{\psi}\g^\m\psi) e_\m$, que é preservada, ou seja, o campo espinorial $\psi$ satisfaz à equação de Dirac\cite{Mei2}.

Seja uma m\'etrica geral para um buraco negro axissim\'etrico e estacion\'ario em cinco dimens\~oes \cite{Mei2}
\beq 
ds^2&=&-f_t(dt+f_1d\phi_1+f_2d\phi_2)^2+f_rdr^2+f_\theta d\theta^2+g_{22}(d\phi_2-\omega_2dt)^2\nonumber
,\\ &+&g_{11}\left[d\phi_1-\omega_1dt+g_{12}(d\phi_2-\omega_2dt)\right],
\eeq
 onde $f_t,f_1,f_2,f_r,f_\theta, g_{11},g_{12},g_{22},\omega_1$ e $\omega_2$ são funções apenas de $r$ e $\theta$. No referencial do buraco negro, $t$ representa a coordenada temporal, $r$ denota a distância radial, $\theta$ é o ângulo latitudinal, e $\phi_1$ e $\phi_2$ são ângulos azimutais. Nesse sentido, $\omega_1$ e $\omega_2$ são velocidades angulares referentes à essas direções azimutais, respectivamente. Como a métrica calculada próximo ao buraco negro não muda de sinal, podemos escrevê-la em termos de {\it fünfbeinen}.
 \be 
 ds^2=\eta_{AB}e^Ae^B,\qquad A,B=0,\ldots, 4,
 \ee 
 tal que as componentes dos $e^A={e^A}_\m dx^\m$ são \cite{Mei2}
\beq
e^0=\sqrt{f_t}(dt+f_1d\phi_1+f_2\phi_2), \quad e^1=\sqrt{f_r}dr,\quad e^2=\sqrt{f_\theta}d\theta,\\
e^3=\sqrt{g_{11}} 
\left[d\phi_1-\omega_1dt+g_{12}(d\phi_2-\omega_2dt)\right], \quad e^4=\sqrt{g_{22}}(d\phi_2-\omega_2 dt)
\eeq
 e podem ser expressos em termos matriciais, como \cite{Mei2}
\be\label{e}
[{e^A}_\m]\equiv\left(\begin{tabular}{ccccc} 
$\sqrt{f_t}$&0&0&$\sqrt{f_t}f_1$&$\sqrt{f_t}f_2$\\
0&$\sqrt{f_r}$&0&0&0\\
0&0&$\sqrt{f_\theta}$&0&0\\
-$\sqrt{g_{11}}(\omega_1+g_{12}\omega_2)$&0&0&$\sqrt{g_{11}}$&$\sqrt{g_{11}}g_{12}$\\
-$\sqrt{g_{22}}\omega_2$&0&0&0&$\sqrt{g_{22}}$
\end{tabular}\right), 
\ee

Considere agora um campo espinorial $\psi$ sobre esse espaço e o campo vetorial $\xi^\m$, construído a partir do vetor densidade de corrente $J^\m$ de $\psi$,
\be \label{ximu}
\xi^\m=b_\psi J_\psi^\m= b_\psi \bar{\psi} \g^\m\psi,
\ee
 onde $b_\psi$ é uma constante. As componentes do campo vetorial $\xi^\m$ são obtidas através da matriz inversa $[{e^A}_\m]^{-1}$. Mas para isso, considere uma mudança de base feita através das matrizes $\g$ e um vínculo no campo espinorial $\psi$
 $$
 \g^\m={e_A}^\m\g^A, \;{\rm onde}\quad {e_A}^\m {e_\m}^B={\d_A}^B,\quad
 {e_\m}^A{e_A}^\n={\d_\m}^\n,\quad e^A={e^A}_\m dx^\m,
$$ onde $\g^A$ denota as matrizes $\g$ sobre um espaço de 4+1 dimensões na base {\it fünfbein}, as quais são representadas por \cite{Mei2}
\be\label{gamma} 
\g^0=i\s^1\otimes {\bf 1}_2,\quad \g^k=-\s^2\otimes \s^k,\quad \g^4=\s^3\otimes{\bf 1}_2. 
\ee

A eq. (\ref{ximu}) torna-se
$$
\xi^\m=b_\psi ({e_A}^\m)\underbrace{(\bar{\psi}\g^A\psi)}_{J^A}
.$$
Considere o campo espinorial $\psi$ regido pelo vínculo
$$
(\g^0\pm \g^5)\psi=0
,$$ onde $\g^5=-\g^1\g^2\g^3\g^4\g^0=-i{\bf 1}_4$ é o elemento de volume. Ent\~ao, $\psi$ é simplificado, quando escrito na base {\it fünfbein} \cite{Mei2},
\be 
\psi=\left(\begin{tabular}{c}$\vec{\a}$\\ $\vec{\a}$
\end{tabular}\right)=\left(\begin{tabular}{c}1\\1
\end{tabular}\right)\otimes\vec{\a},\text{ onde}\quad\vec{\a}=\left(\begin{tabular}{c}$\a_1$\\ $\a_2$
\end{tabular}\right) \quad
{\rm e \ \ }\a_1,\a_2 {\rm \ \ s\tilde{a}o\ \ constantes\ \ complexas}.
\ee
Resumindo, o campo vetorial $\xi^\m$ pode ser escrito em uma base local
\be
\xi^\m= [{({e^{-1})}^\m}_A][\bar{\psi}\g^A\psi]=\left(\begin{tabular}{c}1\\0\\0\\
$\omega_1$\\
$\omega_2$
\end{tabular}\right)\equiv \p_t+\omega_1\p_{\phi_1}+\omega_2\p_{\phi_2}
,\ee
onde denotamos por $[\bar{\psi}\g^A\psi]$ a matriz coluna dos escalares $\bar{\psi}\g^A\psi$. 

Como foi feito em 
\cite{BR2016} para campos espinoriais sobre espa\c cos de $6+1$ dimens\~oes, o campo vetorial densidade de corrente  é escrito em termos das componentes de {\it fünfbein}:
\be
J^\m\p_\m=(\bar{\psi}\g^\m\psi)\p_\m=2(|\a_1|^2+|\a_2|^2)\p_t,
\ee
onde utilizamos a representação (\ref{gamma}) para as matrizes $\g$. Essa expressão é muito similar à encontrada para a densidade de corrente sobre espaços Lorentzianos de sete dimensões 
 \cite{BR2016}.

Além desse bilinear vetorial $\varphi_1=J^\m e_\m$, em nossa classificação de campos espinorias sobre espaços Lorentzianos de cinco dimensões, há apenas um outro bilinear covariante não-nulo, que denotamos por $\tilde{\varphi}_2$ e que possui estrutura quaterniônica.  
Se tal estrutura é representada da maneira mais direta 
$$
J_i=i\s_i\otimes 1_2,  \text{ onde $\s_i$ são matrizes de Pauli,}
$$
então a forma bilinear covariante  $\tilde{\varphi}_2=\f{1}{2}\left(\bar{\psi}J_k\g_{\m\n}\psi\right)e^{\m\n}$ pode ser facilmente calculada no referencial próprio da base de {\it f\"unfbein}. Dessa forma, para o campo espinorial $\psi=(\a,\a)^\intercal$ dado acima, obtemos
\beq
\tilde{\varphi}^1_2&=&-1_2\otimes\left[\left(\a^\dagger_1\a_2+\a^\dagger_2\a_1\right)e_{01}+i\left(\a^\dagger_2\a_1-\a^\dagger_1\a_2\right)e_{02}+\left(|\a_1|^2-|\a_2|^2\right)e_{03}\right],\\ \nonumber
\tilde{\varphi}^2_2&=&-i1_2\otimes\left[\left(\a^\dagger_1\a_2+\a^\dagger_2\a_1\right)e_{01}+i\left(\a^\dagger_2\a_1-\a^\dagger_1\a_2\right)e_{02}+\left(|\a_1|^2-|\a_2|^2\right)e_{03}\right]=i\tilde{\varphi}^1_2,\\
\tilde{\varphi}^3_2&=&-i1_2\otimes\left[\left(\a^\dagger_1\a_2+\a^\dagger_2\a_1\right)e_{23}+i\left(\a^\dagger_2\a_1-\a^\dagger_1\a_2\right)e_{31}+\left(|\a_1|^2-|\a_2|^2\right)e_{12}\right]=i(\star\tilde{\varphi}^1_2).\nonumber
\eeq 
Concluímos, então, que as formas bilineares $\tilde{\varphi}^1_2, \tilde{\varphi}^2_2$ e $\tilde{\varphi}^3_2$ são equivalentes entre si, pois possuem as mesmas componentes funcionais e diferem apenas por uma dualidade. Juntamente com $\varphi_1$, essas componentes totalizam todos os invariantes que podem ser construídos com o campo espinorial $\psi$:
  \be
 \a^\dagger_1\a_2\pm\a^\dagger_2\a_1\quad{\rm e}\quad|\a_1|^2\pm|\a_2|^2.
 \ee
Esses novos espinores podem desempenhar um importante papel ao explorar localização de férmions sobre o {\it bulk} em modelos de branas. Espinores sobre o {\it bulk} têm sua dinâmica governada pela equação de Dirac, no entanto aqui exploramos um pouco além desse contexto. Investigar a localização de férmions sobre a brana está além dos objetivos deste trabalho. No entanto, ele nos permitiu uma compreensão mais profunda a respeito da dinâmica de campos espinoriais sobre o {\it bulk} $AdS_5$, tanto ao classificar espinores sobre esses espaços, quanto ao expressar todos os bilineares covariantes na base {\it f\"unfbein}. É importante ressaltar que essa classificação também pode ser muito útil na busca por novas partículas.

 Resumindo, o que fizemos foi classificar campos espinoriais sobre espaços Lorentzianos de cinco dimensões, entre eles o {\it bulk} e buracos axissimétricos em cinco dimensões, e encontramos três novas classes de espinores regulares e outras três de singulares. Além disso, notamos que os únicos bilineares covariantes que podem ser não-nulos são as formas $\varphi_1$, identificada com a densidade de corrente, e $\tilde{\varphi}_2$, que é o spin dual com estrutura quaterniônica. Verificamos que sobre o {\it bulk}, esses bilineares não nulos são constantes quando expressos na base {\it fünfbein} e totalizam todos os invariantes possíveis que podem ser escritos com o campo espinorial $\psi=(\a, \a)^\intercal$:
\be
|\a_1|^2\pm|\a_2|^2\text{\;\; and\;\; }\a^\dagger_1\a_2\pm\a^\dagger_2\a_1.
\ee
Se é feita uma classificação de espinores sobre espaços Riemannianos de cinco dimensões, em particular $S^5$, é possível explorar a dinâmica de campos espinoriais em supergravidade $AdS_5\times S^5$, no entanto isso não foi realizado neste trabalho. Vale ressaltar que a natureza do bilinear $\tilde{\varphi}_2$ é algo que precisa ser bastante explorado. Uma compreensão maior a respeito desses campos sobre o {\it bulk} pode ser obtida através das equações de movimento, no entanto isso foge do escopo de nosso objeto de estudo.

\myclearpage
\par

\chapter{Conclusões e perspectivas}
\indent Como vimos, o mote de nosso trabalho é a classificação de campos espinoriais sobre espaços Riemannianos de sete dimensões, em particular, sobre o espaço compacticado $AdS_4\times\underline{S^7}$ \cite{BBR} e sobre espaços Lorentzianos de cinco dimensões, em particular, o {\it bulk} $\underline{AdS_5}\times S^5$ \cite{BR}. O primeiro \'e um estudo das propriedades de campos fermiônicos sobre a esfera $S^7$ \cite{BBR} e no segundo trabalho verificamos que todos os invariantes sobre o {\it bulk} construídos com espinores correspondem aos obtidos por meio da classificação em classes \cite{BR}. 
 
Para que esse estudo pudesse ser iniciado, foi preciso revisar alguns  dos conceitos preliminares, o que foi feito no segundo capítulo. 
Dessa forma, a primeira seção foi dedicada ao estudo das \'algebras de Clifford e grupos ortogonais e espinoriais, na segunda seção alguns fundamentos geométricos foram revisitados. Na terceira seção, revisamos alguns conceitos de compactificações maximalmente simétricas. A seguir, no terceiro capítulo, apresentamos uma classifica\c c\~ao de espinores feita por Lounesto, sobre o espa\c co de Minkowski segundo bilineares covariantes, que \'e o m\'etodo de classifica\c c\~ao de espinores \cite{lou2}.
Em seguida, no quarto cap\'itulo, que \'e o cap\'itulo central deste trabalho, classificamos espinores em espaço Riemannianos de sete dimensões e em espaços Lorentzianos de cinco dimensões. Na primeira seção, foram listados todos os bilineares covariantes que podem ser constru\'idos em dimens\~oes e assinaturas arbitr\'arias, bem como foram encontrados v\'inculos bem gerais para os campos espinoriais $\psi$ sob a forma de identidade de Fierz. Um agregado de Fierz mais geral pode ser escrito em espa\c cos de dimens\~oes arbitr\'arias. Ent\~ao, na segunda seção, uma classifica\c c\~ao dos espinores de Majorana sobre uma variedade Riemanniana $(M,g)$ de sete dimens\~oes foi dada \cite{BBR}, seguindo os mesmos crit\'erios que o Lounesto. Uma restri\c c\~ao no n\'umero de classes de espinores, atrav\'es dos v\'inculos das identidades de Fierz, nos d\'a somente uma classe de espinores em uma classifica\c c\~ao para espinores reais de Majorana $\psi\in \G(S^+)$. Vimos que, nesse caso, espinores $\psi$ s\~ao se\c c\~oes do fibrado espinorial $S$ sobre um espaço Riemanniano de sete dimens\~oes $M^7$, em particular $S^7$. A partir disso, atrav\'es de um extensão do bilinear admiss\'ivel $B$ encontramos uma classifica\c c\~ao de espinores mais geral em $\G(S)$, a qual apresenta três classes não-triviais. Um agregado de Fierz graduado \'e tamb\'em encontrado sobre $M^7$, o que \'e de grande utilidade, pois agrega todos os invariantes do campo espinorial $\psi$. Al\'em disso, uma din\^amica para esses campos espinoriais de mat\'eria \'e dada sobre tal variedade $(M,g)$, atrav\'es de uma realiza\c c\~ao quaterni\^onica dos campos $\psi$ \cite{Top}. A classifica\c c\~ao dos espinores, juntamente com a Lagrangiana que constru\'imos para esses espinores, nos garante potenciais possibilidades em f\'isica e candidatos mais gerais a novas solu\c c\~oes sobre a compactificação $AdS_4\times S^7$ e suas vers\~oes ex\'oticas. \\
\indent Na terceira seção do quarto capítulo, classificamos campos espinoriais sobre variedades Lorentzianas 5-dimensionais, como por exemplo o {\it bulk}, seguindo o mesmo crit\'erio de classifica\c c\~ao em classes acima, dando continua\c c\~ao aos trabalhos, onde foram classificados espinores sobre variedades Riemannianas \cite{BBR} e Lorentziana \cite{BR2016} 7-dimensionais. Com isso, obtemos seis novas classes não-triviais de campos espinoriais, sendo três regulares e as outras três, singulares. Por fim, na última seção, calculamos explicitamente os bilineares sobre o {\it bulk} e concluímos que suas componentes perfazem todos os invariantes possíveis da teoria \cite{BR}.

\section*{Perspectivas}

\qquad Essas classificações dos espinores em classes em espaços arbitrários,
tais como: o espaço de Minkowski, o {\it bulk} e o espaço compactificado $S^7$, 
dão margem a discussões bem interessantes e a novos resultados com férmions que
podem ser encontrados. Assim, podemos explorar esses campos em outros contextos,
entre eles: uma caracterização de espinores no {\it bulk}, uma classificação de
espinores sobre o espaço de Minkowski em segunda quantização, modos de Kaluza-Klein
espinoriais na brana, extensões octoniônicas de espinores em $S^7$, entre outros.

\subsection*{Caracterização de campos espinoriais no {\it bulk}}

\qquad No momento, estamos trabalhando com a caracterização de espinores no {\it bulk}, pois este estudo trata-se de um detalhamento da classificação em classes e poderá ser bastante útil no estudo prático do comportamento e da dinâmica de tais campos fermiônicos e, consequentemente, também em sua compreensão \cite{BR,brane}. Além disso, possibilitará o cálculo de ações efetivas envolvendo interações desses campos com campos de espinores na brana e com a própria brana, através de uma curvatura induzida na mesma \cite{Duff}. Isso nos possibilitará, também, uma compreensão mais abrangente acerca de localização de espinores sobre a brana \cite{Kuerten:2016cho,Jardim:2014xla}.

Seguindo uma linha de raciocínio análoga a caracterização de espinores no espaço de Minkowski \cite{Cavalcanti:2014wia}, caracterizaremos os campos 
espinoriais no {\it bulk}, restringindo o número de componentes desses através 
dos vínculos com bilineares covariantes obtidos da classificação dos campos
espinoriais sobre o {\it bulk} \cite{BR}. Dessa forma, estamos buscando por 
expressões gerais para os campos espinorias em cada classe. Para isso, 
os bilineares covariantes são expressos em termos das componentes complexas de cada espinor.

\subsection*{Classifica\c c\~ao de espinores sobre o espa\c co de Minkowski em $2^a$ quantiza\c c\~ao}

\qquad A classifica\c c\~ao de campos espinoriais feita por Lounesto sobre o espa\c co-tempo de Minkowski e as classifica\c c\~oes de campos espinoriais sobre espa\c cos Riemannianos de sete dimens\~oes \cite{BBR} e Lorentzianos de sete \cite{BR2016} e de cinco dimens\~oes \cite{BR} 
 seguem um formalismo de primeira quantiza\c c\~ao, ou seja, descrevem apenas uma part\'icula. Se procurarmos por uma classifica\c c\~ao de espinores no espa\c co de Minkowski para campos espinoriais em um formalismo de segunda quantiza\c c\~ao, uma tentativa ingênua seria basear nossa classifica\c c\~ao nos valores esperados de v\'acuo dos bilineares covariantes segundo-quantizados de uma maneira direta. Tais VEV's para o espinor de Dirac podem ser escritos em termos do propagador de Feymann $:\langle 0|T(\bar{\psi}_a(x)\psi_b(y))|0 \rangle:= :{S_F}_{ab}(x-y):$. Vejamos, então, como fazer essa classificação, que é um tanto quanto rudimentar, em segunda quantização, mas que serve de exemplo para uma classificação mais consistente e realista que poderá ser feita futuramente. Seja, dessa forma, o seguinte VEV: 
\beq 
:\langle 0|T(\bar{\psi}(x)\g_A\psi(y))|0 \rangle :&=&:\g_A^{ab}{S_F}_{ab}(x-y):=:Tr[\g_A{S_F}(x-y)]:,
\eeq
onde $$
{S_F}_{ab}(x-y)=\int \f{d^4p}{(2\pi)^4}\f{e^{-ip(y-x)}(\slashed{p}+m)}{p^2-m^2+i\e}.
$$
{\bf Espinores de Dirac} \quad A nulidade desses VEV's para espinores de Dirac podem ser determinadas atrav\'es do tra\c co de matrizes gama \cite{Pesk} 
\beq
Tr(\g_\m)=0,& Tr(\g_\m\g_\n)=4g_{\m\n},&Tr(\g_5\g_\m\g_\n)=0,\\
Tr(\g_5\g_\m)=0,&Tr(\g_5)=0,&
\eeq
de onde se segue que
\beq 
:\langle 0|T(\bar{\psi}(x)\g_A\psi(y))|0 \rangle:=\int \f{d^4p}{(2\pi)^4}\f{e^{-ip(y-x)}}{p^2-m^2+i\e}
:Tr[\g_A(\g_\m p^\m +m)]:
\eeq
se anula apenas para certos valores do índice composto $A=\{\phi,\m,\m\n,\m\n\s,5\}$. Logo, apenas os seguintes VEV's s\~ao n\~ao-nulos para espinores de Dirac
\beq
:\langle 0|T(\bar{\psi}(x)\psi(y))|0 \rangle:=\int \f{d^4p}{(2\pi)^4}\f{me^{-ip(y-x)}}{p^2-m^2+i\e}\neq 0,\\
:\langle 0|T(\bar{\psi}(x)\g_\m\psi(y))|0 \rangle:=\int \f{d^4p}{(2\pi)^4}\f{4p_\m e^{-ip(y-x)}}{p^2-m^2+i\e}\neq 0,\\ :\langle 0|T(\bar{\psi}(x)\g_\m\g_\n\psi(y))|0 \rangle:=\int \f{d^4p}{(2\pi)^4}\f{4mg_{\m\n}e^{-ip(y-x)}}{p^2-m^2+i\e}\neq 0.
\eeq

{\bf Espinores de Weyl} \quad Uma constru\c c\~ao an\'aloga pode ser feita para espinores de Weyl, utilizando tamb\'em os resultados da subse\c c\~ao anterior, quando escrevemos os espinores de Weyl em termos de espinores de Dirac
\be
\psi_W=\f{1}{2}(1+\g^5)\psi_D. 
\ee
Ent\~ao, usando o fato que $\left[\f{1}{2}(1+\g^5)\right]^2=\f{1}{2}(1+\g^5)$
e que $(\g^5)^2=1$,
\beq 
:\langle 0|T(\bar{\psi}_W(x)\g_A\psi_W(y))|0 \rangle:=\f{1}{4}:\langle 0|T(\bar{\psi}_D(x)(1+\g^5)\g_A(1+\g^5)\psi_D(y))|0 \rangle: ,\\
=\f{1}{4}\int \f{d^4p}{(2\pi)^4}\f{e^{-ip(y-x)}}{p^2-m^2+i\e}
:Tr[(1+\g^5)\g_A(1+\g^5)(\g_\m p^\m +m)]:.
\eeq
Se $\g^A$ tem grau par, e como $\g^5$ anticomuta com todas as matrizes gama,
\beq 
:\langle 0|T(\bar{\psi}_W(x)\g_A\psi_W(y))|0 \rangle:
=\f{1}{2}\int \f{d^4p}{(2\pi)^4}\f{e^{-ip(y-x)}}{p^2-m^2+i\e}
:Tr[(1+\g^5)\g_A(\g_\m p^\m +m)]:,\\
=\f{1}{2}\int \f{d^4p}{(2\pi)^4}\f{e^{-ip(y-x)}}{p^2-m^2+i\e}
:Tr[\g_A(\g_\m p^\m +m)]:,
\eeq
ent\~ao os VEV's assumem os mesmos valores que no caso de espinores de Dirac, ou seja, os VEV's s\~ao n\~ao-nulos apenas quando $\g_A$ tem grau 0 ou 2.  
Se o grau de $\g_A$ \'e impar, ent\~ao
\be
(1+\g^5)\g_A(1+\g^5)=(1+\g^5)(1-\g^5)\g_A=0.
\ee
Logo, todos os VEV's para correspondentes a $\g_A$ de grau \'impar s\~ao nulos. Portanto, os \'unicos VEV's n\~ao-nulos, sobre os quais basear\'iamos uma classifica\c c\~ao de espinores, s\~ao os de grau 0 e 2, citados acima. 

Uma classificação de espinores mais realista sobre o espaço de Minkowski em segunda quantização exige um maior cuidado.
\subsection*{Modos de Kaluza-Klein espinoriais na brana} 
\qquad Da mesma forma que um potencial efetivo para um campo escalar pode ser
calculado sobre a brana no {\it bulk}, juntamente com seus modos de Kaluza-Klein,
podemos fazer algo análogo para campos fermiônicos sobre o {\it bulk}. Então, a partir da equação de Dirac sobre $AdS_5$ e usando a métrica   não-fatorizável 
 \cite{Gimb}:
\be
ds^2= \phi^2(y)\g_{\m\n}dx^\m dx^\n +dy^2,
\ee podemos calcular o fator de dobra $\phi(y)$ 
 para cada classe desses campos e também suas densidade de energia correspondentes. Consequentemente, os modos de Kaluza-Klein também podem ser obtidos \cite{brane,Gimb}.
 
Como perspectiva, também calcularemos
os modos e as massas de Kaluza-Klein sobre $S^7$ para cada classe de espinores,
as quais foram obtidas nesse trabalho \cite{duff86}. Abaixo, falamos brevemente sobre a álgebra dos octônions e algo mais que poderemos explorar por meio da álgebra octoniõnica.
\subsection*{Oct\^onions}
%

\qquad John T. Graves, inspirado pela descoberta dos quat\'ernions feita por seu amigo William Hamilton, descobriu, em 1843, os oct\^onions, que ele mencionou em uma carta a Hamilton em 16 de dezembro de 1843 e os chamou de oitavas ({\it octaves}). No entanto, Arthur Cayley descobriu, independentemente, essa mesma \'algebra e publicou seu trabalho em 1845, um pouco antes da publica\c c\~ao de Graves no mesmo ano \cite{Hilb}. 
 O produto que define tal álgebra \'e dado pela seguinte tabela de multiplica\c c\~ao
\be
{e}_a\circ {e}_b=\e_{ab}^c{e}_c-\d_{ab},
\ee
onde $\e_{ab}^c$ \'e igual ao sinal da permuta\c c\~ao c\'iclica $(abc)\in \{(126),(237),(341),(452),(563),$ $(674),(715)\}$, tal que tais permutações surgem mnemonicamente a partir do plano de Fano. Pode ser obtido ainda, de forma equivalente, usando a \'algebra de Clifford ${\cal C}\ell_{0,7}$
\be
A\circ B=\langle AB(1-\phi)\rangle_{0\oplus 1}, \quad A,B\in \mathbb{R}\oplus \mathbb{R}^{0,7},
\ee
onde $\phi =e_{126}+e_{237}+e_{341}+e_{452}+e_{453}+e_{674}+e_{715} $ e denotamos 
$$
e_{ijk}=\f{1}{3!}\sum_{perm(ijk)}e_ie_je_k,
$$
tal que a justaposi\c c\~ao \'e o produto de Clifford \cite{trae,trae1,ced}.

Elementos octoni\^onicos n\~ao satisfazem a identidade de Jacobi $J(x,y,z):= [x,[y,z]]+[y,[z,x]]+[z,[x,y]]=0$ como o fazem os quat\'ernions e, por isso, os oct\^onions imagin\'arios n\~ao formam uma \'algebra de Lie. No entanto, eles possuem as seguintes relações de comutação: $[e_a, e_b]=2\epsilon_{ab}^c e_c$ e, além disso,  satisfazem às identidades de Malcev 
\be 
J(x,y,[x,z])=[J(x,y,z),x]
,\ee formando uma espécie de álgebra de Lie generalizada, denominada álgebra de Malcev. Para o caso particular de oct\^onions imagin\'arios, as identidades de Malcev identificam-se com as identidades de Jacobi estendidas, expressas por:
\be\label{jacobiext}
J(e_i,e_j,e_k) =3\e_{ijkl}{e}_l,
\ee
onde $\e_{ijkl}=-\e_{mij}\e_{mkl}-\d_{il}\d_{jk}-\d_{ik}\d_{jl}$.

É interessante considerar ainda alguns produtos octoniônicos que serão úteis futuramente:
\subsubsection*{* Produto $X$}
\qquad Dados $X,Y\in \mathbb{R}\oplus \mathbb{R}^{0,7}$ unit\'arios, isto \'e, tal que $X\bar{X}=\bar{X}X=1$ e $Y\bar{Y}=\bar{Y}Y=1$, onde a  conjuga\c c\~ao octoni\^onica $\bar{X}$ \'e um antiautomorfismo equivalente à conjuga\c c\~ao de Clifford restrita a $\mathbb{R}\oplus \mathbb{R}^{0,7}$ (paravetores), o produto-$X$ pode ser definido por \cite{trae1}
\be
A\circ_XB:=(A\circ X)\circ(\bar{X}\circ B).
\ee
As seguintes identidades de Moufang podem ser escritas com esse produto 
\be\label{Mouf}
(A\circ X)\circ(\bar{X}\circ B)=X\circ((\bar{X}\circ A)\circ B)=
(A\circ(B\circ X))\circ\bar{X}.
\ee

\subsubsection*{Produto $\bullet$}
\qquad Dado um multivetor $u=u_1\dots u_k \in \wedge^k(\mathbb{R}^{0,7})\subset {\cal C}\ell_{0,7}$, um paravetor $A\in \mathbb{R}\oplus \mathbb{R}^{0,7}$ pode ser multiplicado à esquerda ou à direita por $u$, resultando tamb\'em em paravetores definidos, respectivamente, por
\beq
A\bullet_\llcorner u=(\cdots((A\circ u_1)\circ u_2)\circ\cdots)\circ u_k,\\
u\bullet_\lrcorner A=u_1\circ(u_2\circ(\cdots\circ(u_k\circ A)\cdots)).
\eeq Este produto pode ser estendido trivialmente para $\wedge(\mathbb{R}^{0,7})\approx {\cal C}\ell_{0,7}$.

\subsubsection*{* Produto $u$} 
\qquad Se utilizamos o produto-$\bullet$, uma extens\~ao do produto $X$ pode ser feita para $X$ pertencente a \'algebra de Clifford, tal extens\~ao \'e chamada de produto-$u$ e definida como se segue
\be
A\circ_u B=(A\bullet_\llcorner u)\circ(\bar{u}\bullet_\lrcorner B),
\ee
onde $A,B \in \mathbb{R}\oplus \mathbb{R}^{0,7}$. Como consequência, $A\circ_u B\in \mathbb{R}\oplus \mathbb{R}^{0,7}$. Como \'e referido em \cite{trae1}, as identidades equivalentes \`as eqs.  (\ref{Mouf}) n\~ao s\~ao satisfeitas para o produto-$u$.
\be
(A\bullet u)\circ(\bar{u}\bullet B)\neq u\bullet((\bar{u}\bullet A)\circ B)\neq
(A\circ(B\bullet u))\bullet\bar{u}\neq (A\bullet u)\circ(\bar{u}\bullet B).
\ee
De qualquer forma, n\~ao sabemos se o produto-$X$ ou produto-$u$ satisfazem as identidades de Moufang,  ou seja, \'e um problema em aberto se, para cada $u\in {\cal C}\ell_{0,7}$, a \'algebra $(\mathbb{R}\oplus\mathbb{R}^7,[\;,\;]_{\circ_u})$ \'e uma \'algebra de Malcev, ou não. Apenas um caso particular foi considerado, o caso em que $u=e_4e_2e_1$, para o qual  mostramos que, ao menos para uma dada componente, a \'algebra $(\mathbb{R}\oplus\mathbb{R}^7,[\;,\;]_{\circ_u})$ comporta-se como uma \'algebra de Malcev. Se queremos uma solu\c c\~ao completa
, \'e suficiente mostrar que as constantes de estrutura ${^u}{c^k}_{ji}$ do produto-$u$ satisfazem a seguinte identidade \footnote{Verificaremos se a \'algebra octoni\^onica do produto-$u$ \'e uma \'algebra de Malcev ou tem uma sub-\'algebra que seja de Malcev.}
$$
{^u}c^l_{mi} {^u}c^m_{n[j} {^u}c^n_{ki]}= {^u}c^l_{m[i} {^u}c^m_{nj]} {^u}c^n_{ki}
,$$
ou ainda, que em uma dada representa\c c\~ao matricial, as constantes ${[^u{c^k}_j]}_i$ devem satisfazer:
 \be
  [c]_i\left[[c]_i,[c]_k\right]=c^n_{ki}\left([c]_n[c]_i+c^m_{ni}[c]_m\right).
 \ee

\subsection*{Extensões octoniônicas para espinores sobre $S^7$}

\qquad Futuramente, alguns c\'alculos realizados sobre $S^7$ poderão ser estendidos 
através da \'algebra dos oct\^onions e, posteriormente, para produtos
octoni\^onicos mais gerais, como por exemplo o produto-$u$, que carrega
algo das álgebras de Clifford \cite{trae,trae1,ced}. Dessa forma, a classifica\c c\~ao de espinores
em sete dimens\~oes, abordada no cap\'itulo quatro, poderá ser considerada
neste contexto da \'algebra octoni\^onica \cite{mart} e assim poderemos procurar por uma
classificação de campos espinorias que tomam valores na \'algebra dos oct\^onions \cite{Top, Top1}. Pois, a estrutura octoniônica pode ser utilizada para se obter uma geometria paralelizável sobre $S^7$. Dessa forma, a torção é considerada proporcional às constantes de estrutura da álgebra octoniônica, das quais resultam uma curvatura nula sobre $S^7$ \cite{eng4,SUSYBr,akiv,logi07}.
Indo mais além, buscaremos por relações e interconexões 
entre o formalismo octoniônico e as extensões $Z_3$ das álgebras de Clifford, ambos
no contexto dos campos espinoriais \cite{Baez,Abramov, Kerner, Kerner1}.

\subsection*{Extensões octoniônicas das famílias de geometrias sobre $S^7$}

\qquad Buscaremos explorar um pouco mais, tanto no aspecto geom\'etrico quanto em propriedades dos campos, as extens\~oes das fam\'ilia de geometrias sobre $S^7$ a um par\^ametro constru\'ida por Loginov \cite{eng4} atrav\'es de produtos  octoni\^onicos mais gerais. Akivis \cite{akiv} e Loginov \cite{eng4,logi09,logi07} estenderam os tr\^es tipos de geometrias em $S^7$ existentes: a geometria Riemanniana induzida do $\mathbb{R}^8$ e as multiplica\c c\~oes à esquerda e à direita, e encontraram uma fam\'ilia de geometrias a um par\^ametro sobre $S^7$ que possui essas tr\^es geometrias como casos particulares. Pelo uso das extens\~oes do produto octoni\^onico $u$, 
generalizamos ainda mais essa fam\'ilia de geometrias sobre $S^7$, as quais de certa forma são descritas pelo tensores torção e curvatura, sendo que estes estão estritamente relacionados ao campo de calibre que descreve o spin e ao tensor eletromagnético estendido \cite{spincalibre,eng4}.

\appendix
\chapter{}
\section{Compactificação}\label{apen1}
\qquad Aqui escrevemos explicitamente os cálculos de compactificação realizados por Freund e Rubin \cite{Freundrubin}. Para isso, considere a cisão 
$M=M_2\times M_{d-2}$, cujo tensor métrico é dado por 
\be\label{gdec}
g_{ab}=\left(\begin{tabular}{cc}
$g_{mn}(x^p)$&$0$\\$0$&$g_{\bar{m} \bar{n}}(x^{\bar{p}})$
\end{tabular}\right)
.\ee
Assim, obtemos os tensores de Ricci e os escalares de curvatura através das equações de Einstein-Maxwell, que são expressas por
\be
G^{ab}:=R^{ab}-\f{1}{2}g^{ab}R=
-8\pi G({F_\rho}^a F^{\rho b}-\f{1}{4}F_{\a\b}F^{\a\b}g^{ab}),
\ee
onde $F^{ab}=\f{\e^{ab}}{\sqrt{|g_2|}}f$, $f$ é uma constante com dimensão de massa quadrada e $g_2$ é a métrica sobre o espaço $M_2$
. Para isso, denotemos $R_2=g^{mn}R_{mn}$ e $R_{d-2}=g^{\bar{m},\bar{n}}R_{\bar{m},\bar{n}}$.
Isso resulta que, para $a,b=m,n$, é obtida a equação
\beq\nonumber
g_{mn}G^{mn}=g_{mn}(R^{mn}-\f{1}{2}g^{mn}R)&=&R_2-R=g_{mn}(-8\pi G)f^2(\overbrace{{\e_\rho}^m\e^{\rho n}}^{g^{mn}}-\f{1}{4}\overbrace{\e_{\a\b} \e^{\a\b}}^2 g^{mn}),\\
&\Rightarrow &-R_{d-2}\;=-8\pi Gf^2\f{\det(g_{mn})}{|\det(g_{mn})|}=:-\l.
\eeq
Enquanto que, para $a,b=\bar{m},\bar{n}$, segue
\beq 
g_{\bar{m}\bar{n}}G^{\bar{m}\bar{n}}=&g_{\bar{m}\bar{n}}(R^{\bar{m}\bar{n}}-\f{1}{2}g^{\bar{m}\bar{n}}R) =g_{\bar{m}\bar{n}}(-8\pi G)({\e_\rho}^{\bar{m}} \e^{\rho\bar{n}} -\f{1}{4}\e_{\a\b}\e^{\a\b} g^{\bar{m}\bar{n}})f^2,\\ \Rightarrow \;& R_{d-2}-\f{d-2}{2}(R_2+R_{d-2})=\underbrace{8\pi Gf^2\f{\det(g_{mn})}{|\det(g_{mn})|}}_\l\f{1}{2}(d-2)=\f{\l}{2}(d-2).
\eeq
Então, como $R_{d-2}=\l$, os escalares de curvatura para os respectivos espaços $M_2$ e $M_{d-2}$ são expressos a seguir 
\be
2\l-(d-2)(R_2+\l)=\l(d-2)
\ee
\beq
\Rightarrow\quad\left\{\begin{tabular}{ccc}
$R_2=$&$-2\f{d-3}{d-2}\l$, \\
$R_{d-2}=$&$\l$
\end{tabular}\right.
\eeq
Este resultado pode ser estendido, se consideramos o tensor eletromagnético $F^{\m\n}$ para ordens superiores
\be
 F^{\a_1\ldots\a_s}=\left\{\begin{tabular}{c}
 $\f{f}{\sqrt{|g_s|}}\e^{\a_1\ldots\a_s},$\quad $1\leq \a_i \leq s$\\ 
 $0,$\quad caso contrário.
 \end{tabular}\right.
\ee
 Dessa forma, o tensor energia-momento pode ser escrito como
\be
\theta^{ab}={F_{\a_1\ldots\a_{s-1}}}^a F^{\a_1\ldots\a_{s-1}b}-\f{1}{2s}F_{\a_1\ldots\a_s}F^{\a_1\ldots\a_s}g^{ab}
.\ee
Isso nos dá uma cisão do espaço $M$ em $M_s\times M_{d-s}$, de onde segue também uma decomposição da métrica, como em \ref{gdec}, onde os parâmetros assumem os valores
$
1\leq m,n \leq s\text{\;\;e\;\;}s+1\leq \bar{m},\bar{n} \leq d.
$
Daí, o escalar de Ricci é calculado em cada espaço.
\begin{itemize}
\item Para $a,b=m,n$, a equação de Einstein é escrita como
\beq
\hspace{-1cm}g_{mn}G^{mn}&=&-8\pi Gf^2\left(\f{1}{(s-1)!}\overbrace{{\e_{\a_1\ldots \a_{s-1}}}^m \e^{\a_1\ldots\a_{s-1}n}}^{ (s-1)!g^{mn}}\right.\\&-&\left. \f{1}{2 s!}\overbrace{\e_{\a_1\ldots\a_s} \e^{\a_1\ldots\a_s}}^{s!}g^{mn}\right)g_{mn}\nonumber 
\eeq
\be
\Rightarrow R_s-\f{1}{2}s(R_s+R_{d-s})=-8\pi Gf^2\f{1}{2}s=-\f{s}{2}\l.
\ee

\item Para $a,b=\bar{m},\bar{n}$,
\beq
g_{\bar{m}\bar{n}}G^{\bar{m}\bar{n}}&=&-8\pi Gf^2(\f{1}{(s-1)!}\overbrace{{\e_{\a_1\ldots\a_{s-1}}}^{\bar{m}} \e^{\a_1\ldots\a_{s-1}\bar{n}}}^{0}\nonumber
\\&-&\f{1}{2s!}\overbrace{\e_{\a_1\ldots\a_s} \e^{\a_1\ldots\a_s}}^{s!}g^{\bar{m}\bar{n}})g_{\bar{m}\bar{n}}\nonumber
\eeq
\be
\Rightarrow R_{d-s}-\f{1}{2}(d-s)(R_s+R_{d-s})= 4\pi Gf^2(d-s)=\f{d-s}{2}\l.\nonumber
\ee
\end{itemize}

Portanto, os valores para os escalares de Ricci são os seguintes:
\beq
\begin{tabular}{c}
$-sR_{d-s}+(2-s)R_s=-s\l$\\
$(2-d+s)R_{d-s}-(d-s)R_s=(d-s)\l$
\end{tabular}\Rightarrow \begin{tabular}{c}
$R_{d-s}=\f{(s-1)(d-s)}{d-2}\l$\\$R_s=-\f{s(d-s-1)}{d-2}\l$.
\end{tabular}
\eeq
Diretamente, temos $R=R_s+R_{d-s}=\f{2s-d}{d-2}\l$. A seguir, o tensor de Ricci pode ser calculado, o qual segue de maneira natural, pois $R^{ab}=\f{1}{2}g^{ab}R-8\pi G\theta^{ab}$. Então, obtemos os seguintes espaços de Einstein:
\beq 
R^{mn}&=&\f{1}{2}g^{mn}\f{2s-d}{d-2}\l-\f{1}{2}\l g^{mn}=-\f{d-s-1}{d-2}\l g^{mn},\\
R^{\bar{m}\bar{n}}&=&\f{1}{2}g^{\bar{m}\bar{n}}\f{2s-d}{d-2}\l+\f{1}{2}\l g^{\bar{m}\bar{n}}=\f{s-1}{d-2}\l g^{\bar{m}\bar{n}}.
\eeq
Como $d>s$, concluímos que se $d>2$ e $M_s,M_{d-s}$ são espaços maximalmente simétricos, então eles são exatamente os espaços $AdS_s$ e $S^{d-s}$, respectivamente.

Vejamos o caso particular em que $d=11$. O espaço é decomposto em $M^4\times M^7$, se escolhemos $s=4$, ou seja, a $4$-forma $F_{abcd}$. Então, temos as seguintes curvaturas escalares e tensores de Ricci :
\beq 
 R=-\f{4\cdot 6}{9}\l=-\f{8\l}{3},&\quad R_7=\f{3\cdot 7}{9}\l=\f{7\l}{3},\\
R^{mn}=-\f{2}{3}\l g^{mn},&\quad R_7^{\bar{m}\bar{n}}=\f{1}{3}\l g^{\bar{m}\bar{n}}.
\eeq
Neste caso, como mencionamos acima, $M^4$ é um espaço Lorentziano de curvatura escalar negativa (espaço de anti-de Sitter) e $M^7$, um espaço Riemanniano de curvatura escalar positiva (espaço esférico), ou seja, o espaço $M$ é decomposto de forma maximalmente simétrica como $AdS_4\times S^7$.
\section{Invariantes sobre buracos negros em cinco dimensões}
\qquad Seja a representação das matrizes $\g$ para um espaço de assinatura $4+1$:
\be
\g^0=i\s^1\otimes 1_2,\quad\g^4=\s^3\otimes 1_2,\quad\g^j=-\s^2\otimes\s^j, \text{ onde }\quad j=1,2,3. 
\ee
Sabendo-se que $\s^i\s^j=\d^{ij}+i\e^{ijk}\s^k$ e que o vínculo $(\g^0\pm \g^5)\psi=0$ é satisfeito para o espinor $\psi$, ou seja, 
\be 
\psi=\left(\begin{tabular}{c}
$1$\\$1$\end{tabular}\right)\otimes\left(\begin{tabular}{c}
$\a_1$\\$\a_2$
\end{tabular}\right),\ee
então podemos calcular os bilineares covariantes $\varphi_1,\,\tilde{\varphi}_2$ em termos das coordenadas de $\psi$:
\beq
\nonumber\varphi_1 &=&(\bar{\psi}\g^\m\psi)e_\m=\left[\left(\begin{tabular}{cc}$1$&$1$
\end{tabular}\right)\ot\left(\begin{tabular}{cc}$\a_1$&$\a_2$
\end{tabular}\right)\right](i\s^1\ot 1_2)\left[\left(\begin{tabular}{c}$1$\\$1$
\end{tabular}\right)\ot\left(\begin{tabular}{c}$\a_1$\\$\a_2$
\end{tabular}\right)
\right]e_0+\\
\nonumber &-&\left[\left(\begin{tabular}{cc}$1$&$1$
\end{tabular}\right)\ot\left(\begin{tabular}{cc}$\a_1$&$\a_2$
\end{tabular}\right)\right](\s^2\ot\s^1)\left[\left(\begin{tabular}{c}$1$\\$1$
\end{tabular}\right)\ot\left(\begin{tabular}{c}$\a_1$\\$\a_2$
\end{tabular}\right)
\right]e_1+\\ \nonumber
&-&\left[\left(\begin{tabular}{cc}$1$&$1$
\end{tabular}\right)\ot\left(\begin{tabular}{cc}$\a_1$&$\a_2$
\end{tabular}\right)\right](\s^2\ot\s^2)\left[\left(\begin{tabular}{c}$1$\\$1$
\end{tabular}\right)\ot\left(\begin{tabular}{c}$\a_1$\\$\a_2$
\end{tabular}\right)
\right]e_2+\\ \nonumber
&-&\left[\left(\begin{tabular}{cc}$1$&$1$
\end{tabular}\right)\ot\left(\begin{tabular}{cc}$\a_1$&$\a_2$
\end{tabular}\right)\right](\s^2\ot\s^3)\left[\left(\begin{tabular}{c}$1$\\$1$
\end{tabular}\right)\ot\left(\begin{tabular}{c}$\a_1$\\$\a_2$
\end{tabular}\right)
\right]e_3+\\ \nonumber
&+&\left[\left(\begin{tabular}{cc}$1$&$1$
\end{tabular}\right)\ot\left(\begin{tabular}{cc}$\a_1$&$\a_2$
\end{tabular}\right)\right](\s^3\ot 1_2e_4)
\left[\left(\begin{tabular}{c}$1$\\$1$
\end{tabular}\right)\ot\left(\begin{tabular}{c}$\a_1$\\$\a_2$
\end{tabular}\right)
\right]
\\&\Rightarrow &\varphi_1=(2i)\ot(|\a_1|^2+|\a_2|^2)e_0
.\eeq
A partir dos outros bilineares que carregam as estruturas quaterniônicas, temos
{\footnotesize{\beq 
\tilde{\varphi}^k_2&=&\f{1}{2}\left(\bar{\psi}J_k\g_{0i}\psi\right)e^{0i}+\f{1}{2}\left(\bar{\psi}J_k\g_{ij}\psi\right)e^{ij},\; \text{ tal que a tripla é ordenada: }(ijk)=(1,2,3)\\&=&\f{1}{2}\left[\left(\begin{tabular}{c}
$1$\\$1$\end{tabular}\right)^\intercal\otimes\left(\begin{tabular}{c}
$\a_1$\\$\a_2$
\end{tabular}\right)^\dagger\right]\left(i\s^1\ot 1_2\right)\left(i\s^{4-k}\ot 1_2\right)\left(i\s^1\ot 1_2\right)\left(-\s^2\ot\s^i\right)\left[\left(\begin{tabular}{c}
$1$\\$1$\end{tabular}\right)\otimes\left(\begin{tabular}{c}
$\a_1$\\$\a_2$
\end{tabular}\right)\right]e_{0i}+\nonumber\\
\nonumber &+&\f{1}{2}\left[\left(\begin{tabular}{c}
$1$\\$1$\end{tabular}\right)^\intercal\otimes\left(\begin{tabular}{c}
$\a_1$\\$\a_2$
\end{tabular}\right)^\dagger\right]\left(i\s^1\ot 1_2\right)\left(i\s^{4-k}\ot 1_2\right)\left(-\s^2\ot\s^i\right)\left(-\s^2\ot\s^j\right)\left[\left(\begin{tabular}{c}
$1$\\$1$\end{tabular}\right)\otimes\left(\begin{tabular}{c}
$\a_1$\\$\a_2$
\end{tabular}\right)\right]e_{ij}\\
\nonumber &=&\f{1}{2}\left[\left(\begin{tabular}{cc}
$1$&$1$\end{tabular}\right)\left(\begin{tabular}{cc}
$0$&$i$\\$i$&$0$\end{tabular}\right)i\s^{4-k}i\s^1(-\s^2)\left(\begin{tabular}{c}
$1$\\$1$\end{tabular}\right)\right]\ot\left[\left(\begin{tabular}{cc}
$\a^\dagger_1$&$\a^\dagger_2$\\$1$\end{tabular}\right)\s^i\left(\begin{tabular}{c}
$\a_1$\\$\a_2$\end{tabular}\right)\right]e_{0i}+\\ \nonumber
&+&\f{1}{2}\left[\left(\begin{tabular}{cc}
$1$&$1$\end{tabular}\right)\left(\begin{tabular}{cc}
$0$&$i$\\$i$&$0$\end{tabular}\right)i\s^{4-k}(-\s^2)^2\left(\begin{tabular}{c}
$1$\\$1$\end{tabular}\right)\right]\ot\left[\left(\begin{tabular}{cc}
$\a_1^\dagger$&$\a_2^\dagger$\end{tabular}\right)\s^i\s^j\left(\begin{tabular}{c}
$\a_1$\\$\a_2$\end{tabular}\right)\right]e_{ij}\\
&=&\f{1}{2}\left[\left(\begin{tabular}{cc}
$1$&$1$\end{tabular}\right)
\s^{4-k}\left(\begin{tabular}{c}
$-1$\\$1$\end{tabular}\right)\right]\ot\left[\left(\a_1^\dagger\a_2+\a^\dagger_2\a_1\right)e_{01}+\left(-i\a_1^\dagger\a_2+i\a_2^\dagger\a_1\right)e_{02}+\left(\a_1^\dagger\a_1-\a_2^\dagger\a_2\right)e_{03}\right]+\nonumber\\\nonumber
&-&\f{i}{2}\left[\left(\begin{tabular}{cc}
$1$&$1$\end{tabular}\right)\s^{4-k}\left(\begin{tabular}{c}
$1$\\$1$\end{tabular}\right)\right]\ot \left[\left(\a_1^\dagger\a_2+\a^\dagger_2\a_1\right)e_{23}+\left(-i\a_1^\dagger\a_2+i\a_2^\dagger\a_1\right)e_{31}+\left(\a_1^\dagger\a_1-\a_2^\dagger\a_2\right)e_{12}\right].\eeq}}
Se levamos em conta as seguintes expressões:
\beq&
\left(\begin{tabular}{cc}
$1$&$1$\end{tabular}\right)\s^1\left(\begin{tabular}{c}
$\pm 1$\\$1$\end{tabular}\right)=\left(\begin{tabular}{cc}
$1$&$1$\end{tabular}\right)\left(\begin{tabular}{cc}
$0$&$1$\\
$1$&$0$\end{tabular}\right)
\left(\begin{tabular}{c}
$\pm 1$\\$1$\end{tabular}\right)=1\pm 1,\\&
\left(\begin{tabular}{cc}
$1$&$1$\end{tabular}\right)\s^2\left(\begin{tabular}{c}
$\pm1$\\$1$\end{tabular}\right)=\left(\begin{tabular}{cc}
$1$&$1$\end{tabular}\right)\left(\begin{tabular}{cc}
$1$&$-i$\\$i$&$0$\end{tabular}\right)\left(\begin{tabular}{cc}
$\pm 1$\\$1$\end{tabular}\right)=-i\pm i,\\&
\left(\begin{tabular}{cc}
$1$&$1$\end{tabular}\right)\s^3\left(\begin{tabular}{c}
$\pm 1$\\$1$\end{tabular}\right)=\left(\begin{tabular}{cc}
$1$&$1$\end{tabular}\right)\left(\begin{tabular}{cc}
$0$&$1$\\$0$&$-1$\end{tabular}\right)\left(\begin{tabular}{c}
$\pm 1$\\$1$\end{tabular}\right)=\pm 1-1,\eeq
o espinor dual $\tilde{\varphi}^i_2$ pode ser simplificado como segue:
\beq
\tilde{\varphi}_2^1=-1_2\ot\left[\left(\a_1^\dagger\a_2+\a^\dagger_2\a_1\right)e_{01}-i\left(\a_1^\dagger\a_2-\a_2^\dagger\a_1\right)e_{02}+\left(|\a_1|^2-|\a_2|^2\right)e_{03}\right],\\
\tilde{\varphi}_2^2=-i1_2\ot\left[\left(\a_1^\dagger\a_2+\a^\dagger_2\a_1\right)e_{01}-i\left(\a_1^\dagger\a_2-\a_2^\dagger\a_1\right)e_{02}+\left(|\a_1|^2-|\a_2|^2\right)e_{03}\right],\\
\tilde{\varphi}_2^3=-i1_2\ot\left[\left(\a_1^\dagger\a_2+\a^\dagger_2\a_1\right)e_{23}-i\left(\a_1^\dagger\a_2-\a_2^\dagger\a_1\right)e_{31}+\left(|\a_1|^2-|\a_2|^2\right)e_{12}\right].\eeq
Portanto, $\tilde{\varphi}_2^1,\tilde{\varphi}_2^2$ e $\tilde{\varphi}_2^3$ são equivalentes entre si.
Dessa forma, os bilineares covariantes $\varphi_1$ e $\tilde{\varphi}^k_2$ contém todos os invariantes possíveis: 
\beq
\a^\dagger_1\a_2\pm \a_2^\dagger\a_1,\qquad |\a_1|^2\pm |\a_2|^2.
\eeq
\section{Espinores Elko}\label{elko}

 \qquad Os campos espinoriais do Elko descrevem férmions do tipo spin-$\f{1}{2}$ e dimensão de massa $1$, sendo também um possível candidato à matéria escura.  Eles são definidos como autoespinores do operador de conjugação de carga com helicidade dual, ou seja, os campos do Elko satisfazem $C\l(p^\m):=\g^2\l^*(p^\m)=\pm \l(p^\m)$ e são determinados, em seus referenciais próprios, por \cite{bht,Cavalcanti:2015nna}:
\beq
\l^S_\pm(k^\m)&=&\left(\begin{tabular}{c}
$i\Theta[\phi^\pm(k^\m)^*]$\\
$\phi^\pm(k^\m)$
\end{tabular}\right),\\
\l^A_\pm(k^\m)&=&\pm\;\left(\begin{tabular}{c}
$-i\Theta[\phi^\mp(k^\m)]^*$\\
$\phi^\mp(k^\m)$
\end{tabular}\right),
\eeq
onde $\Theta$ é o operador de reversão temporal e $k^\m=(m,{\bf 0})$. Dessa forma, os campos do Elko 
são determinados em termos dos autoespinores $\left(\phi_{\pm}^{S,A}\right)$ do operador helicidade $\left(\s\cdot{\bf p}\right)$ e suas reversões temporais, os quais são expressos a seguir:
\beq
{\bf \s}\cdot {\bf p}\,\phi^\pm(k^\m)&=&\pm p\;\phi^\pm(k^\m),\\
{\bf \s}\cdot {\bf p}\,\left[\Theta(\,\phi^\pm(k^\m))^*\right]&=&\mp p\;\Theta(\phi^\pm(k^\m))^*,
\eeq
tal que $p=|{\bf p}|$ e $\hat{\bf p}=({\rm sen}\,\theta\,\cos\,\phi,\,{\rm sen}\,\theta\,{\rm sen}\,\phi,\,\cos\, \theta)$. Esses espinores podem ser escritos explicitamente, como \cite{bht,Cavalcanti:2015nna}:
 \beq
 \phi^+(k^\m)&=&\sqrt{m}\left(\begin{tabular}{c}
 $\cos(\theta/2)e^{-i\phi/2}$\\ sen$(\theta/2)e^{i\phi/2}$
 \end{tabular}\right),\\
  \phi^-(k^\m)&=&\sqrt{m}\left(\begin{tabular}{c}
 $-$sen$(\theta/2)e^{-i\phi/2}$\\ $\cos(\theta/2)e^{i\phi/2}$
 \end{tabular}\right).
 \eeq
Um espinor $\xi$ de momento arbitrário $p^\m$ pode ser obtido a partir do referencial próprio
\be
\xi(p^\m)=e^{i{\bf \kappa}\cdot{\bf \varphi}}\xi(k^\m),
\ee
se escrevemos o operador de transformação em termos do operador helicidade ${\bf \s}\cdot {\bf p}$ \cite{bht,Cavalcanti:2015nna}:
\be e^{i\kappa\cdot \varphi} =\sqrt{\f{E+m}{2m}}\left(\begin{tabular}{cc}
${\bf 1}_2+\f{\bf \s\cdot p}{E+m}$&${\bf 0}_2$\\
${\bf 0}_2$&${\bf 1}_2-\f{\bf \s\cdot p}{E+m}$
\end{tabular}\right),
\ee
%
onde $\cosh\,(\varphi)=E/m,\; \sinh\,(\varphi)=p/m,$ $\hat{ \bf \varphi}=\hat{\bf p}$ e $\kappa=\left(\begin{tabular}{cc}
$-i{\bf \s}/2$&${\bf 0}$\\
${\bf 0}$&$i{\bf \s}/2$
\end{tabular}\right).$ Calculamos isso para o Elko e obtemos após o {\it boost}, que:
\beq
\l_\pm^S(p^\m)=\sqrt{\f{E+m}{2m}}\left(1\mp\f{p}{E+m}\right)\l_\pm^S(k^\m),\\
\l_\pm^A(p^\m)=\sqrt{\f{E+m}{2m}}\left(1\pm\f{p}{E+m}\right)\l_\pm^A(k^\m).
\eeq

As equações de movimento do Elko são um sistema de equações do tipo Dirac acopladas \cite{Elkodirac}:
\beq
\g_\m p^\m\l^S_+(p^\m)&=&im\l^S_-(p^\m),\\
\g_\m p^\m\l^S_-(p^\m)&=&-im\l^S_+(p^\m),\\
\g_\m p^\m\l^A_-(p^\m)&=&im\l^A_+(p^\m),\\
\g_\m p^\m\l^A_+(p^\m)&=&-im\l^A_-(p^\m).
\eeq 
Se combinadas essas equações, resulta que o Elko é aniquilado pelo operador de Klein-Gordon \cite{bht,Cavalcanti:2015nna}. Além disso, ele pode ser expresso em componentes \cite{bht,Cavalcanti:2015nna}:
\be
\l^S_+=\left(\begin{tabular}{c}
$-i\b^*$\\$-\a^*$ \\$ \a$\\$\b$
\end{tabular}\right), \quad 
\l^S_-=\left(\begin{tabular}{c}
$-i\a$ \\$-i\b$ \\ $-i\b^*$\\ $\a^*$
\end{tabular}\right) ,\quad 
\l^A_+=\left(\begin{tabular}{c}
$i\a$\\$i\b$\\$-\b^*$\\$\a^*$
\end{tabular}\right) ,\quad 
\l^A_-=\left(\begin{tabular}{c}
$-i\b^*$\\ $i\a^*$\\ $-\a$\\ $-\b$
\end{tabular}\right) ,\quad 
\ee
onde $\a$ e $\b$ são campos escalares funções do espaço-tempo.

\end{document}